\documentclass[%
 reprint,
superscriptaddress,
 amsmath,amssymb,
 aps,
 onecolumn,
pra,
floatfix,
]{revtex4-2}
\usepackage{graphicx}
\usepackage{dcolumn}
\usepackage{bm}
\usepackage{subcaption}
\usepackage{float}
\usepackage{xcolor}
\usepackage[hidelinks,colorlinks=true, linkcolor = blue,
            urlcolor  = blue,
            citecolor = blue,
            anchorcolor = blue]{hyperref}

\begin{document}

\preprint{APS/123-QED}

\title{Flow-induced vibration of a flexible cantilever in tandem configuration}

\author{Shayan Heydari}
\email[Corresponding author. Email address: ]{sheydari@mail.ubc.ca}
\affiliation{Department of Mechanical Engineering, The University of British Columbia, Vancouver, BC Canada V6T 1Z4}

\author{Rajeev K. Jaiman}
\email[Email address: ]{rjaiman@mech.ubc.ca}
\affiliation{Department of Mechanical Engineering, The University of British Columbia, Vancouver, BC Canada V6T 1Z4}

\date{\today}

\begin{abstract}
The present work investigates the fluid-structure interaction (FSI) of a flexible cylindrical cantilever in a tandem configuration. A fully coupled fluid-structure solver based on the three-dimensional incompressible Navier-Stokes equations and Euler-Bernoulli beam equation is employed to numerically examine the coupled dynamics of the cantilever. We assess the extent to which such a flexible cylindrical structure could sustain oscillations in both subcritical and post-critical regimes of Reynolds number ($Re$). Spatio-temporal power transfer patterns, response amplitudes, and vorticity dynamics are quantified and compared between isolated and tandem configurations. Results of our analysis indicate that the cantilever in tandem configuration is prone to sustained oscillations dependent on the Reynolds number and the reduced velocity parameter ($U^*$). In the subcritical $Re$ regime, the cantilever exhibits sustained oscillations with peak transverse oscillation amplitudes occurring within a specific range of $U^*$. Within this range, the transverse oscillations demonstrate lock-in behavior and synchronization with the vortex shedding frequency. The vorticity dynamics in the subcritical $Re$ regime reveal that in the tandem configuration, the presence of the upstream cylinder significantly modifies the wake structure, delaying vortex formation and extending the near-wake region. In the post-critical regime of $Re$, the cantilever shows a broader range of sustained oscillations in terms of $U^*$, with single- and multi-frequency response dynamics driven by vortex-body interactions. The power transfer analysis shows cyclic energy exchange patterns between the fluid and flexible structure, with significant variations in the hydrodynamic loading along the length of the cantilever. The findings of this work aim to broaden the understanding of sustained oscillations in flexible cantilevers and provide physical insights into the design of artificial cantilever flow sensors for various engineering applications.
\end{abstract}

\keywords{Flow-induced vibration, Lock-in, Vortex-induced vibration, Wake-induced vibration, Tandem cylinder configuration}
\maketitle
\section{\label{sec:introduction}Introduction\protect}
Fluid-structure interactions are ubiquitous in nature and various engineering applications. One intriguing manifestation of FSIs occurs when a flexible body with a bluff cross-section is exposed to a steady incident flow perpendicular to its length. When the Reynolds number, which is based on the body's characteristic dimension ($D$) and free-stream velocity ($U_0$), exceeds a critical value, often approximated as $Re_{c}\approx45$~\cite{jackson_1987}, the shedding of von Kármán vortices can induce sustained oscillations in the body, a phenomenon widely known as vortex-induced vibrations (VIVs). The study of VIVs has garnered substantial attention over the past few decades, primarily due to the intricate vortex dynamics and nonlinear physics involved. 
%
During VIVs, vibration amplitudes are known to exhibit bell-shaped trends as functions of the reduced velocity parameter $U^*$, with peak amplitudes reaching the order of $O(D)$ in the transverse direction~\cite{Williamson2004, Sarpkaya2004}. These peak amplitudes often occur within a range of \(U^*\), known as the synchronization or lock-in regime, where fluid forces and the body's motion oscillate at the same frequency, which is generally close to one of the body's natural frequencies.
%
In marine and offshore engineering, VIVs are particularly important due to their impact on the lifespan of various installations. Structures such as risers, pipelines, and mooring cables are susceptible to VIVs, leading to fatigue damage and structural failure in the long run. Consequently, substantial efforts have been directed toward understanding VIVs and developing effective vibration mitigation strategies~\cite{Yao_Jaiman_2017} to ensure the safety and reliability of these structures. 

The dynamics of VIVs become markedly more intricate when multiple structures are placed in close proximity and allowed to vibrate freely as a coupled system. Several studies have investigated the flow dynamics around circular cylinders in a group arrangement to explore the resultant flow patterns across various configurations. The pioneering works in this topic include the research by Igarashi~\cite{IGARASHI1981,IGARASHI1984}, who investigated the flow field of two geometrically similar stationary circular cylinders in a tandem arrangement, a common configuration for studying nonlinear wake-cylinder interference effects, and Zdravkovich~\cite{ZDRAVKOVICH1987}, who conducted extensive research on the flow morphologies of stationary circular cylinders in a group arrangement. Various flow patterns that emerge due to hydrodynamic interference between two stationary cylinders are generally classified into three categories~\cite{ZDRAVKOVICH1987}: for separation distances \(1.0 < x_{0}/D < 1.8\), where \(x_{0}\) is the distance between cylinder centers, the cylinders behave as a single body, forming a single wake known as the extended bluff-body regime. When the spacing increases to \(1.2 \sim 1.8 < x_{0}/D < 3.8\), the shear layers from the upstream cylinder reattach to the downstream cylinder, creating a force opposite to the free stream on the downstream cylinder. This regime is termed the reattachment regime. For distances \(x_{0}/D > 3.8\), known as the co-shedding regime, vortex shedding occurs in the gap between the cylinders and interacts with the vortices from the downstream cylinder. 

The body of research on the flow dynamics of circular cylinders in group arrangements, as reviewed in~\cite{SUMNER2010, ZHOU2016}, extends to the flow-induced vibration (FIV) of freely oscillating circular cylinders subjected to hydrodynamic interference. The studies on this topic include investigations where a flexibly mounted rigid cylinder is positioned in the wake of an identical stationary cylinder, as well as configurations involving two freely vibrating rigid cylinders in tandem arrangement. Numerous studies, including~\cite{MITTAL2001, PAPAIOANNOU2008833, PRASANTH2009731, ASSI_BEARMAN_MENEGHINI_2010, XU2019375}, among others, have collectively underscored the critical interplay between cylinder spacing, structural flexibility, and vortex-shedding dynamics in determining the vibrational behavior of cylinders. The hydrodynamic interference between two cylinders is typically classified into two distinct categories: VIVs and wake-induced vibrations (WIVs). In the case of WIVs, coherent vortical structures generated by the upstream cylinder impinge upon the downstream body, inducing fluctuations in the fluid forces. A key distinguishing feature of WIVs is that these force fluctuations exhibit a lack of synchronization with the motion of the downstream body~\cite{ASSI_BEARMAN_MENEGHINI_2010}, thus presenting a more intricate fluid-structure interaction scenario compared to the well-understood VIV phenomenon. Research by Mysa et al.~\cite{MYSA201676} investigated the origin of wake-induced vibrations in tandem circular cylinders and pinpointed the origin of vibrations induced by wake vortices.

The published works have made considerable efforts toward understanding the fundamental physical mechanisms of freely vibrating rigid cylinders in a group arrangement. As for the flexible multi-cylinder systems, however, the research is much more limited. Some early experimental works~\cite{BRIKA1997, BRIKA1999} on flow-induced oscillations in a long flexible circular cylinder immersed in the wake of an identical stationary cylinder have shown that the flexible cylinder shows a steady-state oscillatory response with a wider synchronization regime that narrows with increased spacing between the cylinders. Experimental studies~\cite{HUERAHUARTE2011-2, HUERAHUARTE2011} on the FIV response of two tandem flexible cylinders have shown that the response of the downstream cylinder could be VIV, WIV, or combinations of both, depending on the gap distance between the cylinders and the reduced velocity parameter $U^*$. The amplitude response of a flexible cylinder towed behind a stationary cylinder was investigated in~\cite{HUERAHUARTE2016571}. They showed that the amplitude response of the flexible cylinder is largely influenced by synchronization regions at different structural modes, depending on $U^*$. 
The dynamics of a flexible cantilever cylinder when subjected to the unsteady flow and vortices, generated by a stationary cylinder placed upstream, was studied in~\cite{ALAM2023103901}. The cantilever's vibration regime was mapped on a plane defined by the two cylinders' diameter ratio and spacing. Additionally, the vibrations were shown to significantly impact the wake structure behind the downstream cylinder, generating two distinct scales of vortices associated with the vortex shedding frequency and the cylinder's vibration frequency.

\begin{figure}
\includegraphics[width=0.55\linewidth]{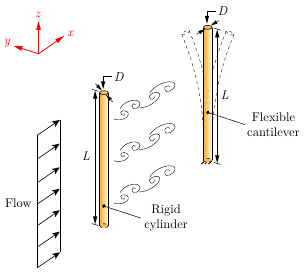}
\caption{\label{Schematic_Problem_Statement}Schematic of the flexible cantilever placed in the wake of a rigid stationary cylinder in tandem configuration.}
\end{figure}

Understanding the fluid-structure dynamics of flexible structures in a group arrangement is relevant to many modern industries, such as an array of wind turbines, long-span bridges, and industrial chimneys, and holds promise within the context of biological flow sensing. Many biological species employ mechanosensory hairs and other flow-sensing mechanisms, which generally take the form of long flexible cantilevers, to detect minute changes in flow patterns and differentiate flow features~\cite{Beem2015}. Investigating the complexities of flow-induced oscillations in such flexible structures makes it possible to inform the design of artificial flow-sensing instruments, tasked with measuring flow information and differentiating various flow patterns. In this context, it is essential to have a comprehensive knowledge of the coupled dynamics of flexible cantilevers in various fluid-structure contexts. Building upon existing research in flow-induced vibrations, our current study investigates the coupled dynamics of a long, flexible cylindrical cantilever positioned downstream of an identical stationary cylinder. This configuration allows us to provide novel physical insights into the coupled dynamics of flexible cylinders subject to hydrodynamic interference effects.
%
To examine the cantilever's response, we employ a three-dimensional numerical framework. We take the cantilever as a flexible cylindrical structure of length $L$ and a constant circular cross-section of diameter $D$. A rigid stationary cylinder of the same geometry is positioned in the upstream region in a tandem arrangement, as shown in Fig.~\ref{Schematic_Problem_Statement}. The non-dimensional parameters considered in this work include mass ratio ($m^\mathrm{*}$), Reynolds number, and reduced velocity given by:
\begin{align}
	m^\mathrm{*} = \frac{4 m}{{\pi D^{2}\rho^\mathrm{f}}}, \qquad
	Re = \frac{\rho^\mathrm{f}U_\mathrm{0}D}{\mu^\mathrm{f}}, \qquad
	U^\mathrm{*} &= \frac{U_\mathrm0}{f_\mathrm{n}D},
\end{align}
where $m$ is the cantilever's mass per unit length, $\rho^\mathrm{f}$ and $\mu^\mathrm{f}$ are the density and dynamic viscosity of the fluid, respectively, $U_0$ is the magnitude of the free-stream flow velocity, and $f_\mathrm{n}$ is cantilever's first-mode natural frequency, taking into account the effect of added fluid mass. We take the mass ratio as $m^*=1$, and the Reynolds number is kept within $20\leq Re\leq100$. For a given Reynolds number, the reduced velocity $U^*$ is varied between $U^*\in[1,19]$ by adjusting the free-stream flow velocity.
%

Our focus on the coupled dynamics of the flexible cantilever at low Reynolds numbers stems from recent advancements in understanding the fluid-structure interaction of isolated cylinders in the laminar subcritical regime of $Re$, i.e., $Re < Re_{c} \approx 45$. While flow-induced oscillations of cylinders have been thoroughly investigated at moderate to high Reynolds numbers, recent studies have revealed intriguing phenomena in the laminar subcritical $Re$ regime. Notably, several studies on the FIV of isolated flexible cylinders~\cite{Heydari_Jaiman_2022, bourguet_2020, buffoni} and elastically-mounted rigid cylinders~\cite{Miyanawala2019, sadeghi_2021, cossu_instability_2000, mittal_singh_2005, dolci_carmo_2019, meliga_chomaz_2011} have revealed the emergence of lock-in behavior and periodic vortex shedding in the laminar subcritical regime of $Re$, challenging the conventional understanding that vortex shedding occurs only in the post-critical regime of $Re$. Further investigation is required to fully understand the impact of hydrodynamic interference on the coupled response of flexible cylinders at such low Reynolds numbers. To address this knowledge gap, the present study investigates the flow-induced oscillation of a flexible cylindrical cantilever in a tandem configuration, aiming to provide a broader understanding of flow dynamics and structural vibrations in flexible cylinders. The investigations are conducted in both subcritical and post-critical Reynolds number regimes and are compared with an isolated flexible cantilever counterpart.
%
Three key questions are addressed in the present study: (i) How do Reynolds number $Re$ and reduced velocity $U^*$ influence the cantilever's oscillatory response? (ii) What are the characteristics of fluid flow in the spacing between two cylinders and the wake of the cantilever? and (iii) How does fluid-structure energy exchange contribute to the cantilever's dynamic response? 

The paper is structured as follows: Section~\ref{sec:numericalMethodology} discusses the governing equations for modeling the cantilever's coupled dynamics and presents the results of the grid convergence study. In Sec.~\ref{sec:results}, we detail the results and analyze the cantilever's response characteristics and wake behavior. Finally, Sec.~\ref{sec:conclusions} provides the conclusions and summarizes the key findings of our study.
%
\section{\label{sec:numericalMethodology} Problem description and numerical methodology}
In this section, we first introduce the physical setup for studying the cantilever's coupled dynamics. We then outline the employed numerical methodology, including the governing equations and the fluid-structure interaction framework. Finally, we present the grid convergence study results.
\subsection{Physical setup}
\begin{figure*}
\begin{subfigure}{0.9\textwidth}
\centering
\includegraphics[width=1\linewidth]{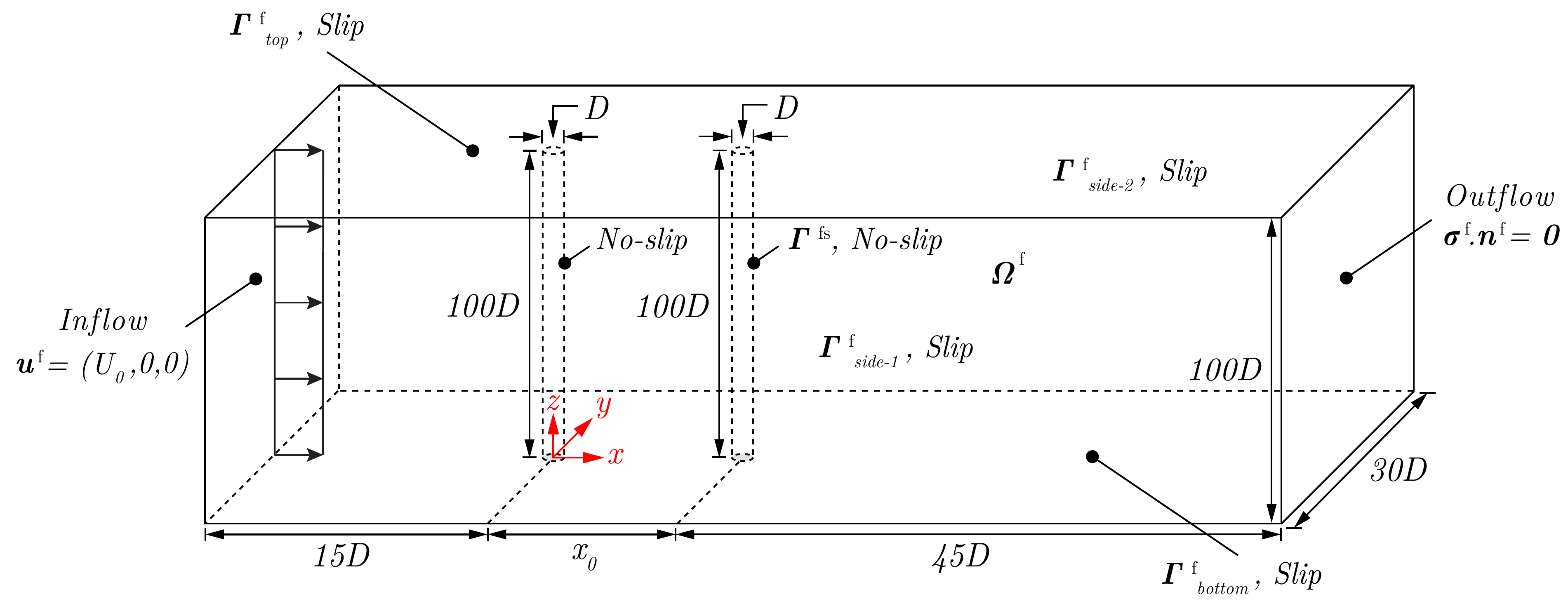}
\caption{}
\label{Beam_Schematic}
\end{subfigure}
\begin{subfigure}{0.3\textwidth}
\centering
\includegraphics[width=1\linewidth]{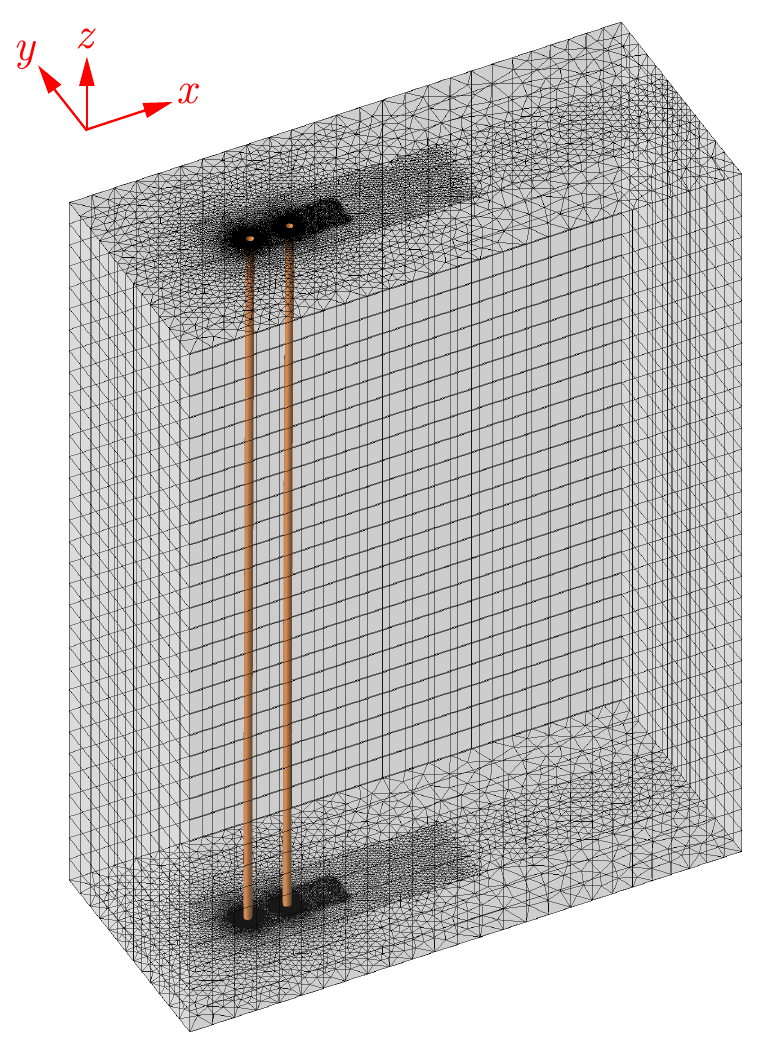}
\caption{}
\label{isoMesh}
\end{subfigure}%
\begin{subfigure}{0.6\textwidth}
\centering
\includegraphics[width=1\linewidth]{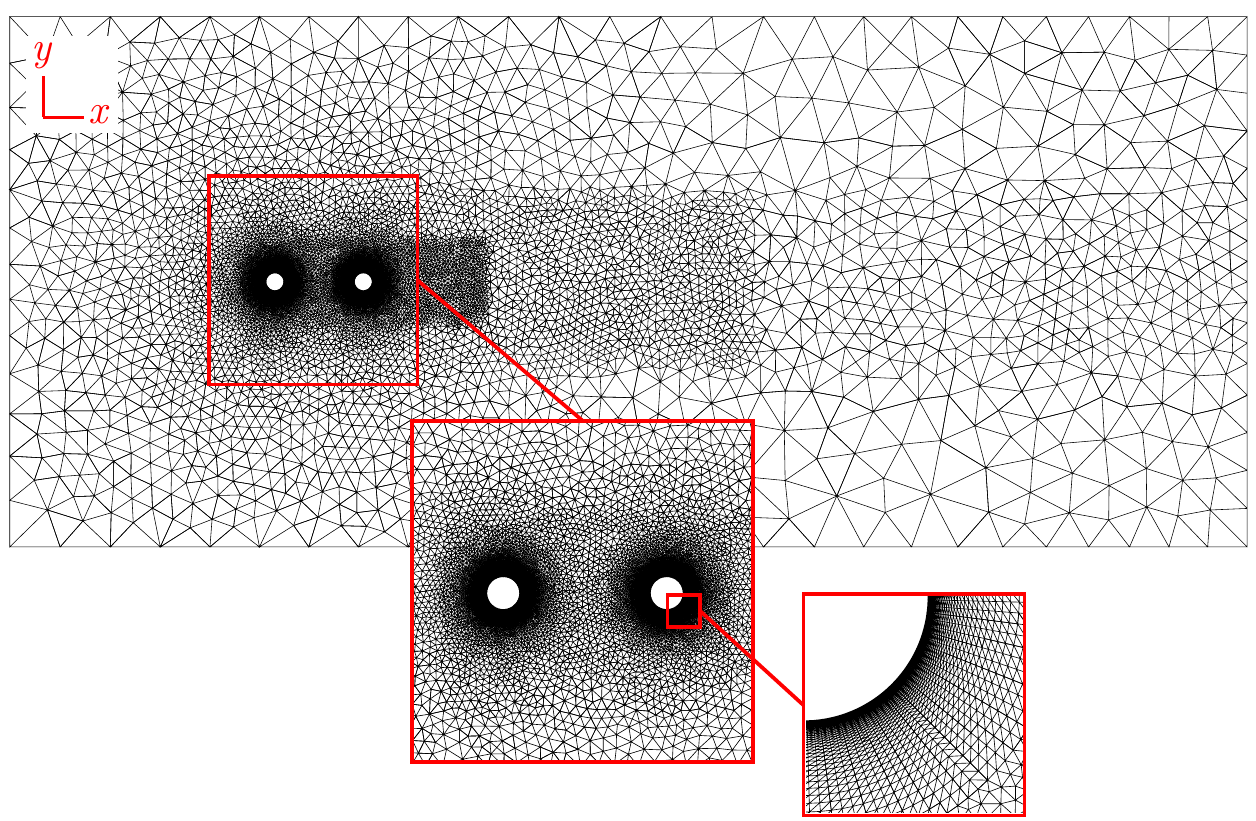}
\caption{}
\label{XYMesh}
\end{subfigure}
\caption{(a) Schematic of the computational domain, including domain size and boundary conditions; (b) isometric view of the computational grid, highlighting the tandem cylinders within the domain; (c) representative $z$-plane slice of the unstructured finite element grid with a close-up view of the boundary layer mesh.}
\label{mesh}
\end{figure*}

A schematic of the computational domain with details of the domain size and boundary conditions is given in Fig.~\ref{mesh}a. The cantilever is taken as a circular cylinder of diameter $D$ and length $L=100D$. A rigid stationary cylinder with a similar geometry is placed in the upstream region at a streamwise distance $x_{0}$ from the cantilever. The separation distances are set at \(x_{0}=5D\) or \(10D\), which exceed the critical threshold for bistable shear layer reattachment. These distances ensure that vortex shedding can develop in the gap between the cylinders. The rigid cylinder is positioned at a distance of $15D$ from the inflow surface, and the flexible cantilever is placed at a distance of $45D$ from the outflow surface. A fixed structural support condition is imposed at one end of the cantilever, corresponding to the position $z/L=0$, where the $z$-axis is oriented along the cantilever's length. The no-slip boundary condition is enforced on the surface of the cantilever, denoted as $\Gamma^\mathrm{fs}$, and the surface of the rigid stationary cylinder. A uniform flow of velocity $\boldsymbol{u}^\mathrm{f} = (U_{0},0,0)$ is applied at the inflow surface, and the slip boundary condition is imposed at the side, top, and bottom surfaces of the computational domain, as illustrated in Fig.~\ref{mesh}a. For the outflow surface, the traction-free boundary condition, given as $\boldsymbol{\sigma}^{\mathrm{f}}.\boldsymbol{n}^\mathrm{f} = \boldsymbol{0}$, is specified, where $\boldsymbol{\sigma}^{\mathrm{f}}$ is the Cauchy stress tensor for a Newtonian fluid and $\boldsymbol{n}^\mathrm{f}$ is the unit normal vector to the outflow surface.

The employed computational framework has been extensively verified and validated in previous studies across various FSI problems, including the flow past a three-dimensional flexible cylindrical beam experiencing a turbulent wake at high Reynolds numbers~\cite{Joshi2017} and flow over an isolated flexible cantilever~\cite{Heydari_Jaiman_2022}.
In the following, we present the governing equations for modeling the dynamics of the flexible cantilever and discuss the strategy used for coupling the fluid and structural solvers. We conclude the section with the results of the grid convergence study.
\subsection{Numerical methodology}
To study the dynamics of the flexible cantilever, we use the three-dimensional incompressible Navier-Stokes equations in combination with the Euler-Bernoulli beam equation. The Euler-Bernoulli beam equation is formulated in a Lagrangian reference frame, while the viscous incompressible fluid flow is addressed using a body-fitted moving boundary approach based on the arbitrary Lagrangian-Eulerian (ALE) description~\cite{Hughes1981}, ensuring accurate modeling of the boundary layer phenomena over the deformable surface of the structure.
\subsubsection{Unsteady Navier–Stokes equations in ALE framework}
The unsteady three-dimensional Navier–Stokes equations are employed to predict the flow dynamics. Using an ALE reference frame on the fluid domain $\Omega^\mathrm{f}(t)$, the Navier–Stokes equations are written as:
\begin{align} \label{DNS}
	\rho^\mathrm{f}\frac{\partial \boldsymbol{u}^\mathrm{f}}{\partial t}\bigg|_{\hat{x}^\mathrm{f}} + \rho^\mathrm{f}(\boldsymbol{u}^\mathrm{f} - {{\boldsymbol{u}^\mathrm{m}}})\cdot\nabla\boldsymbol{u}^\mathrm{f} &= \nabla\cdot \boldsymbol{\sigma}^\mathrm{f} + \boldsymbol{b}^\mathrm{f}\ \ \ \mathrm{on\ \ \Omega^\mathrm{f}(t)},\\
	\nabla\cdot\boldsymbol{u}^\mathrm{f} &= 0\ \ \ \mathrm{on\ \ \Omega^\mathrm{f}(t)},
\end{align}
where $\boldsymbol{u}^\mathrm{f} = \boldsymbol{u}^\mathrm{f}(\boldsymbol{x}^\mathrm{f},t)$ and $\boldsymbol{u}^\mathrm{m}=\boldsymbol{u}^\mathrm{m}(\boldsymbol{x}^\mathrm{f},t)$ are the fluid and mesh velocities related to every spatial point $\boldsymbol{x}^\mathrm{f} \in \Omega^\mathrm{f}$, respectively, $\boldsymbol{b}^\mathrm{f}$ represents the body force exerted on the fluid and $\boldsymbol{\sigma}^\mathrm{f}$ is the Cauchy stress tensor, given as:
\begin{align}
	\boldsymbol{\sigma}^\mathrm{f} = -p\boldsymbol{I} + \mu^\mathrm{f}( \nabla\boldsymbol{u}^\mathrm{f} + (\nabla\boldsymbol{u}^\mathrm{f})^T),
\end{align}
where $p$ denotes the fluid pressure. The first term in Eq.~\ref{DNS} represents the partial derivative of $\boldsymbol{u}^\mathrm{f}$ with respect to time while the ALE referential coordinate $\hat{x}^\mathrm{f}$ is kept fixed.
The fluid forcing acting on the cantilever's surface is calculated by integrating the surface traction on the fluid-solid interface $\Gamma^\mathrm{fs}$. The instantaneous coefficients of the lift and drag forces are then computed as:
\begin{align}
	C_\mathrm{L} = \frac{1}{\frac{1}{2}\rho^\mathrm{f}U_\mathrm{0}^\mathrm{2}DL}\int_{\Gamma^\mathrm{fs}} (\boldsymbol{\sigma}^\mathrm{f}\cdot \boldsymbol{n})\cdot \boldsymbol{n}_\mathrm{y} \mathrm{d\Gamma}, \quad
C_\mathrm{D} = \frac{1}{\frac{1}{2}\rho^\mathrm{f}U_\mathrm{0}^\mathrm{2}DL}\int_{\Gamma^\mathrm{fs}} (\boldsymbol{\sigma}^\mathrm{f}\cdot \boldsymbol{n})\cdot \boldsymbol{n}_\mathrm{x} \mathrm{d\Gamma},
\end{align}
where $\boldsymbol{n}_\mathrm{x}$ and $\boldsymbol{n}_\mathrm{y}$ are the Cartesian components of the unit outward normal vector $\boldsymbol{n}$.
A stabilized Petrov-Galerkin finite element method is employed to discretize the fluid domain $\Omega^f$ into $n_{el}$ non-overlapping finite elements $\Omega^e$ in space such that $\Omega^f = \bigcup_{e=1}^{n_{el}} \Omega^e$. The employed computational grid consists of unstructured prism elements, with a boundary layer mesh around the cylinders, as shown in Figs.~\ref{mesh}b and \ref{mesh}c. The results for the grid convergence study are detailed in Sec.~\ref{sec:gridConvergence}.
%
\subsubsection{\label{subsec:Euler-Bernoulli beam eqn}Modal analysis for a linear flexible body}
We model the structural dynamics through a linear modal analysis by solving the Euler-Bernoulli beam equation. Let $\Omega^\mathrm{s}$ be the structural domain consisting of structure coordinates $\boldsymbol{x}^\mathrm{s} = (x,y,z)$. The transverse displacements $\boldsymbol{w}^\mathrm{s}(z,t)$ are solved using the Euler-Bernoulli beam equation excited by a distributed unsteady fluid loading per unit length $\boldsymbol{f}^\mathrm{s}$. Neglecting the damping and shear effects, the equation of motion of the beam is written as:
\begin{align}\label{eq:beam_eqn}
	{m}\frac{\partial^2 \boldsymbol{w}^\mathrm{s}(z,t)}{\partial t^2} + EI\frac{\partial^4 \boldsymbol{w}^\mathrm{s}(z,t)}{\partial z^4}  = \boldsymbol{f}^\mathrm{s}(z,t),
\end{align}  
where $m=\rho^\mathrm{s} A$ is the mass per unit length of the beam, with $\rho^\mathrm{s}$ and $A$ being the density of the structure and the cross-sectional area, respectively. The Young's modulus and second moment of area of the beam are denoted by $E$ and $I$, respectively. Considering the cantilever configuration, the boundary conditions at the ends of the beam are given as:
\begin{align}
	\boldsymbol{w}^\mathrm{s}(z,t)|_{z=0} &= 0,\qquad
	\frac{\partial\boldsymbol{w}^\mathrm{s}(z,t)}{\partial z}\bigg|_{z=0} = 0, \\
 	\frac{\partial^2\boldsymbol{w}^\mathrm{s}(z,t)}{\partial z^2}\bigg|_{z=L} &= 0,\qquad
	\frac{\partial^3\boldsymbol{w}^\mathrm{s}(z,t)}{\partial z^3}\bigg|_{z=L} = 0.
\end{align}
To find a solution for Eq.~\ref{eq:beam_eqn}, we use a mode superposition approach, assuming small displacements from the mean position of the beam, characterized by the normal vibration modes. The frequency of the $\mathrm{i}^\mathrm{th}$ mode is given by:
\begin{align} \label{eq:natural_frequency}
	f_\mathrm{i} = \frac{{\lambda_\mathrm{i}}^2}{2\pi L^2}\sqrt{ \frac{EI}{m} },
\end{align}
in which $\lambda_\mathrm{i}$ is the nondimensional frequency parameter dependent on the mode number. For the current configuration, the eigenmodes are taken as the sums of sine, cosine, sinh, and cosh functions such that the eigenmode shape, denoted by $S_{\mathrm{i}}$, associated with the $\mathrm{{i}}^{\mathrm{th}}$ mode is written as:
\begin{align}\label{eq:modal_shape}
S_\mathrm{i}\left( z \right) = \cosh \left( \frac{\lambda_\mathrm{i}z} {L} \right) - \cos \left( \frac{\lambda_\mathrm{i} z} {L} \right) - \sigma_\mathrm{i} \sinh \left( \frac{\lambda_\mathrm{i} z} {L} \right) + \sigma_\mathrm{i} \sin \left( \frac{\lambda_\mathrm{i} z} {L} \right),
\end{align}
where $\sigma_\mathrm{i}$ is a nondimensional parameter dependent on the mode number (see~\cite{Blevins2016} for values of $\lambda_\mathrm{i}$ and $\sigma_\mathrm{i}$).
Equation~\ref{eq:beam_eqn} can be rewritten into a matrix form as:
\begin{align}\label{eq:beam_eqn_matrix}
	{\boldsymbol{M}^{\mathrm{s}}}\frac{\partial^2 \boldsymbol{w}^{\mathrm{s}}}{\partial t^2} + \boldsymbol{K}^{\mathrm{s}}\boldsymbol{w}^\mathrm{s}  = \boldsymbol{f}^{\mathrm{s}},
\end{align}  
where $\boldsymbol{M}^{\mathrm{s}}$ and $\boldsymbol{K}^{\mathrm{s}}$ represent the mass and stiffness matrices, respectively, $\boldsymbol{w}^\mathrm{s}$ is the vector of unknown displacements across the span of the beam, and $\boldsymbol{f}^{\mathrm{s}}$ is the vector of the force acting on the beam. By projecting this equation into the eigenspace using the eigenmodes defined by Eq.~\ref{eq:modal_shape}, we transform Eq.~\ref{eq:beam_eqn_matrix} into a system of linear equations, with '$\mathrm{i}$' degrees of freedom, i.e, modes, as:
\begin{align}\label{eq:beam_eqn_matrix_eigen}
	\widetilde{\boldsymbol{M}^{\mathrm{s}}}\frac{\partial^2 \boldsymbol{\zeta}^{\mathrm{s}}}{\partial t^2} + \widetilde{\boldsymbol{K}^{\mathrm{s}}}\boldsymbol{\zeta}^{\mathrm{s}}  = \widetilde{\boldsymbol{f}^{\mathrm{s}}}, 
\end{align}
where ${\widetilde{\boldsymbol{M}^{\mathrm{s}}}} = \boldsymbol{S^T}\boldsymbol{M}^{\mathrm{s}}\boldsymbol{S}$ and ${\widetilde{\boldsymbol{K}^{\mathrm{s}}}} = \boldsymbol{S^T}\boldsymbol{K}^{\mathrm{s}}\boldsymbol{S}$ represent the projected matrices onto the eigenspace, with $\boldsymbol{S}$ representing the matrix containing the eigenvectors. The vector of the modal responses is denoted by $\boldsymbol{\zeta}^{\mathrm{s}}$. The relationship between the displacement vector $\boldsymbol{w}^{\mathrm{s}}$ and the vector $\boldsymbol{\zeta}^{\mathrm{s}}$ is given by $\boldsymbol{w}^{\mathrm{s}} = \boldsymbol{S\zeta}^{\mathrm{s}}$. The vector ${\widetilde{\boldsymbol{f}^{\mathrm{s}}}}$ in the above equation is the projected force vector given by ${\widetilde{\boldsymbol{f}^{\mathrm{s}}}} = \boldsymbol{S^{T}f}^{\mathrm{s}}$. The uncoupled Eq.~\ref{eq:beam_eqn_matrix_eigen} is solved for the modal amplitudes $\boldsymbol{\zeta}^{\mathrm{s}}$ which is then used to find the structural displacement. The structure is considered to be undeformed at time $t=0$. The modal amplitudes for the subsequent time steps are evaluated using the generalized-$\alpha$ time integration scheme~\cite{Chung1993,Jansen1999}.
%
\subsubsection{Treatment of the fluid-solid interface}
Let $\Gamma^\mathrm{fs} = \partial \Omega^\mathrm{f}(0) \cap \partial \Omega^\mathrm{s}$ be the fluid-solid interface at $t=0$ and $\Gamma^\mathrm{fs}(t) = \boldsymbol{\varphi}^\mathrm{s}(\Gamma^\mathrm{fs},t)$ be the interface at time $t$. We need to satisfy the continuity of velocity and traction at the fluid-solid interface, which requires satisfying the following conditions:
\begin{align}
	\boldsymbol{u}^\mathrm{f}(\boldsymbol{\varphi}^\mathrm{s}(\boldsymbol{x}^\mathrm{s}_0,t),t) = \boldsymbol{u}^\mathrm{s}(\boldsymbol{x}^\mathrm{s}_0,t), \\
	\int_{\boldsymbol{\varphi}^\mathrm{s}(\gamma,t)} \boldsymbol{\sigma}^\mathrm{f}(\boldsymbol{x}^\mathrm{f},t)\cdot \boldsymbol{n} \mathrm{d\Gamma}(\boldsymbol{x}^\mathrm{f}) + \int_\gamma  \boldsymbol{t}^\mathrm{s} \mathrm{d}\Gamma = 0,
\end{align}
where, $\boldsymbol{\varphi}^\mathrm{s}$ represents the position vector that maps the initial position $\boldsymbol{x}^\mathrm{s}_0$ of the structure to its position at time $t$, i.e., $\boldsymbol{\varphi}^\mathrm{s}(\boldsymbol{x}^\mathrm{s},t) = \boldsymbol{x}^\mathrm{s}_0 + \boldsymbol{w}^\mathrm{s}(\boldsymbol{x}^\mathrm{s},t)$, $\boldsymbol{t}^\mathrm{s}$ is the fluid traction vector corresponding to the fluid force as
$\boldsymbol{f}^\mathrm{s}(z,t) = 
\int_{\Gamma^\mathrm{fs}} \boldsymbol{t}^\mathrm{s} \mathrm{d}\Gamma$, and  $\boldsymbol{u}^\mathrm{s}$ represents the structure's velocity at time $t$, defined as $\boldsymbol{u}^\mathrm{s} = \partial\boldsymbol{\varphi}^\mathrm{s}/\partial t$. Here, $\boldsymbol{n}$ denotes the outer normal to the interface, $\gamma$ represents any part of the interface $\Gamma^\mathrm{fs}$ in the reference configuration, $\mathrm{d\Gamma}$ is the differential surface area, and $\boldsymbol{\varphi}^\mathrm{s}(\gamma,t)$ corresponds to the fluid part at time $t$. These conditions ensure that the fluid velocity matches the velocity of the structure at the interface.
%
To account for changes in the fluid-solid interface location, the motion of each spatial point in the fluid domain is explicitly controlled to ensure the kinematic consistency of the discretized interface. The spatial points $\boldsymbol{x}^\mathrm{f} \in {\Omega}^{\mathrm{f}}(t)$ are adjusted within the fluid domain by solving the following equation:
\begin{align}
\nabla \cdot \boldsymbol{\sigma}^{\mathrm{m}} = 0, \label{eq:ALE}
\end{align}
where $\boldsymbol{\sigma}^{\mathrm{m}}$ is the stress experienced by the spatial points due to the strain induced by the interface deformation. Assuming that the fluid mesh behaves as an elastic material, $\boldsymbol{\sigma}^{\mathrm{m}}$ could be expressed as:
\begin{align}
\boldsymbol{\sigma}^{\mathrm{m}} = (1+k_{\mathrm{m}})\left[\nabla \boldsymbol{w}^{\mathrm{f}} + \left(\nabla \boldsymbol{w}^{\mathrm{f}}\right)^T + \left(\nabla \cdot \boldsymbol{w}^{\mathrm{f}}\right)\boldsymbol{I}\right],
\end{align}
where $k_{\mathrm{m}}$ is the local element-level mesh stiffness parameter, chosen based on element sizes, and $\boldsymbol{w}^{\mathrm{f}}$ represents the ALE mesh nodal displacement. This approach ensures the movement of internal finite element nodes maintains mesh quality, even under large structural displacements. Additionally, a Lagrangian finite element method is employed to adjust the mesh according to the updated domain geometry.
A nonlinear partitioned iterative approach~\cite{Jaiman2016,Jaiman2016_2} is used to couple the fluid and structure equations. At each time step, the fluid traction exerted on the structure's surface is projected onto the eigenvectors to calculate the generalized modal forces. These projected modal forces are then used to determine the modal amplitudes and displacements for the subsequent time step. For a comprehensive review of the numerical algorithm and implementation details, interested readers are referred to~\cite{Jaiman2022}.
%
\subsection{\label{sec:gridConvergence}Grid convergence study}
For the convergence study, we begin with a relatively coarse mesh, labeled M1, and progressively double the number of elements to create the M2 and M3 meshes. Figures~\ref{mesh}b and \ref{mesh}c provide an isometric view of the discretized domain and a $z$-plane slice of the unstructured grid, respectively. The grid convergence study examines the coupled dynamics of the cantilever at $Re = 40$, $m^* = 1$, and $U^* = 9$ in the tandem configuration with $x_{0} = 5D$. Grid convergence errors are calculated using the results of the finest mesh, M3, as the reference. Table~\ref{tab:meshConvergence} presents the values for the frequency ratio ($f_\mathrm{y}/f_\mathrm{n}$), mean streamwise deformation ($\overline{A_\mathrm{x}}/D$), root-mean-square (rms) of the nondimensional transverse vibration amplitude ($A_\mathrm{y}^{rms}/D$), mean drag coefficient ($\overline{C_\mathrm{D}}$), and rms of the lift coefficient ($C_\mathrm{L}^{rms}$).
According to Table~\ref{tab:meshConvergence}, the relative errors using the M2 mesh are less than $2.5\%$; thus, the M2 mesh is selected as the suitable grid for our current work. It is also worth noting that in our previous numerical study on the FSI of an isolated flexible cantilever~\cite{Heydari_Jaiman_2022}, we examined the influence of the gap ratio on the cantilever's oscillatory response. The gap ratio, defined as \(\delta_{z}/L\), where \(\delta_{z}\) is the distance from the free end of the cantilever to the top surface of the computational domain, was found to have a negligible effect on the cantilever's response. Therefore, we maintain the length of the domain in the \(z\) direction equal to the length of the cantilever to optimize computational costs.
\begin{table*}
\caption{\label{tab:meshConvergence}%
Grid convergence study results at $Re = 40$, $m^* = 1$, and $U^* = 9$. The values in parentheses indicate the relative error compared to the M3 mesh results. A constant time-step size of $\Delta t = 0.005$ is used.
}
\begin{ruledtabular}
\begin{tabular}{lccc}
 &
\textrm{M1}&
\textrm{M2}&
\textrm{M3}\\
\colrule
Number of nodes & 211,992 & 404,151 & 812,196\\
Number of elements & 407,424 & 776,864 & 1,562,944\\
Frequency ratio $f_{\mathrm{y}}/f_\mathrm{n}$ & 0.9874 (2.00\%)
 & 0.9975 (1.00\%) & 1.0076\\
Mean streamwise deformation $\overline{A_\mathrm{x}}/D$  & 0.7374 (5.64\%) & 0.7631 (2.35\%) & 0.7815\\
rms of transverse vibration amplitude $A_\mathrm{y}^{rms}/D$   & 0.5032
(6.28\%) & 0.5245 (2.31\%) & 0.5369\\
Mean drag coefficient $\overline{C_\mathrm{D}}$ & 0.6026
(3.43\%) & 0.6167 (1.17\%) & 0.6240\\
rms of lift coefficient $C_\mathrm{L}^{rms}$ & 0.0629 (73.00\%) & 0.2277 (2.27\%) & 0.2330\\
\end{tabular}
\end{ruledtabular}
\end{table*}
%
\section{\label{sec:results} Results and discussion}
This section presents a comprehensive analysis of the cantilever's response characteristics and the associated vorticity dynamics in the subcritical and post-critical regimes of $Re$. The results are detailed for both isolated and tandem configurations. Key findings are discussed throughout this section, and a map of the cantilever's response dynamics in terms of \(Re\) and \(U^*\) is provided at the end, emphasizing the key observations and trends identified in this work.
\subsection{\label{subsec:subcritical} Coupled dynamics in subcritical Reynolds number}
\subsubsection{\label{subsubsec:subcritical-responseCharacteristics} Response characteristics in subcritical Re regime}
\begin{figure*}
\begin{subfigure}{0.55\textwidth}
\centering
\includegraphics[width=1\linewidth]{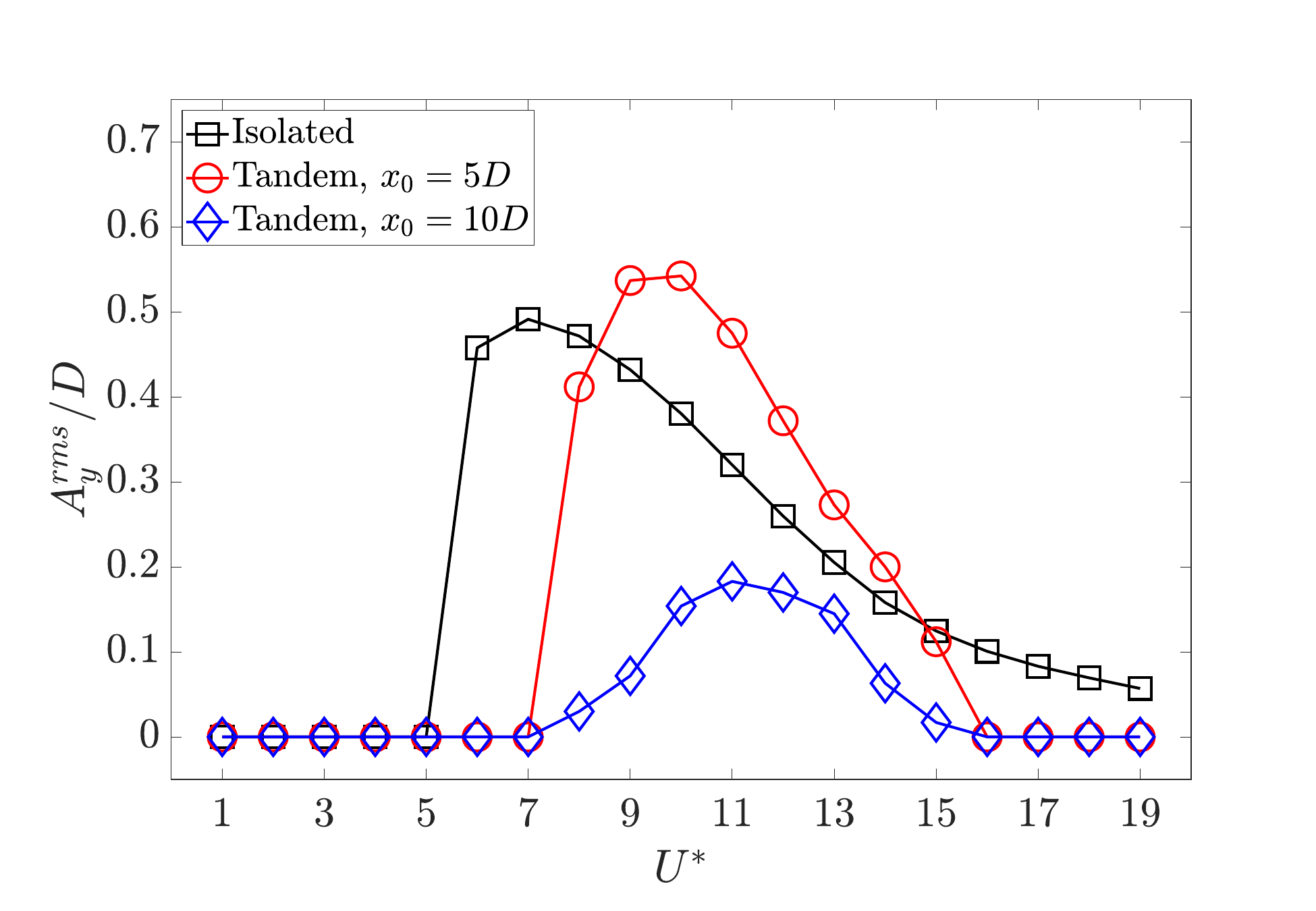}
\caption{}
\end{subfigure}
\begin{subfigure}{0.55\textwidth}
\centering
\includegraphics[width=1\linewidth]{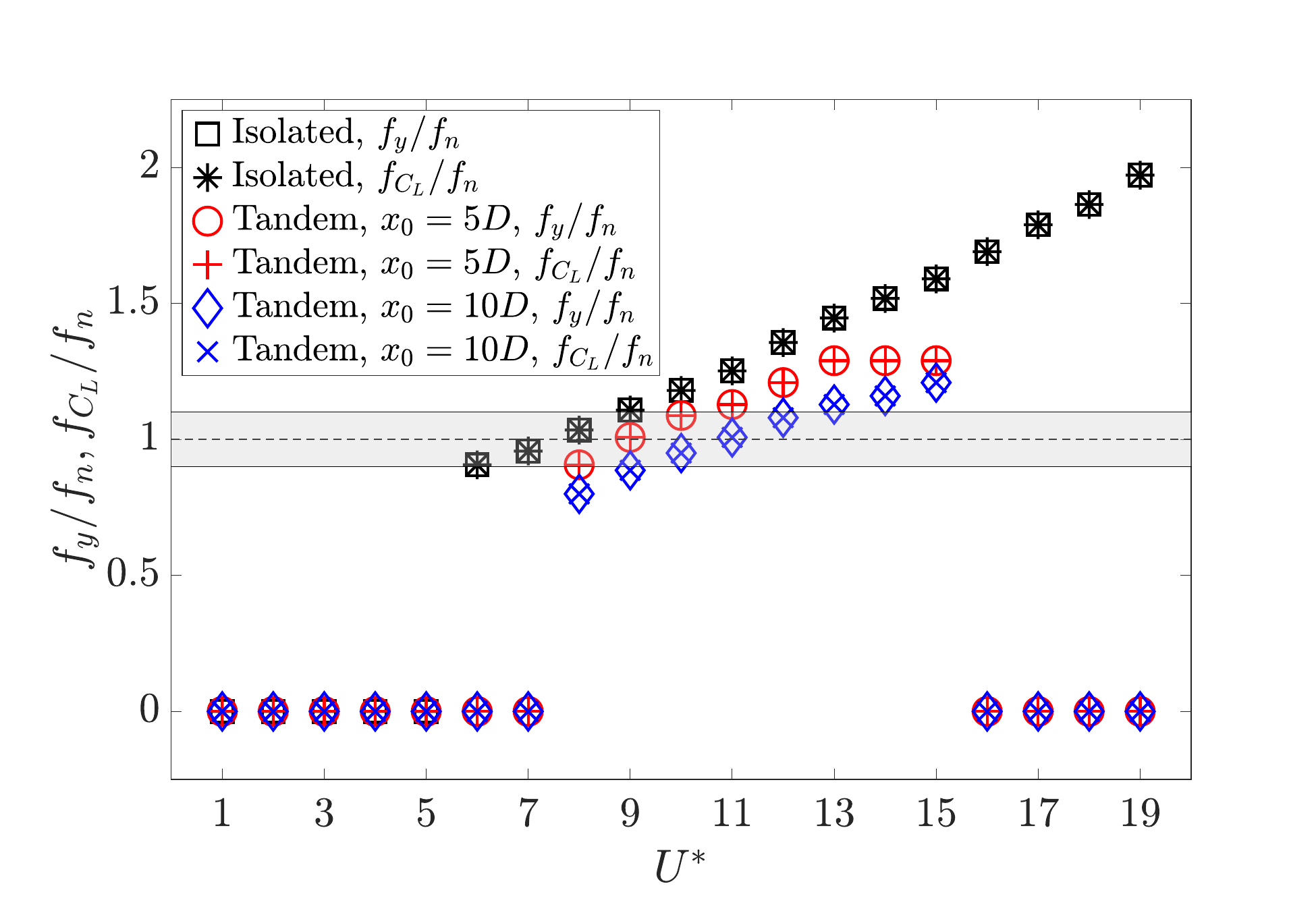}
\caption{}
\end{subfigure}
\caption{\label{Response_Characteristics_subcritical}(a) Comparison between the value of $A_\mathrm{y}^{rms}/D$ at the free end of the cantilever in isolated versus tandem configurations; (b) variations of $f_\mathrm{y}/f_\mathrm{n}$ at the free end of the cantilever and $f_\mathrm{C_L}/f_\mathrm{n}$ with respect to $U^*$. The results for the isolated configuration are derived from~\cite{Heydari_Jaiman_2022}.}
\end{figure*}

We begin by examining the response characteristics of the flexible cantilever at a subcritical Reynolds number $Re=40$ and mass ratio $m^*=1$ for reduced velocities $U^*\in[1,19]$. We focus on characterizing the root-mean-square value of the nondimensional transverse oscillation amplitude $A_\mathrm{y}^{rms}/D$, along with the frequency of the oscillations and the fluid loading frequency, as functions of $U^*$.
Figure~\ref{Response_Characteristics_subcritical}a compares the amplitude of transverse oscillations across three distinct configurations: the isolated configuration, and the tandem configurations with the streamwise distance between the cylinders $x_{0}$ taken either as $x_{0}=5D$ or $x_{0}=10D$. 
In all three configurations, the cylinders have a similar circular cross-section of diameter $D$ and an aspect ratio $L/D=100$. According to Fig.~\ref{Response_Characteristics_subcritical}a, the isolated cantilever exhibits a sustained oscillatory response for $U^*>5$. The amplitude of the transverse oscillation reaches a maximum value of $A_\mathrm{y}^{rms}/D\approx0.50$ at $U^*=7$, gradually decreasing beyond this $U^*$. 
For the tandem case with $x_{0}=5D$, the cantilever remains in its steady deflected position, i.e., no sustained oscillations, when $U^*\leq7$. Within $U^*\in(7,16)$, the cantilever undergoes sustained oscillations, as depicted in Fig.~\ref{Response_Characteristics_subcritical}a. In this configuration, the maximum transverse oscillation amplitude is marginally higher than the isolated case, reaching a value of $A_\mathrm{y}^{rms}/D \approx 0.55$ at $U^*=10$. Compared to the isolated configuration, the cantilever in the tandem cases experiences a sharp drop in the amplitude of oscillations beyond the peak point, and no sustained oscillations are observed for $U^*\geq16$. 
The vibrational amplitudes for the tandem configuration with $x_{0}=10D$ remain relatively low across all examined reduced velocities, reaching a maximum value of $A_\mathrm{y}^{rms}/D \approx 0.20$ at $U^*=11$, as seen in Fig.~\ref{Response_Characteristics_subcritical}a. In this configuration, sustained oscillations are present within $U^*\in(7,16)$, similar to the tandem case with $x_{0}=5D$. 

\begin{figure*}
\begin{subfigure}{0.55\textwidth}
\centering
\includegraphics[width=1\linewidth]{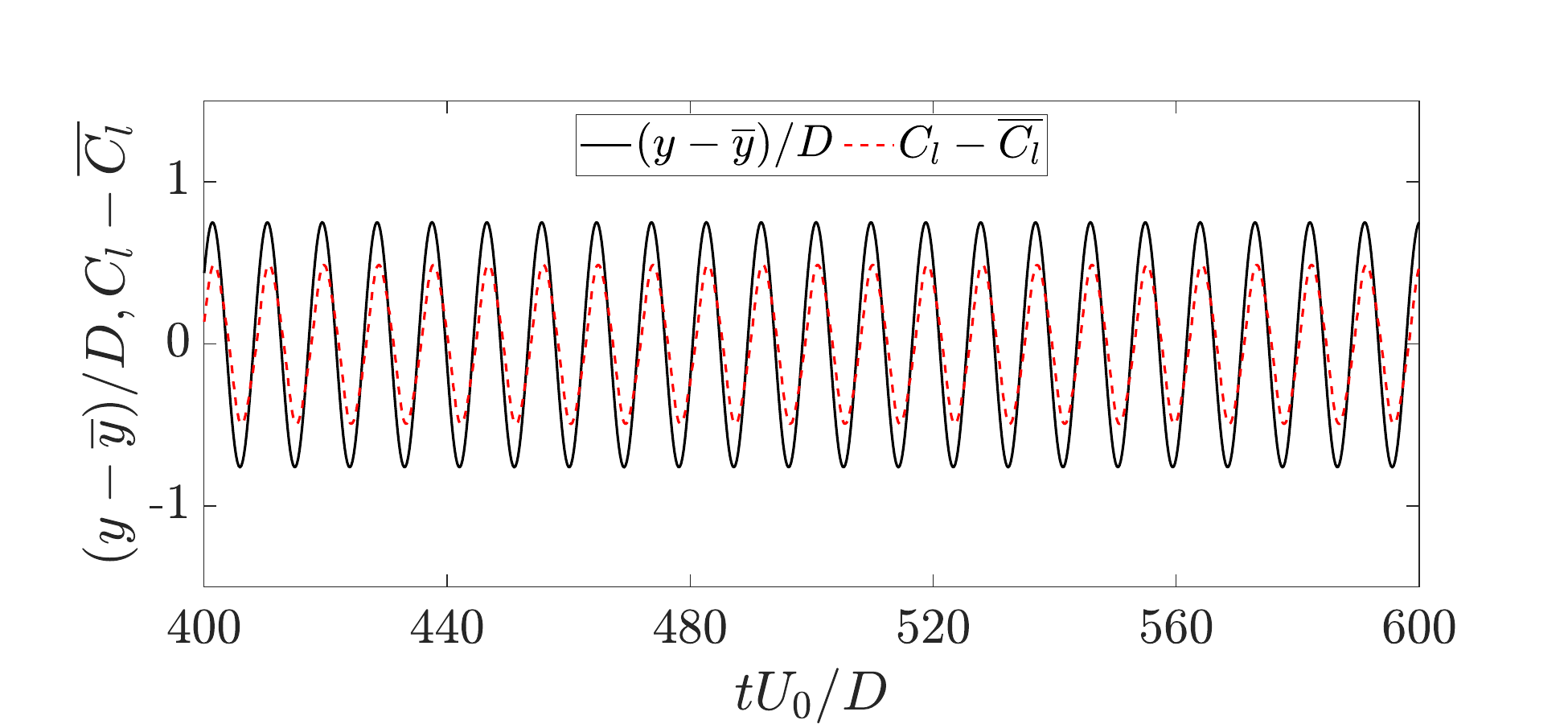}
\caption{}                             
\end{subfigure}
\begin{subfigure}{0.6\textwidth}
\centering
\includegraphics[width=1\linewidth]{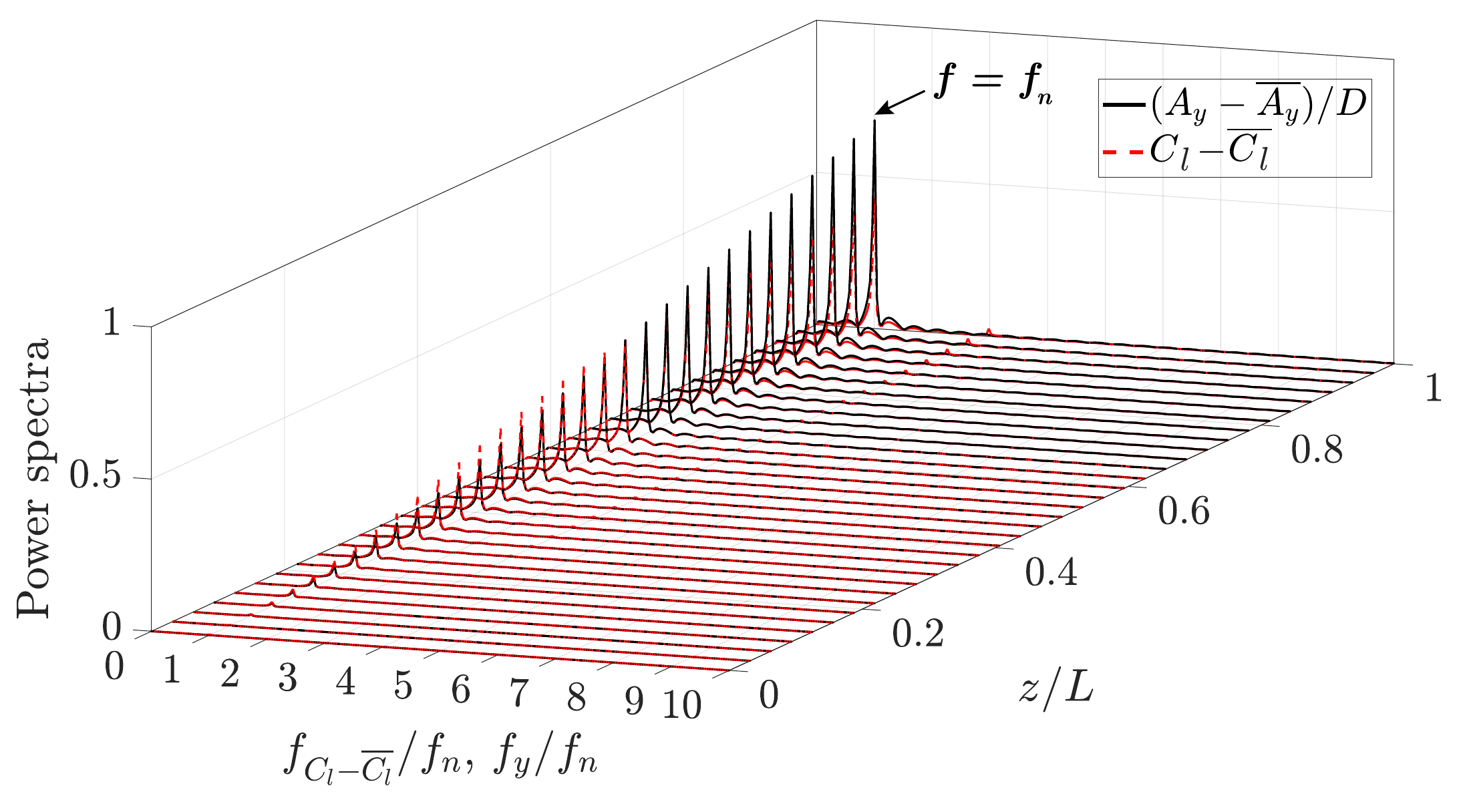}
\caption{}
\end{subfigure}
\caption{(a) Variations of the nondimensional transverse displacement $(y-\overline{y})/{D}$ and the cross-sectional lift coefficient $C_{l}-\overline{C_{l}}$ at the free end of the cantilever in the time domain; (b) power spectra of the nondimensional amplitude of transverse oscillations $(A_\mathrm{y}-\overline{A_\mathrm{y}})/{D}$ and $C_{l}-\overline{C_{l}}$ along the cantilever length in the frequency domain. The results are gathered for the tandem case with $x_{0}=5D$ in the time window $tU_{0}/D\in[400, 600]$ at $Re = 40$, $m^* = 1$, and $U^* = 9$. The overline $\overline{(.)}$ denotes the mean value.}
\label{CLAy-main}
\end{figure*}

The response curves presented in Fig.~\ref{Response_Characteristics_subcritical}a indicate that the flexible cantilever in the tandem configuration is prone to sustained oscillations, similar to an isolated cantilever, at \( Re = 40 \). The amplitude of transverse oscillations exhibits an inverse relationship with the spacing between the tandem cylinders, while the range of sustained oscillations remains within the same \( U^* \) range for both tandem configurations. Figure~\ref{Response_Characteristics_subcritical}b provides the values for the nondimensional transverse vibration frequency ($f_{\mathrm{y}}/f_\mathrm{n}$) and the nondimensional lift coefficient frequency ($f_\mathrm{C_L}/f_\mathrm{n}$) as functions of $U^*$. In all the instances where the flexible cantilever exhibits sustained oscillations, a frequency match is observed between the frequency of the transverse oscillation and the frequency of the lift coefficient. A typical lock-in behavior is found to be present during the cantilever's oscillatory response. 
To gain a more comprehensive understanding of the cantilever's response dynamics, we have presented the time histories of the nondimensional transverse displacement ($(y-\overline{y})/{D}$) and the cross-sectional lift coefficient ($C_{l}-\overline{C_{l}}$) at the free end of the cantilever for the tandem configuration with $x_{0}=5D$ at $U^*=9$ in Fig.~\ref{CLAy-main}a. According to the time series plot, the transverse displacement exhibits a time-periodic behavior in phase with the lift coefficient. For all the studied cases, we find that there is a match between the frequency of the transverse oscillation and the frequency of the cross-sectional lift coefficient across the length of the cantilever, as shown in Fig.~\ref{CLAy-main}b for the representative tandem case at $U^*=9$.

\begin{figure*}
\begin{subfigure}{0.4\textwidth}
\centering
\includegraphics[width=1\linewidth]{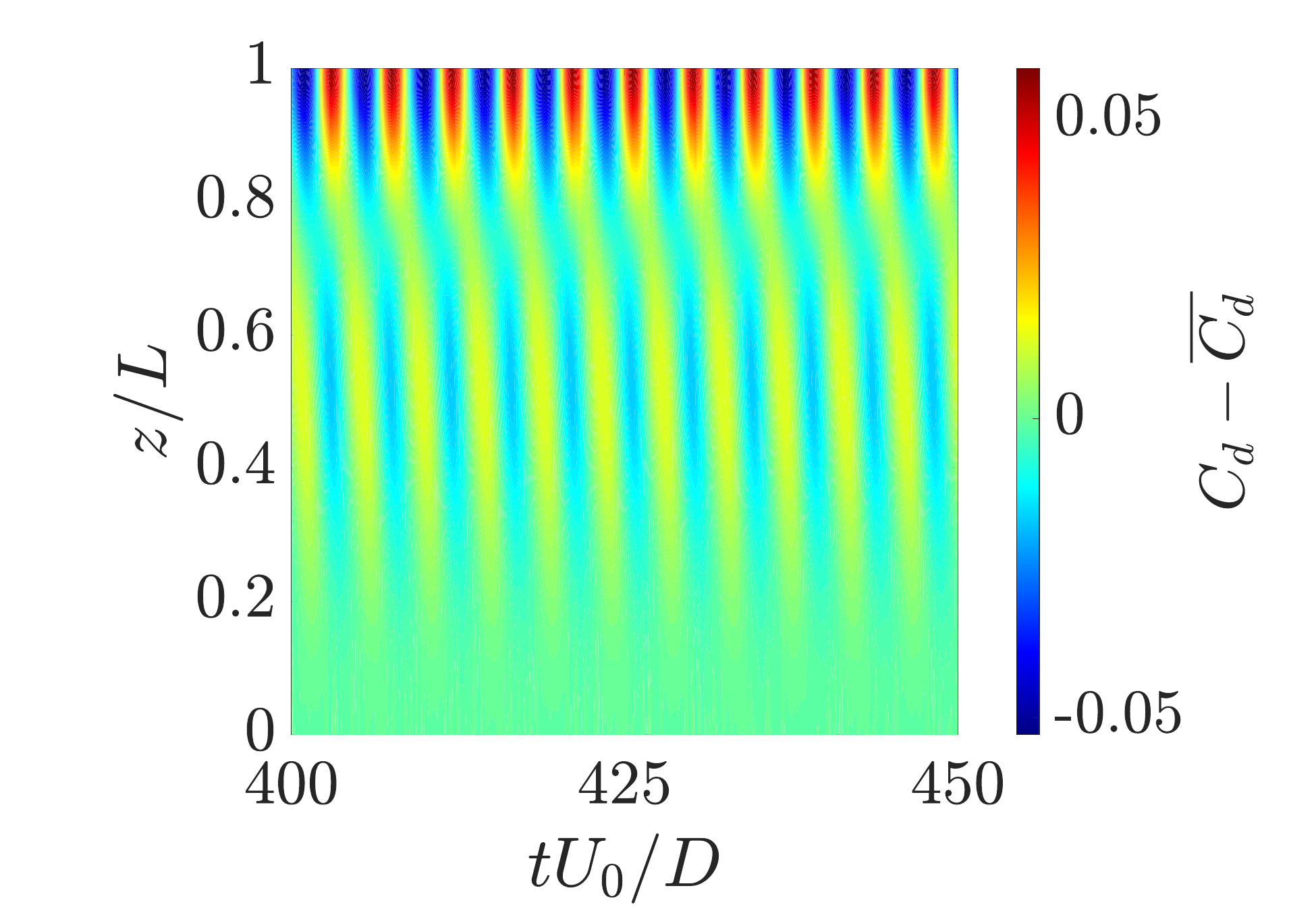}
\caption{}
\end{subfigure}%
\begin{subfigure}{0.4\textwidth}
\centering
\includegraphics[width=1\linewidth]{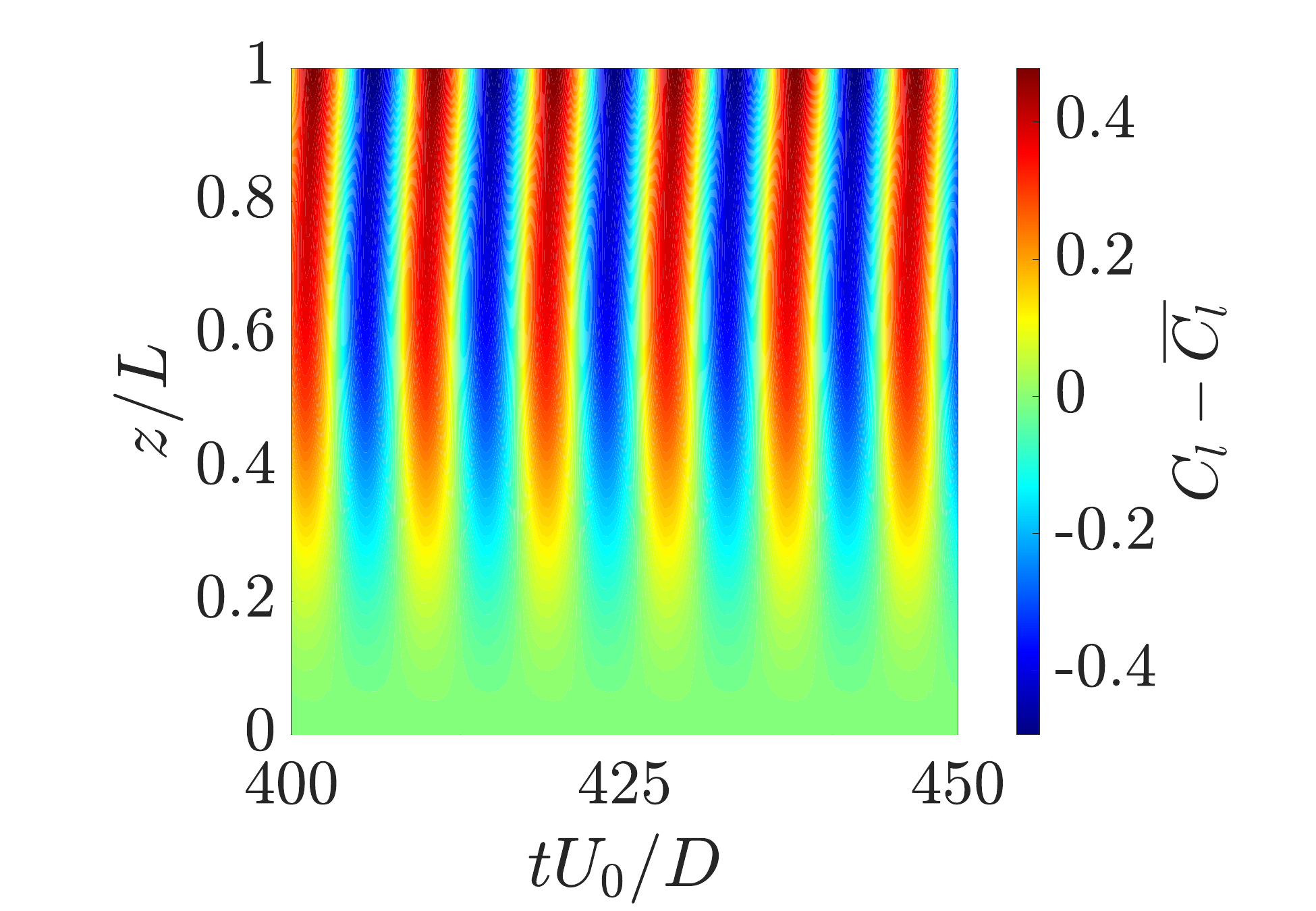}
\caption{}
\end{subfigure}
\begin{subfigure}{0.4\textwidth}
\centering
\includegraphics[width=1\linewidth]{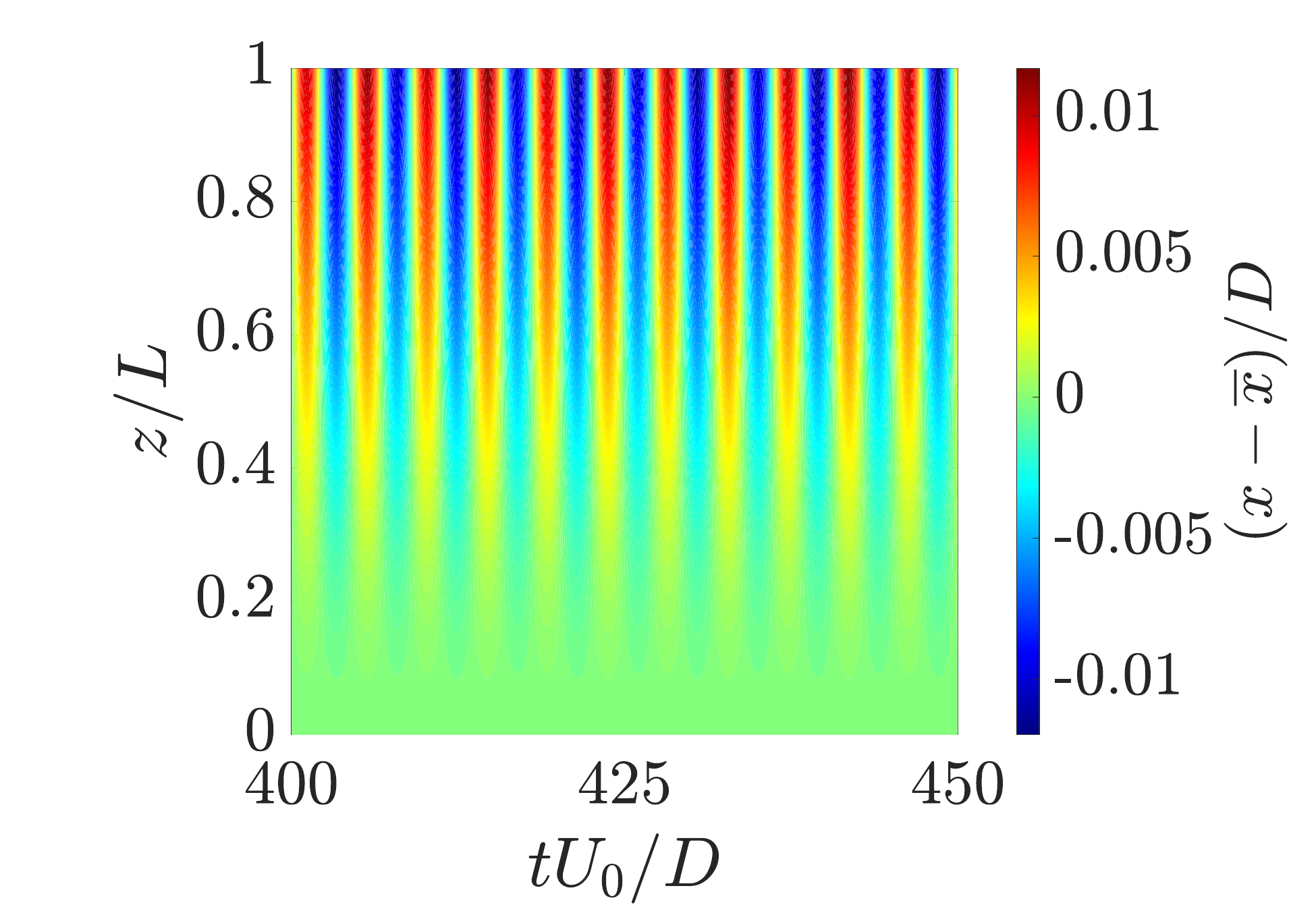}
\caption{}
\end{subfigure}%
\begin{subfigure}{0.4\textwidth}
\centering
\includegraphics[width=1\linewidth]{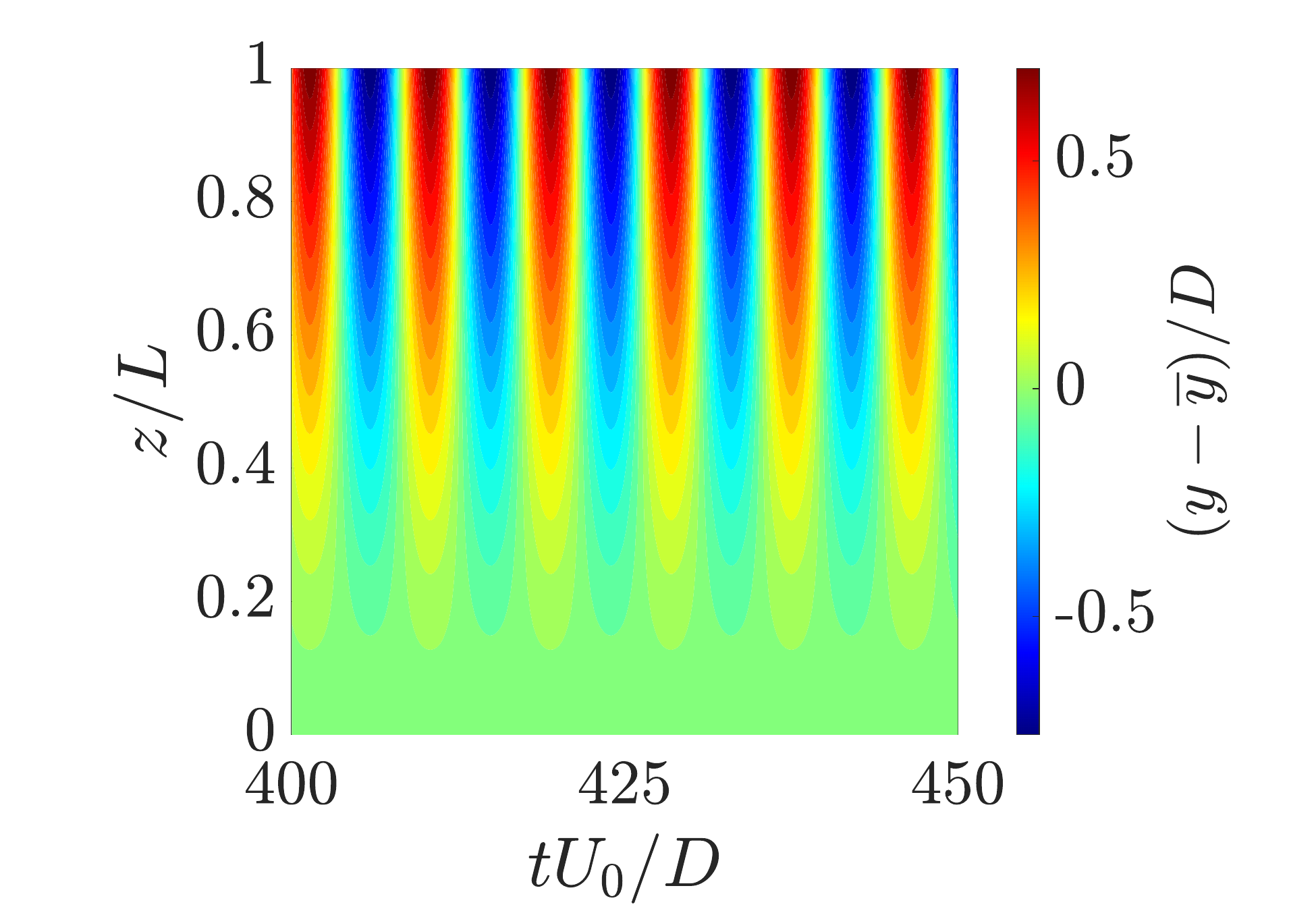}
\caption{}
\end{subfigure}
\caption{(a) Variations of the cross-sectional drag coefficient $C_{d}-\overline{C_{d}}$ and (b) cross-sectional lift coefficient $C_{l}-\overline{C_{l}}$ along the cantilever length in the time domain; (c) variations of the nondimensional streamwise displacement $(x-\overline{x})/{D}$ and (d) nondimensional transverse displacement $(y-\overline{y})/{D}$ across the cantilever length. The results are shown for the tandem case with \(x_{0} = 5D\) in the time window $tU_{0}/D\in[400, 450]$ at $Re = 40$, $m^* = 1$, and $U^* = 9$.}
\label{Scalogram-FFT-Spanwise}
\end{figure*}

\begin{figure*}
    \begin{subfigure}{0.5\textwidth}
    \centering
    \includegraphics[width=1\linewidth]{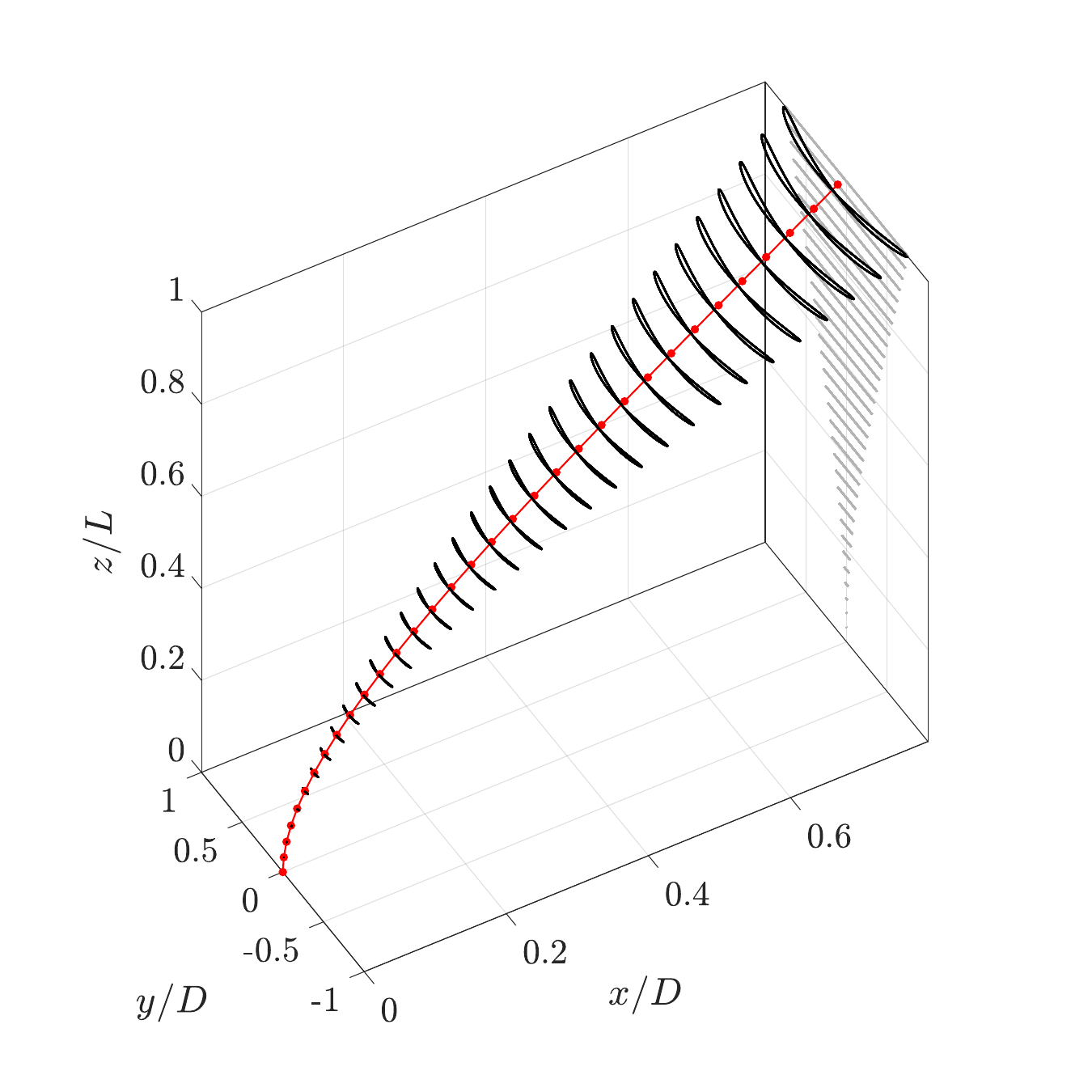}
    \caption{}
    \label{trajectory-5D-U9}
    \end{subfigure}
    \begin{subfigure}{0.25\textwidth}
    \centering
    \includegraphics[width=0.85\linewidth]{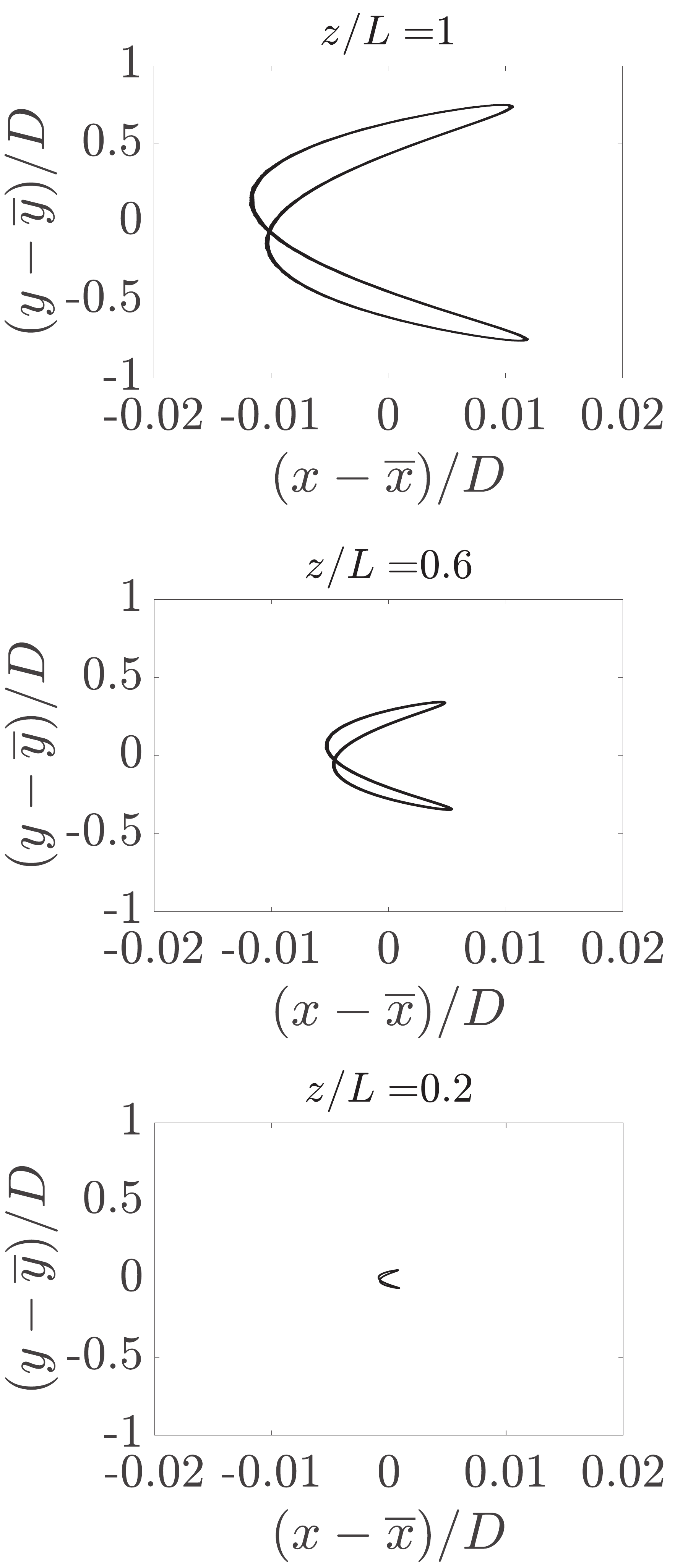}
    \caption{}
    \label{trajectory-5D-U9-zoomed}
    \end{subfigure}%
\caption{\label{trajectory}(a) Isometric view of the cantilever's motion trajectory (shown in solid black lines); the red filled dots represent the mean position of the probing nodes along the cantilever length, and the red solid line shows the cantilever's mean deformed position. The grey line projections into the $yz$-plane show superimposed snapshots of the cantilever's profile in the transverse direction. (b) $z$-plane slices of motion trajectories at different spanwise locations. The results are gathered at $Re = 40$, $m^* = 1$, and $U^* = 9$.}
\end{figure*}

Next, we investigate the spatio-temporal response of the cantilever and the fluid loading coefficients in both the streamwise and transverse directions along the cantilever length. The results for the representative tandem case at \(U^* = 9\) are illustrated in Fig.~\ref{Scalogram-FFT-Spanwise}. Figures~\ref{Scalogram-FFT-Spanwise}a and ~\ref{Scalogram-FFT-Spanwise}b present the variations of the cross-sectional drag and lift coefficients, respectively, in the time domain. According to Figs.~\ref{Scalogram-FFT-Spanwise}a and ~\ref{Scalogram-FFT-Spanwise}b, the variations in the lift coefficient are more pronounced than the drag coefficient, reflecting the stronger influence of transverse fluid forces on the cantilever's oscillatory response. 
The variations of the nondimensional streamwise and transverse displacements in the time domain across the length of the cantilever are shown in Figs.~\ref{Scalogram-FFT-Spanwise}c and ~\ref{Scalogram-FFT-Spanwise}d, respectively. The streamwise displacement shows small amplitude oscillations, reaching peak magnitudes of about \(0.01D\), while the transverse displacement exhibits significant oscillations, with peak magnitudes of approximately \(0.75D\). The larger amplitude of the transverse displacement compared to the streamwise displacement indicates that the oscillatory motion along the cantilever's length is primarily in the transverse direction, with the lift force exerting a significant influence on the cantilever's response.
An isometric view of the cantilever's motion trajectory in the tandem configuration with $x_{0}=5D$ at $U^* = 9$ is illustrated in Fig.~\ref{trajectory}a. A figure-eight motion trajectory is observed along the cantilever's length, characterized by dominant vibration amplitudes in the transverse direction. This pattern is consistent across all cases where the cantilever exhibits sustained oscillations. In these instances, the streamwise vibration frequency is twice the transverse vibration frequency along the cantilever's length, resulting in the observed figure-eight shape trajectories.

\begin{figure*}
\begin{subfigure}{0.3\textwidth}
\centering
\includegraphics[width=1\linewidth]{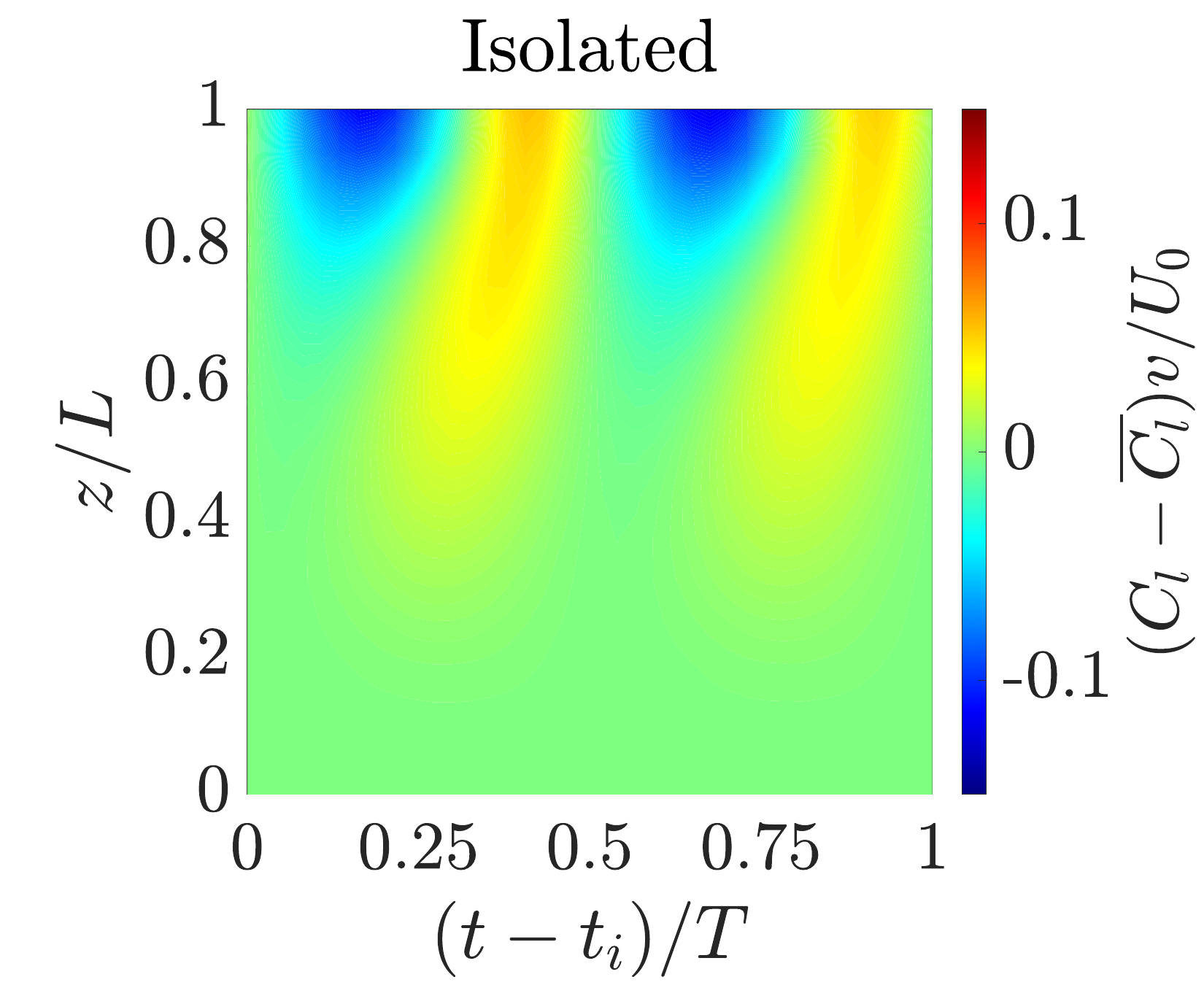}
\caption{}
\end{subfigure}%
\begin{subfigure}{0.3\textwidth}
\centering
\includegraphics[width=1\linewidth]{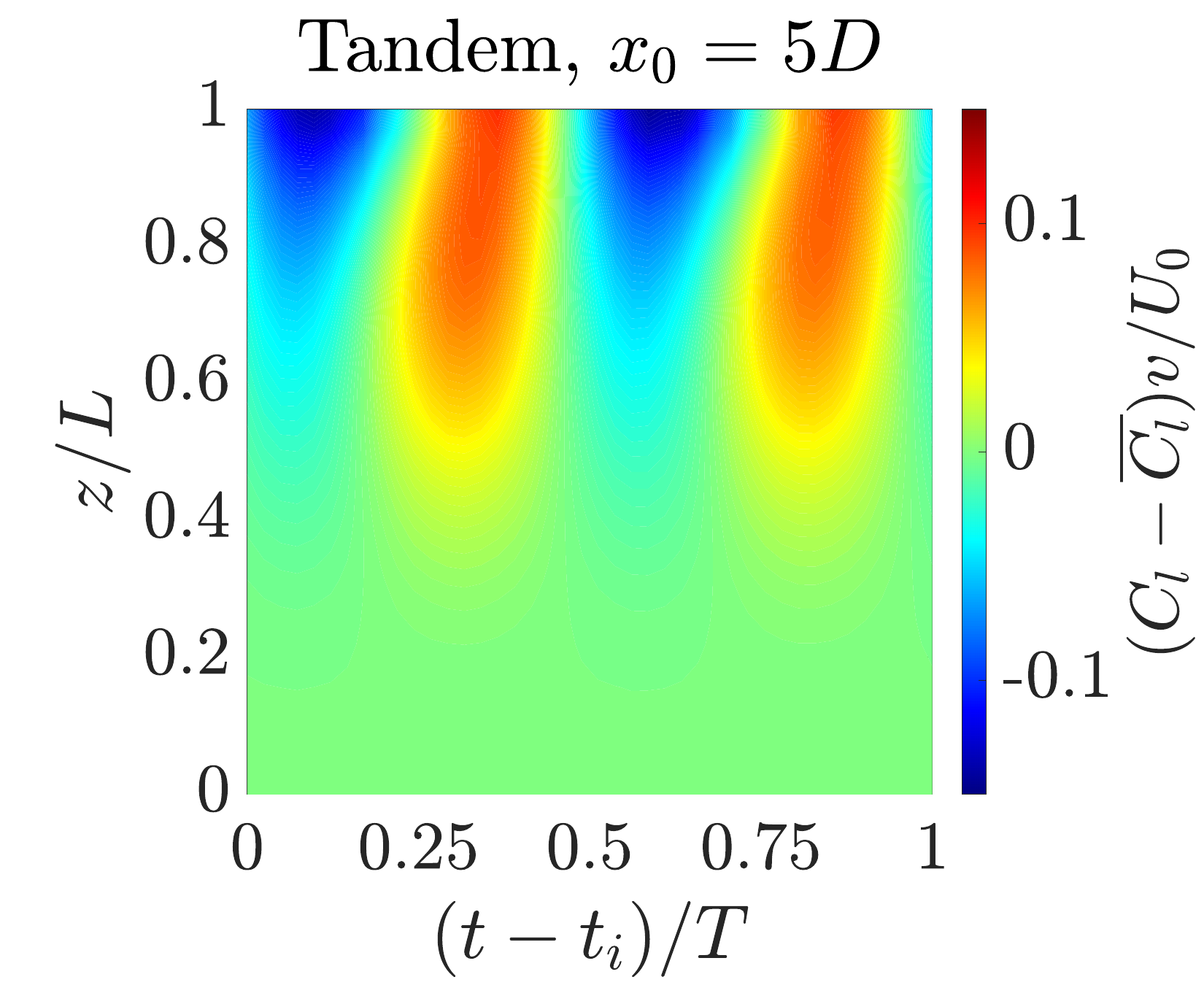}
\caption{}
\end{subfigure}%
\begin{subfigure}{0.3\textwidth}
\centering
\includegraphics[width=1\linewidth]{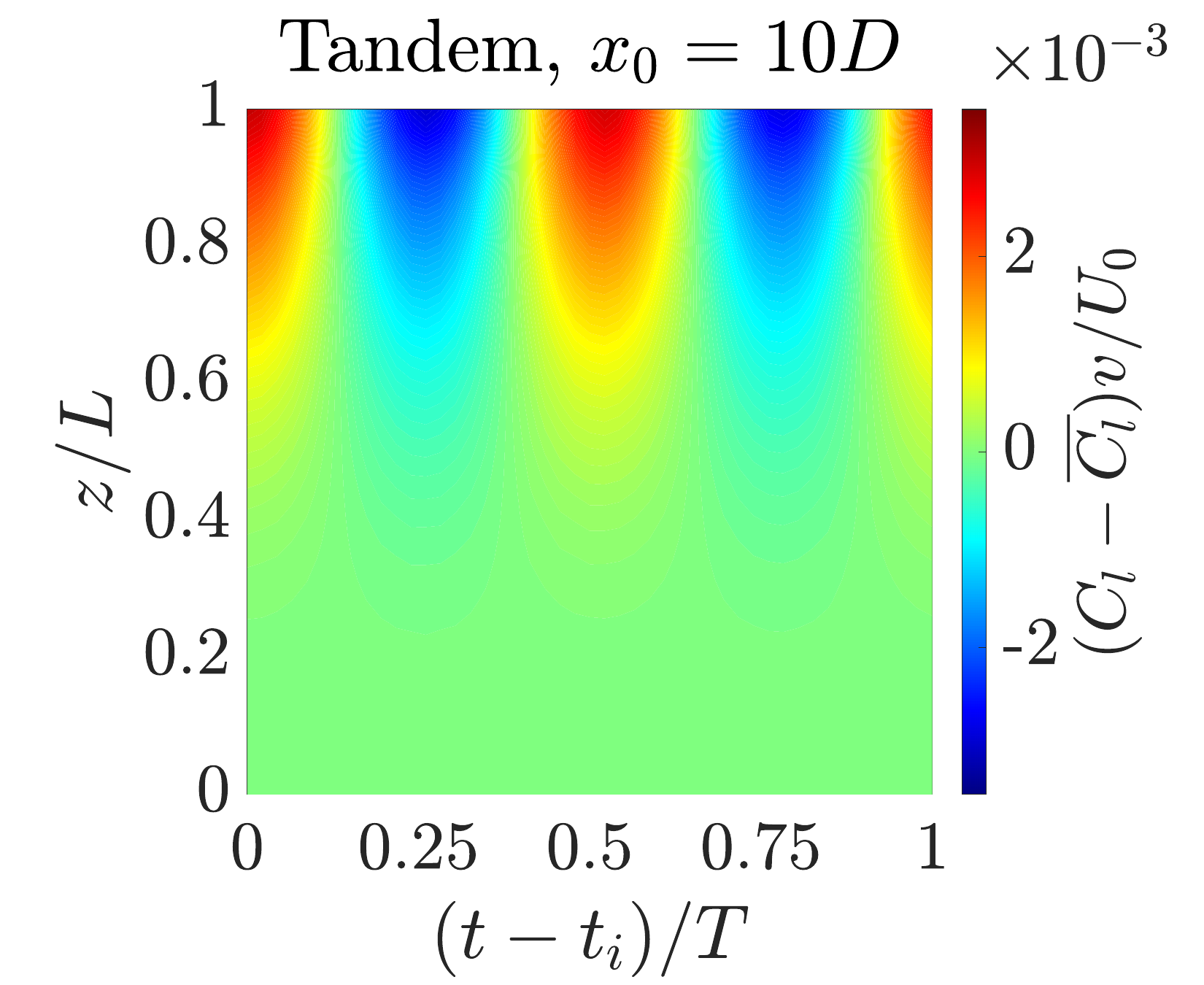}
\caption{}
\end{subfigure}
\begin{subfigure}{0.55\textwidth}
\centering
\includegraphics[width=1\linewidth]{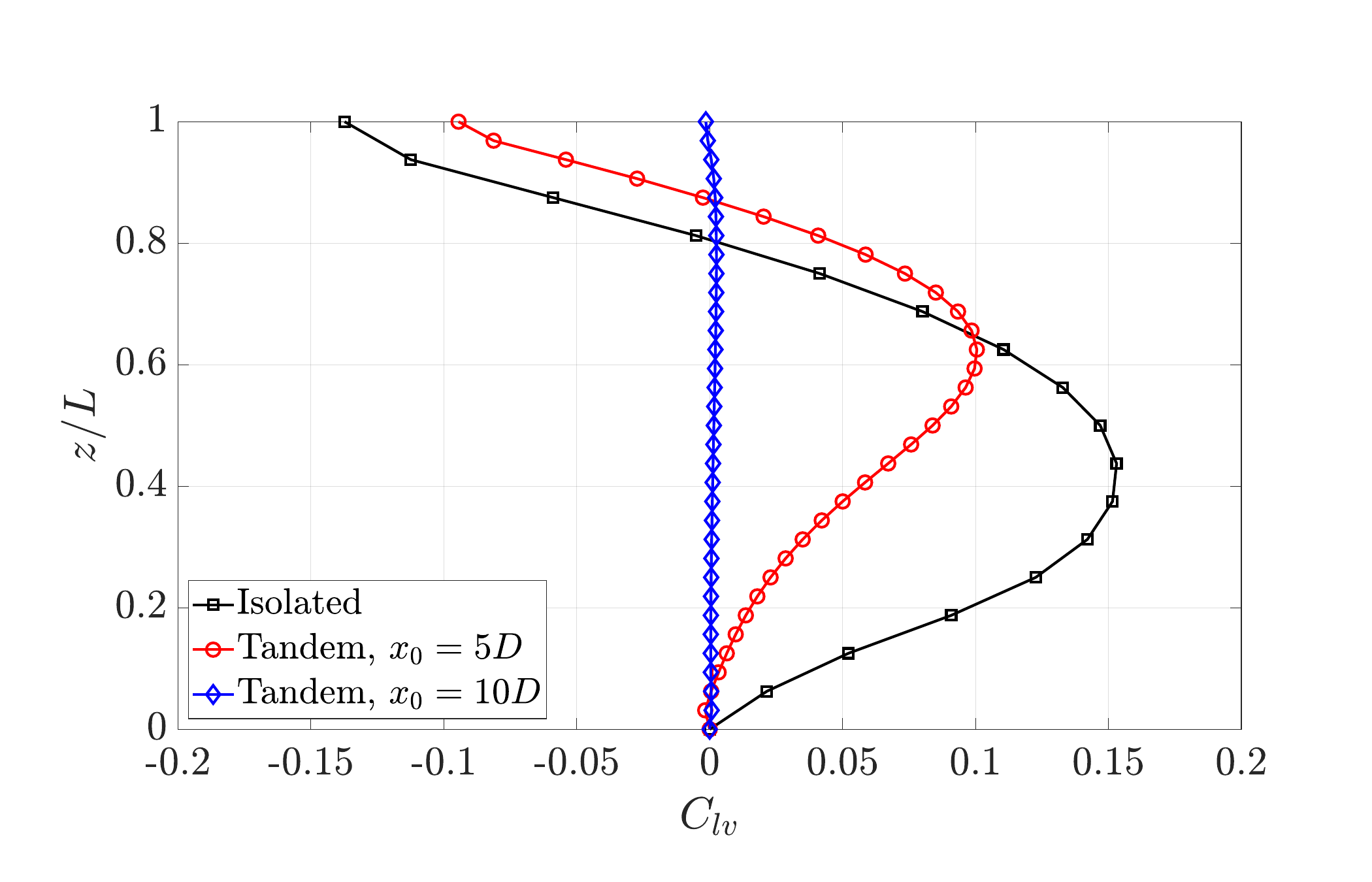}
\caption{}
\end{subfigure}
\caption{(a)-(c) Instantaneous power transfer between the flexible cantilever and fluid flow in the transverse direction in isolated versus tandem configurations; The cantilever is an energy sink (source) in regions with positive (negative) values. (d) Comparison between values of $C_{lv}$ along the cantilever length. The results are gathered at $Re = 40$, $m^* = 1$, and $U^* = 9$ over one period of the cantilever's transverse motion. Here, $t_\mathrm{i}$ is the initial sampling time.}
\label{EnergyTransfer-scalogram}
\end{figure*}

Next, we examine the power transfer between the transverse fluid loading and the flexible cantilever to characterize the energy exchange mechanisms driving the system's dominant oscillatory behavior in the transverse direction. Figure~\ref{EnergyTransfer-scalogram} illustrates the time history of the instantaneous power transfer from the fluid flow to the flexible cantilever over a single period of its transverse oscillation ($T$) for both isolated and tandem configurations at $U^* = 9$. The nondimensional power transfer coefficient in the transverse direction is defined as $(C_{\mathrm{l}}-\overline{C_\mathrm{l}})v/U_0$, where $v$ is the cross-sectional transverse velocity of the structure. A time-periodic interaction between the flexible cantilever and the fluid loading is observed, resulting in a cyclic power transfer in and out of the flexible structure.
Given the fluid loading and the cantilever's vibration response, the hydrodynamic coefficient in phase with the cantilever's transverse velocity is calculated using the following equation:
\begin{align}
C_\mathrm{lv}(z) = \frac{\frac{2}{T}\int_{T}[C_\mathrm{l}(z,t)-\overline{C_\mathrm{l}}(z)]v(z,t)\mathrm{d}t}{\sqrt{\frac{2}{T}\int_{T}{v(z,t)^{2}}\mathrm{d}t}},
\end{align}
where $C_\mathrm{lv}$ represents the hydrodynamic coefficient in phase with the transverse velocity of the structure. Figure~\ref{EnergyTransfer-scalogram}d illustrates the distribution of $C_\mathrm{lv}$ along the cantilever length for the isolated and tandem configurations at $U^* = 9$. We assume that $C_\mathrm{lv} = 0$ at the fixed end of the cantilever.
As shown in Fig.~\ref{EnergyTransfer-scalogram}d, both the isolated configuration and the tandem case with \( x_0 = 5D \) exhibit significant variations in \( C_\mathrm{lv} \) values. Notably, relatively large positive \( C_\mathrm{lv} \) values are observed in regions extending up to approximately 80\% of the cantilever length. These large positive values enhance energy transfer from the fluid to the solid, contributing to the sustenance of oscillations. In the following, we provide insights into the vorticity dynamics and discuss how wake structures evolve and interact with the flexible cantilever during its oscillatory motion.
%
\subsubsection{Vortex formation and wake structures in subcritical Re regime}
\begin{figure}\centering
\begin{subfigure}{0.32\textwidth}
\centering
\includegraphics[width=1\linewidth]{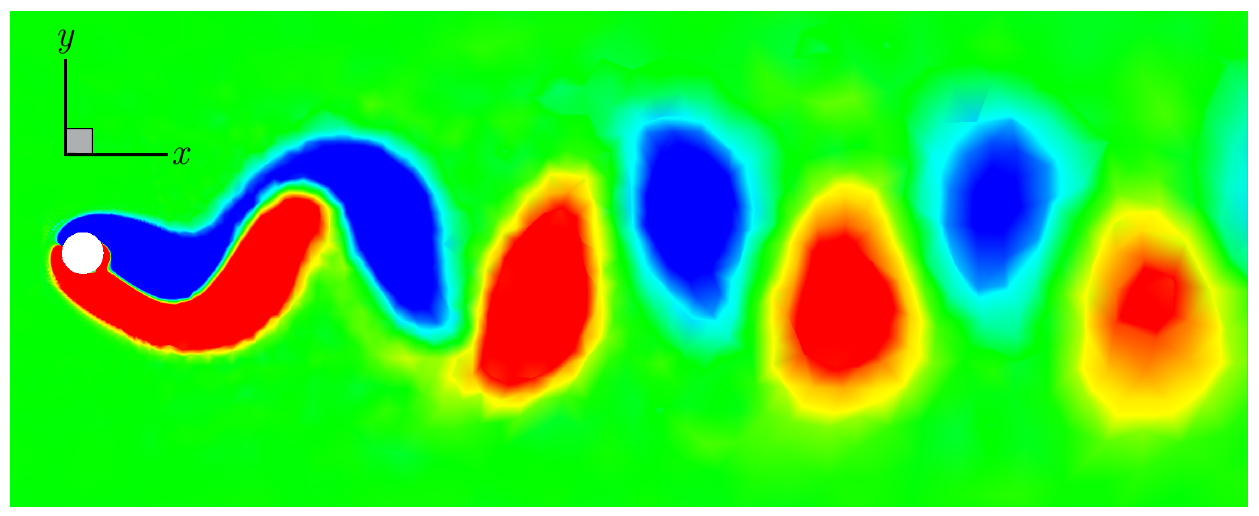}
\caption{}
\end{subfigure}
\begin{subfigure}{0.32\textwidth}
\centering
\includegraphics[width=1\linewidth]{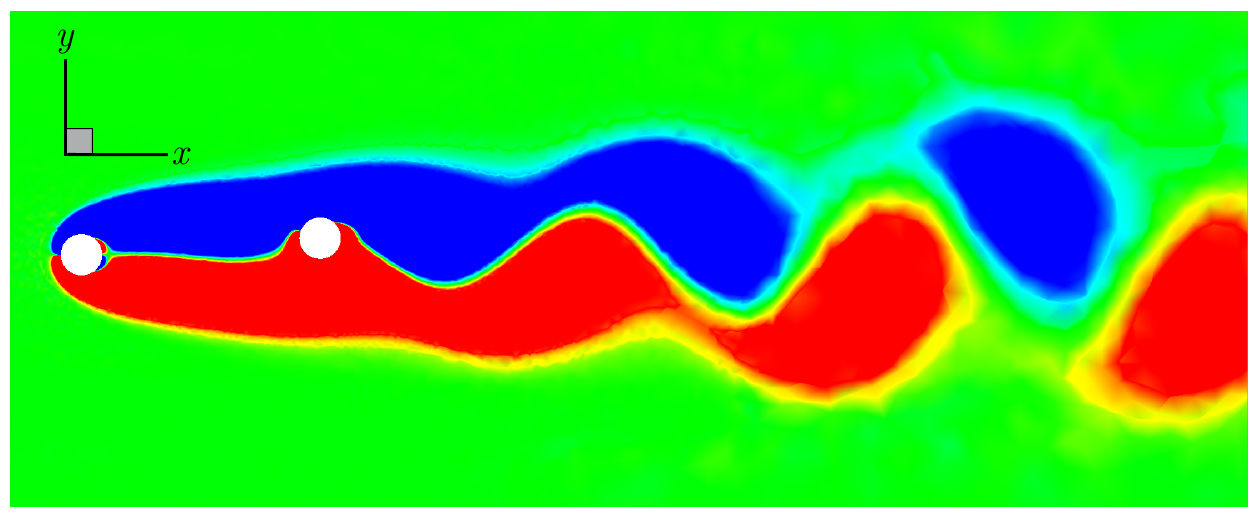}
\caption{}
\end{subfigure}
\begin{subfigure}{0.32\textwidth}
\centering
\includegraphics[width=1\linewidth]{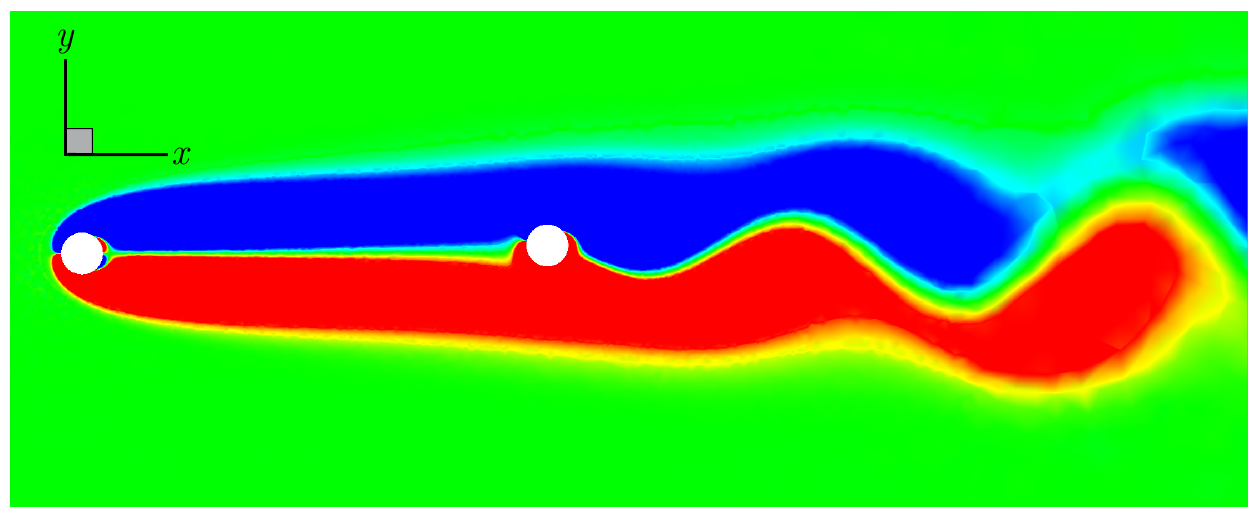}
\caption{}
\end{subfigure}\\
\begin{subfigure}{0.5\textwidth}
\centering
\includegraphics[width=1\linewidth]{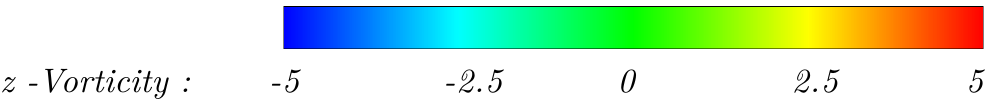}
\end{subfigure}%
\caption{\label{fig:zVort-TandemVsSingle-U11}Comparison of the $z$-plane slices of the $z$-vorticity contour at $z/L=1$ for (a) isolated, (b) tandem with $x_{0}=5D$, and (c) tandem with $x_{0}=10D$ configurations. The results are gathered at $Re = 40$, $m^* = 1$, and $U^* = 11$.}
\end{figure}

The wake dynamics of the flexible cantilever in tandem configuration within the subcritical Reynolds number regime reveal complex fluid-structure interactions that differ from those of an isolated cantilever. At a representative reduced velocity $U^* = 11$, the $z$-vorticity ($\omega_\mathrm{z}$) contours at the cantilever's free end illustrate the profound impact of the upstream cylinder on the wake structure. In the isolated configuration, we observe a classic von Kármán vortex street, characterized by alternating positive and negative vorticity regions. This pattern is indicative of strong transverse oscillations, wherein the periodic shedding of vortices creates an oscillating lift force that excites the cantilever's motion. However, in the tandem configuration, there is a critical modification to this behavior. As shown in Fig.~\ref{fig:zVort-TandemVsSingle-U11}, the upstream cylinder's presence delays vortex formation behind the cantilever, resulting in an extended attached near-wake region. This delay is a consequence of the upstream wake interfering with the shear layers separating from the cantilever, altering the momentum transfer in the boundary layer and postponing the roll-up of vortices. The extended near-wake in tandem configuration significantly impacts the coupled system's dynamics. Specifically, the frequency of the lift force shifts to lower values in the tandem configuration compared to the isolated case, as evidenced previously in Fig.~\ref{Response_Characteristics_subcritical}b.

\begin{figure}
\includegraphics[width=0.65\linewidth]{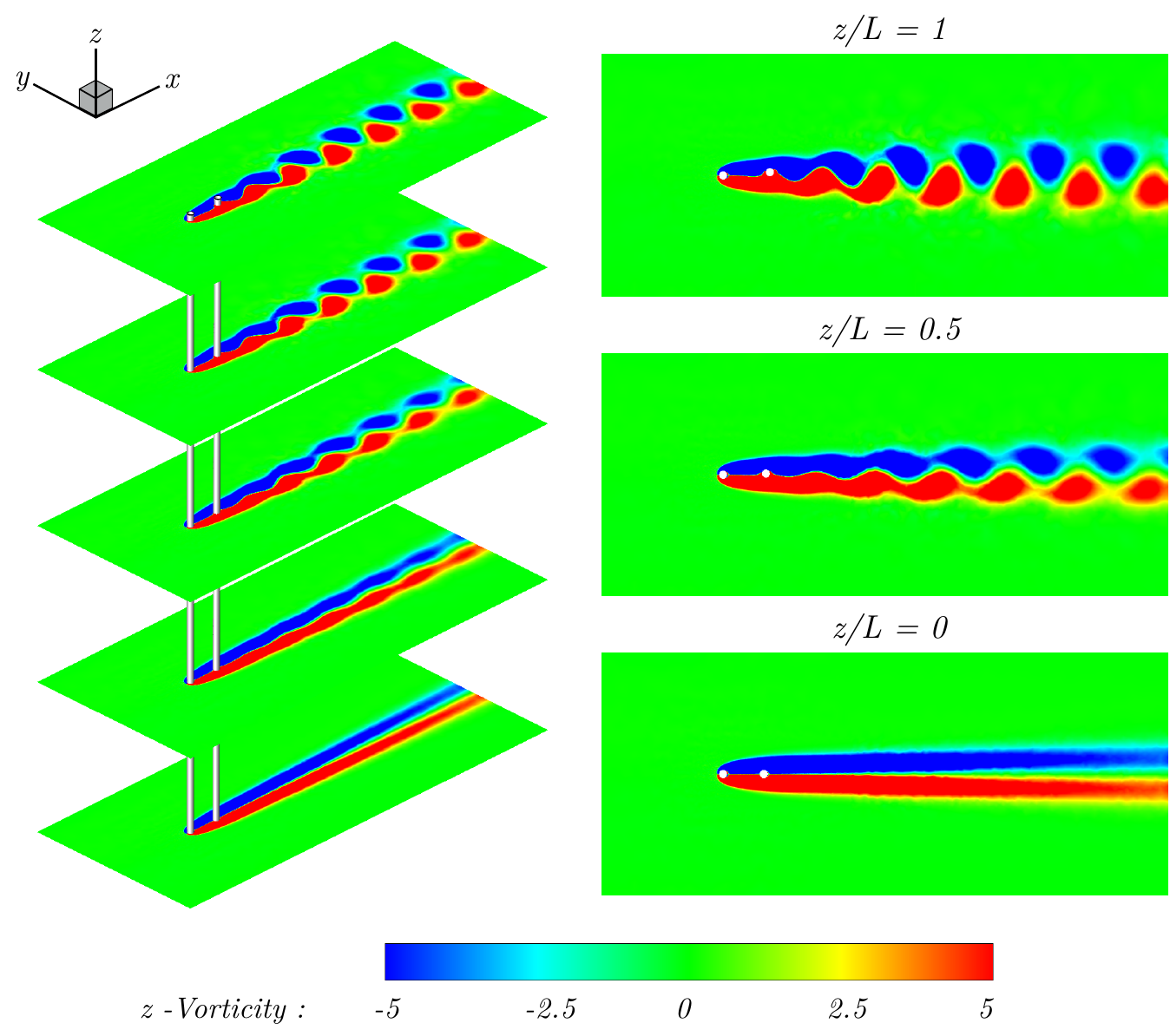}
\caption{\label{Z-Vorticity-New}Isometric view of the $z$-plane slices of the $z$-vorticity contour for the flexible cantilever in the tandem configuration with $x_{0}=5D$ at $Re=40$, $m^*=1$, and $U^* = 11$; The $z$-plane slices of the $z$-vorticity contour at $z/L=1$ (free end), $0.5$ (mid-section), and $0$ (fixed end) are shown in the right-hand side.}
\end{figure}

\begin{figure*}
\begin{subfigure}{0.25\textwidth}
\centering
\includegraphics[width=0.8\linewidth]{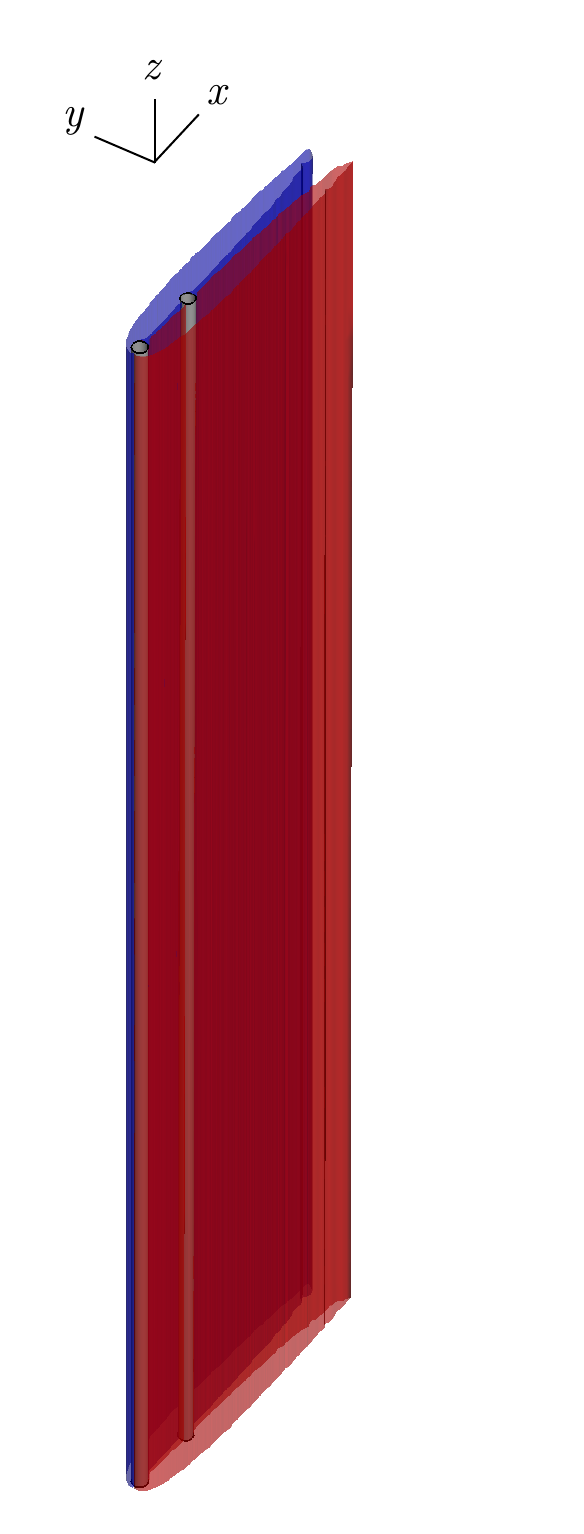}
\caption{$U^*=7$ (pre lock-in)}
\end{subfigure}%
\begin{subfigure}{0.25\textwidth}
\centering
\includegraphics[width=0.8\linewidth]{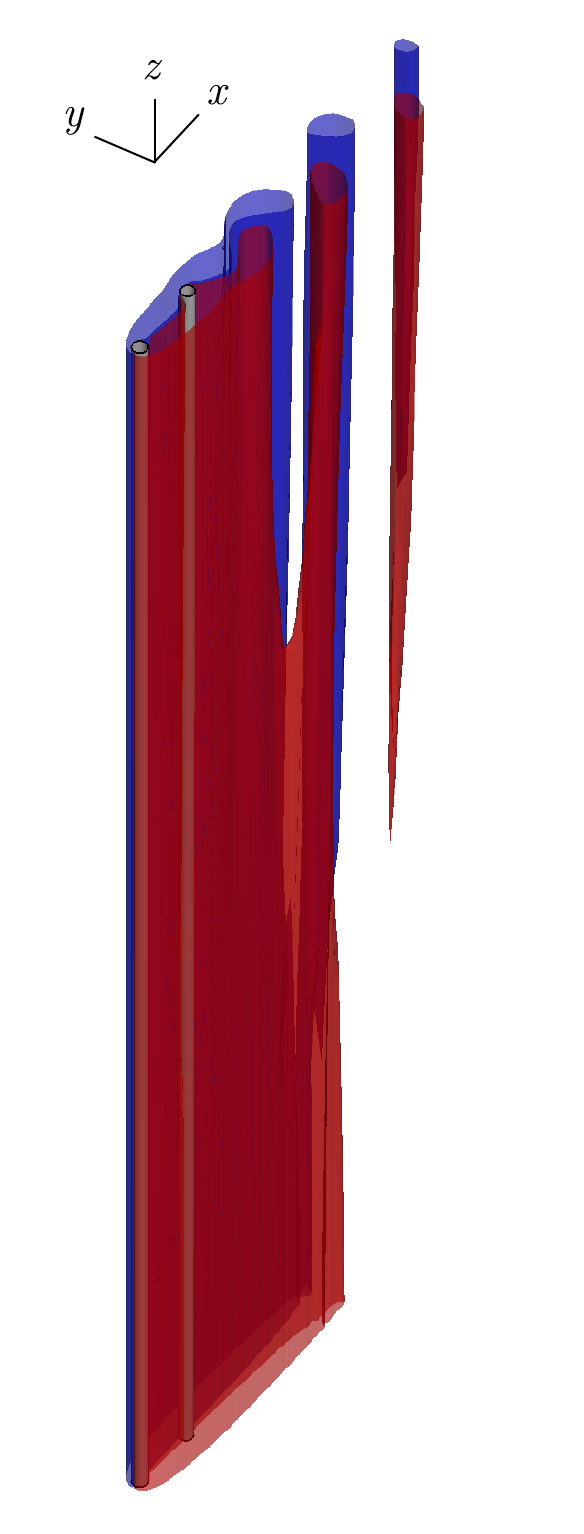}
\caption{$U^*=11$ (lock-in)}
\end{subfigure}%
\begin{subfigure}{0.25\textwidth}
\centering
\includegraphics[width=0.8\linewidth]{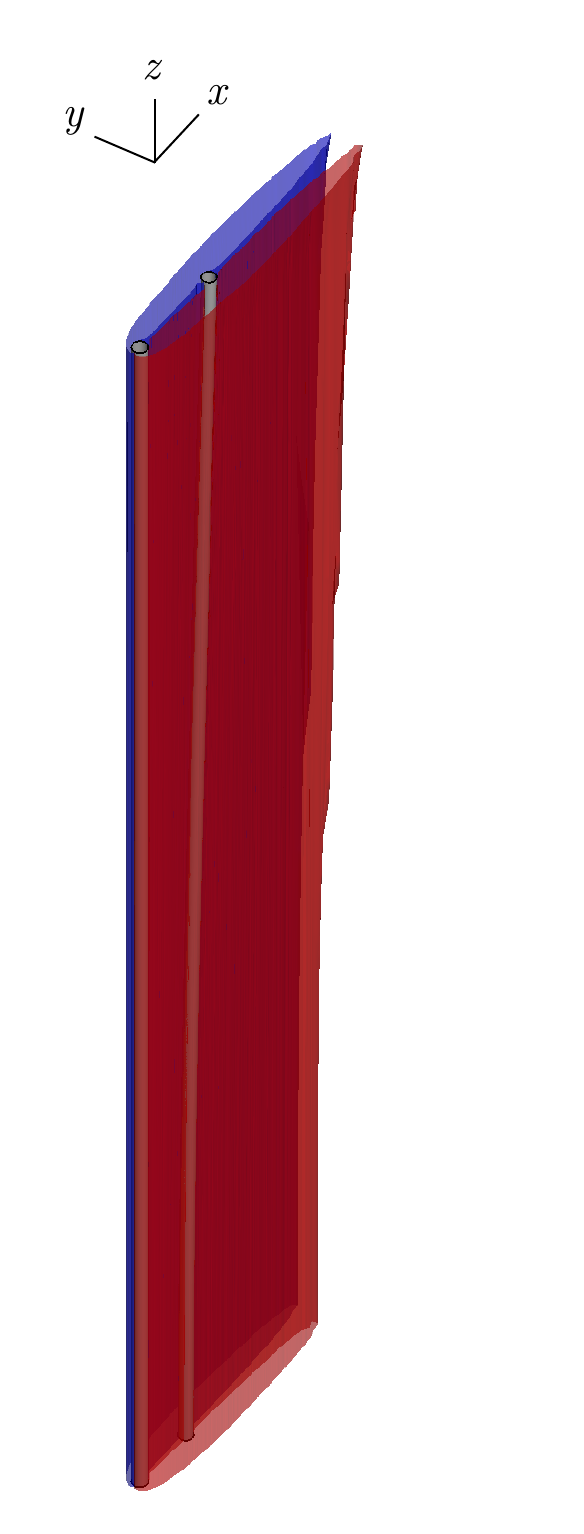}
\caption{$U^*=19$ (post lock-in)}
\end{subfigure}
\caption{\label{zvort-isosurfaces-Re40-5D}Wake structures visualized by the normalized $z$-vorticity iso-surfaces ($\omega_{z}D/U_{0} = -0.224,0.224$) for the tandem case with $x_{0}=5D$ at $Re=40$ and $m^*=1$. Red (blue) indicates regions of positive (negative) vortices.}
\end{figure*}

Examining the $z$-vorticity contours along the cantilever's length for the tandem configuration with $x_{0} = 5D$ at a representative $U^* = 11$ reveals a spatial variation in flow stability, as shown in Fig.~\ref{Z-Vorticity-New}, that is crucial for understanding the system's behavior. At the fixed end, i.e., $z/L = 0$, we observe a steady wake behind the cantilever, indicating dominant viscous forces and weak fluid-structure coupling. As we progress towards the free end, i.e., $z/L = 1$, the wake transitions to an unsteady state with periodic vortex shedding downstream. This spatial variation in wake stability is a result of the cantilever's varying flexibility and vibration amplitude along its length. The free end, experiencing larger oscillations, induces stronger shear layer instabilities, promoting vortex formation even within the laminar subcritical regime of $Re$. The three-dimensional wake structures, visualized through $z$-vorticity iso-surfaces, provide further insight into the coupled system's behavior across different $U^*$ regimes. As shown in Fig.~\ref{zvort-isosurfaces-Re40-5D}, in the pre-lock-in regime ($U^*=7$), the wake remains largely steady, indicating that the kinetic energy of the flow is insufficient to overcome viscous damping and induce unsteady vortex shedding. The lock-in regime ($U^* = 11$) exhibits well-defined periodic vortex shedding patterns, with a strong synchronization process that facilitates maximum energy transfer from the fluid to the structure, resulting in sustained large amplitude oscillations. In the post-lock-in regime ($U^* = 19$), the wake reverts to a steady state. These observations lead us to identify the critical conditions for sustaining wake unsteadiness and oscillations in the tandem configuration at the laminar subcritical regime of $Re$: (i) Sufficient flow inertia to overcome viscous damping and develop instabilities in the shear layer and (ii) system parameters within the lock-in range to allow for sustained energy transfer and overcome the coupled system's damping mechanisms. In the following, we further investigate how system parameters influence the wake dynamics and oscillatory response of the cantilever by studying the coupled dynamics in the post-critical regime of $Re$.
%
\subsection{\label{subsec:postcritical} 
Coupled dynamics in post-critical Reynolds number}
\subsubsection{\label{subsubsec:responseCharacteristics-postcritical} Response characteristics in post-critical Re regime}
\begin{figure}
\begin{subfigure}{0.5\textwidth}
\centering
\includegraphics[width=1\linewidth]{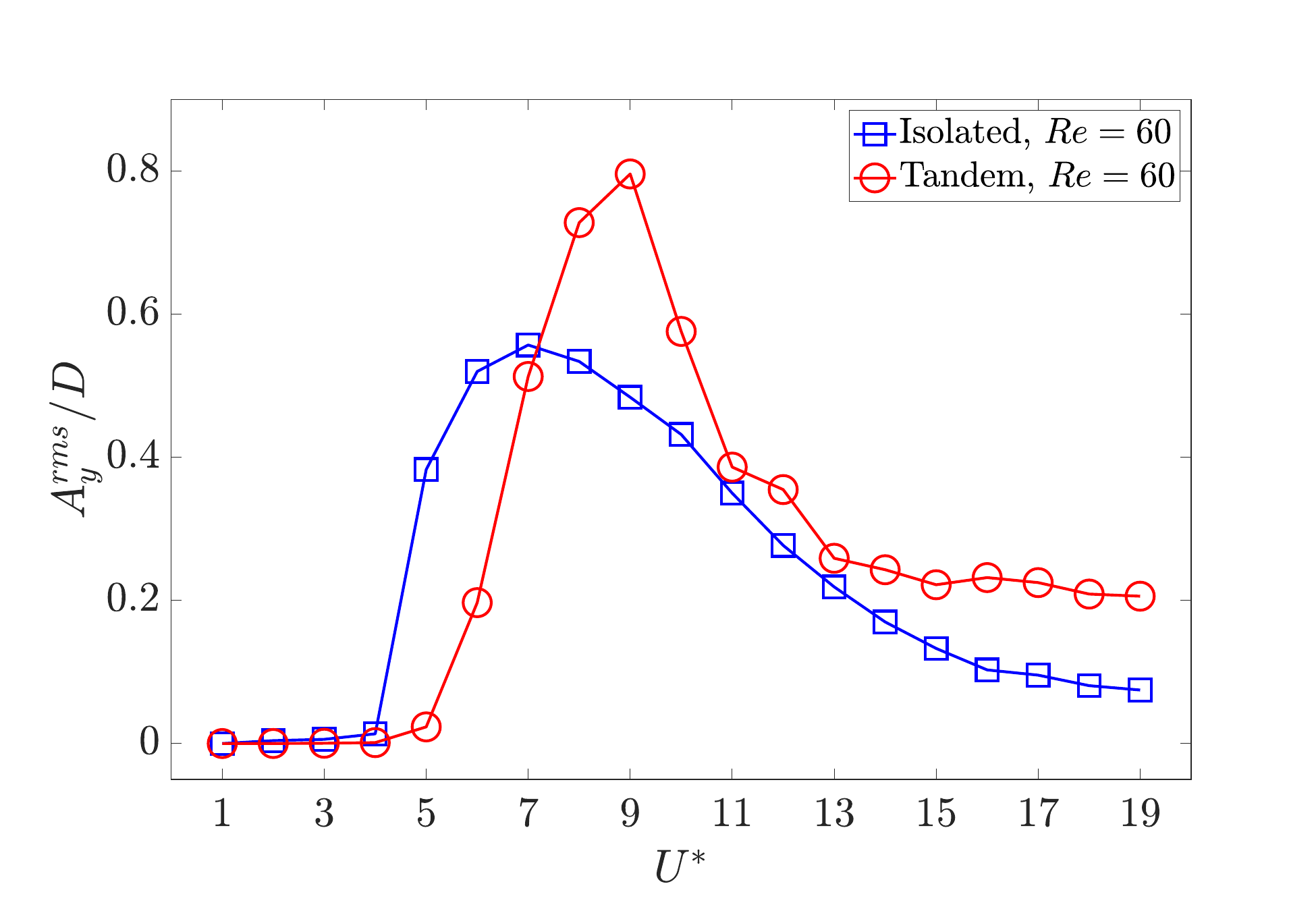}
\caption{}
\label{Ay-Re60}
\end{subfigure}%
\begin{subfigure}{0.5\textwidth}
\centering
\includegraphics[width=1\linewidth]{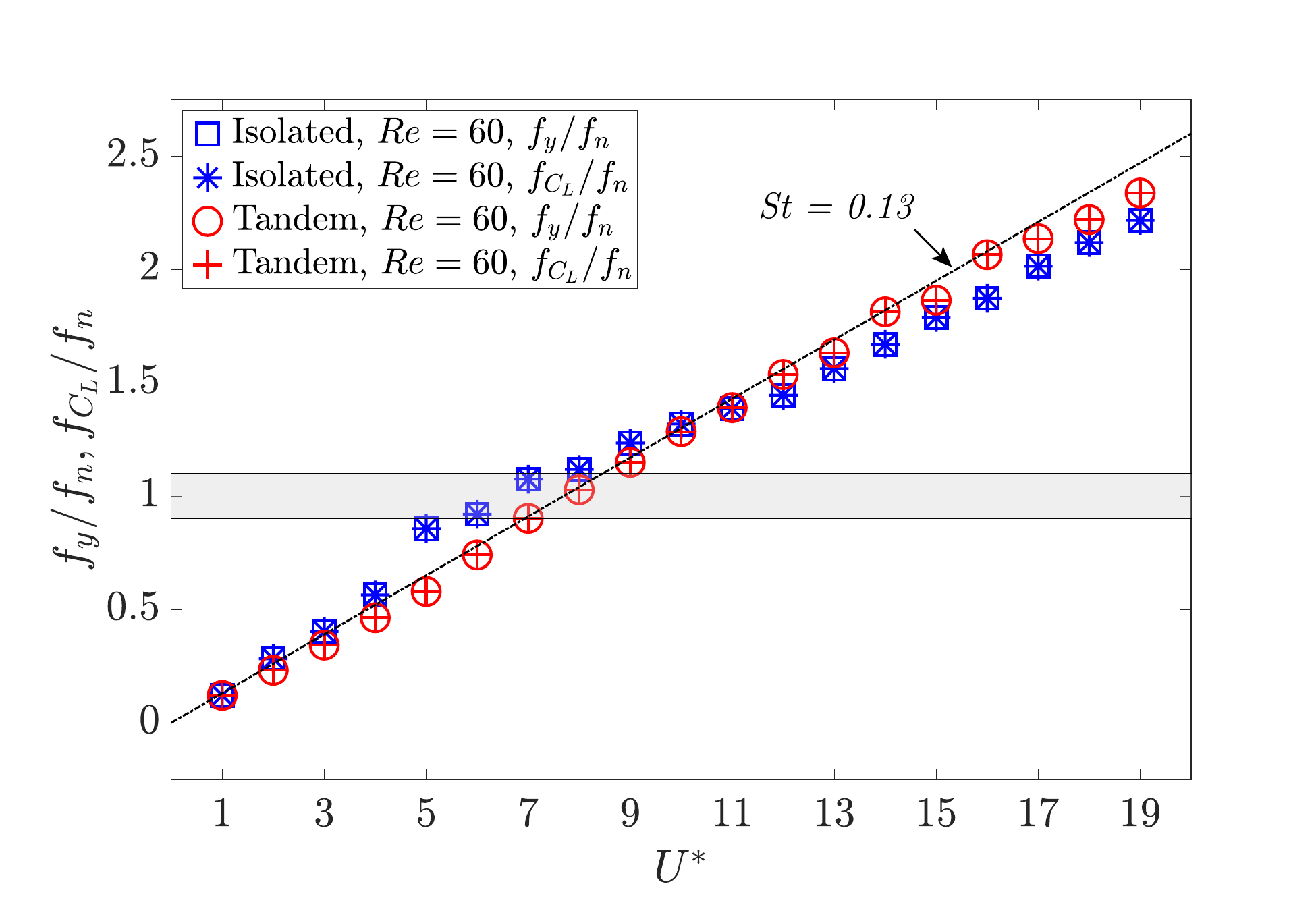}
\caption{}
\label{FFT-CLAy-Re60}
\end{subfigure}
\caption{\label{Response_Characteristics-postcritical-Re60}Comparison between the value of $A_\mathrm{y}^{rms}/D$ at the free end of the cantilever in isolated versus tandem configurations at $Re=60$; (b) variations of $f_\mathrm{y}/f_\mathrm{n}$ at the free end of the cantilever and $f_\mathrm{C_L}/f_\mathrm{n}$ with respect to $U^*$ at $Re=60$; The dash-dotted line represents the vortex-shedding frequency of an isolated rigid stationary cylinder based on the Strouhal ($St$) relationship.}
\end{figure}

Here, we provide a detailed comparison of the cantilever's amplitude response and frequency dynamics in the isolated and tandem configurations at two post-critical Reynolds numbers of $Re=60$ and 100 for $U^*\in[1,19]$ at $m^*=1$. We focus on a single representative separation distance $x_{0}=5D$ for the tandem configuration. Figure~\ref{Response_Characteristics-postcritical-Re60}a presents the values of $A_\mathrm{y}^{rms}/D$ for the isolated and tandem cases at $Re=60$. In the isolated configuration, the amplitude of transverse oscillations is close to zero for $U^*\leq4$. The oscillations reach a peak amplitude at \(U^*=7\), with a value of $A_\mathrm{y}^{rms}/D\approx0.55$. Beyond $U^*=7$, there is a gradual decrease in the amplitude of oscillations, similar to the isolated case at $Re=40$. In the case of the cantilever in tandem configuration, the transverse oscillations reach a peak amplitude at \(U^*=9\), with a value of $A_\mathrm{y}^{rms}/D\approx0.80$. Compared to the tandem cases at $Re=40$ where no sustained oscillations were present in the post-lock-in regime, for the tandem cases at $Re=60$, the cantilever undergoes sustained oscillations for $U^*\geq15$, with values of $A_\mathrm{y}^{rms}/D\approx0.22$, as depicted in Fig.~\ref{Response_Characteristics-postcritical-Re60}a. 
Figure~\ref{Response_Characteristics-postcritical-Re60}b illustrates the nondimensional transverse vibration frequency \(f_\mathrm{y}/f_\mathrm{n}\) at the free end of the cantilever and the nondimensional frequency of the lift coefficient \(f_{\mathrm{C_L}}/f_\mathrm{n}\) as functions of \(U^*\) for the isolated and tandem configurations at \(Re = 60\). In both cases, there is a match between the frequency of the transverse oscillations and the frequency of the lift coefficient for all studied $U^*$. For the isolated case, the transverse oscillation and lift coefficient frequencies closely match the cantilever's first-mode natural frequency within the range $U^*\in(5,8)$, resulting in peak vibration amplitudes in this range. In the tandem configuration, this range of peak amplitude response is shifted to $U^*\in (7,9)$. Based on the results provided in Fig.~\ref{Response_Characteristics-postcritical-Re60}, it is evident that the flexible cantilever in the tandem configuration exhibits sustained oscillations similar to those of an isolated flexible cantilever experiencing VIVs. 

\begin{figure*}
\begin{subfigure}{0.5\textwidth}
\centering
\includegraphics[width=1\linewidth]{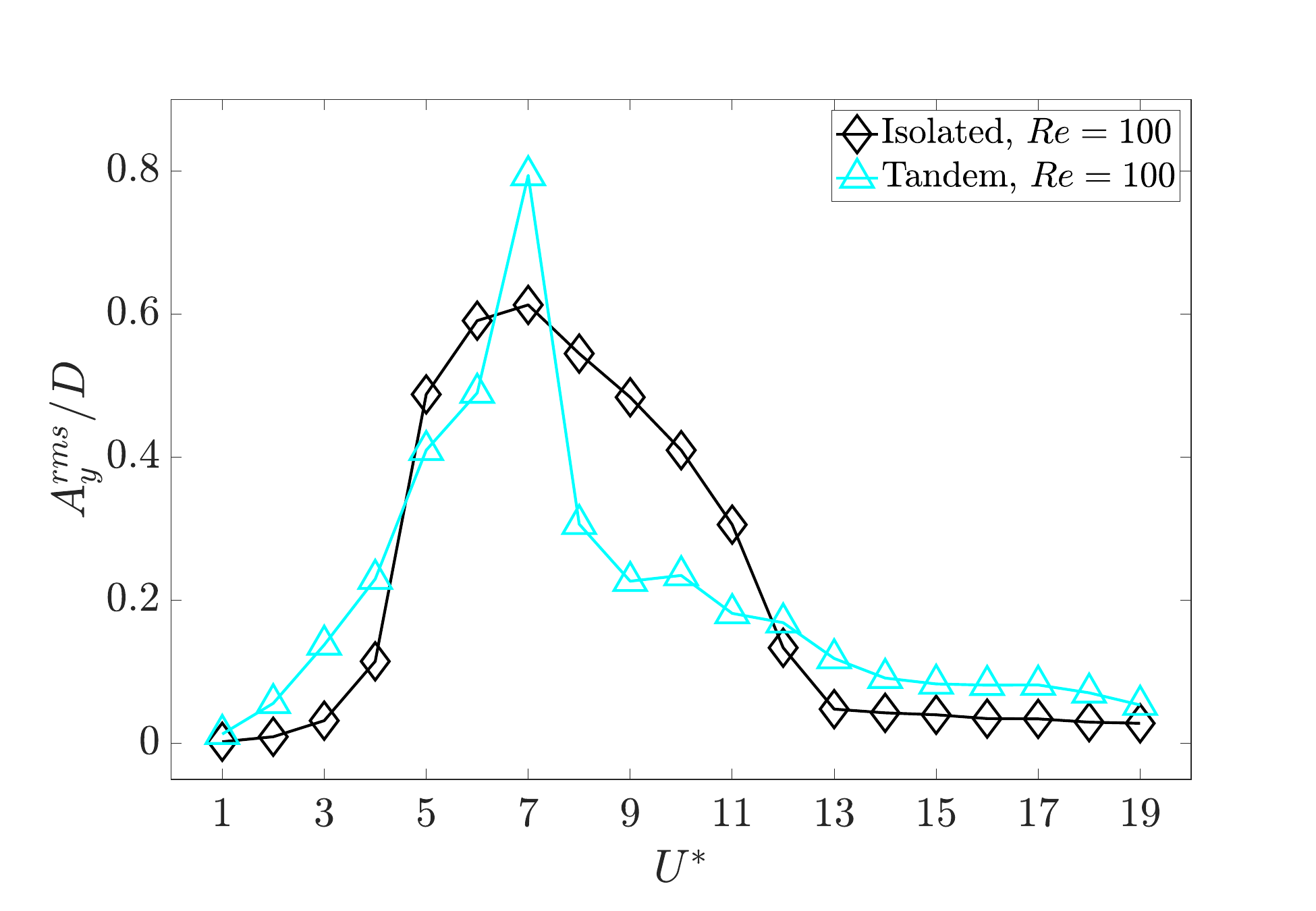}
\caption{}
\label{Ay-Re100}
\end{subfigure}%
\begin{subfigure}{0.5\textwidth}
\centering
\includegraphics[width=1\linewidth]{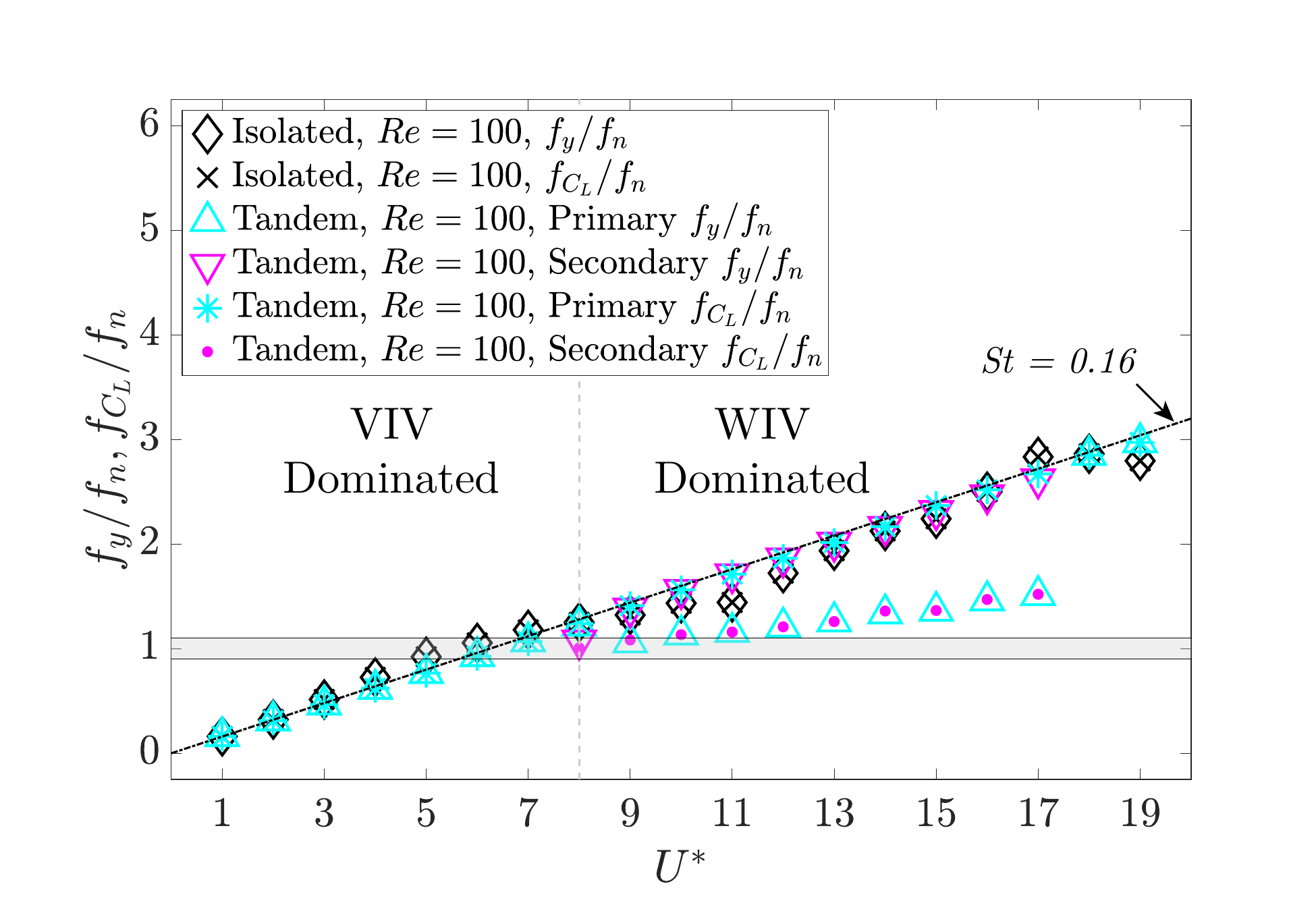}
\caption{}
\label{FFT-CLAy-Re100}
\end{subfigure}
\caption{\label{Response_Characteristics-postcritical-Re100} (a) Comparison between the value of $A_\mathrm{y}^{rms}/D$ at the free end of the cantilever in isolated versus tandem configurations at $Re=100$; (b) variations of $f_\mathrm{y}/f_\mathrm{n}$ at the free end of the cantilever and $f_\mathrm{C_L}/f_\mathrm{n}$ with respect to $U^*$ at $Re=100$.}
\end{figure*}

Next, we compare the values of \(A_\mathrm{{y}}^{rms}/D\) for the isolated and tandem configurations at \(Re = 100\). As shown in Fig.~\ref{Response_Characteristics-postcritical-Re100}a, in the isolated configuration, the amplitude of the transverse oscillation follows a bell-shaped curve as a function of $U^*$. Similar to the isolated case at $Re=60$, the oscillations reach a peak amplitude at \(U^*=7\), with the value of $A_{\mathrm{y}}^{rms}/D\approx0.60$. For the cases in the tandem configuration, the transverse oscillation reaches a peak value at \(U^*=7\), with a value of \(A_\mathrm{{y}}^{rms}/D \approx 0.80\). This peak in the value of the transverse oscillation amplitude is similar to that observed in the tandem configuration at \(Re = 60\). Compared to the tandem cases at $Re=60$, the flexible cantilever at $Re=100$ experiences lower amplitude oscillations for $U^* \geq 15$, with the value of \(A_\mathrm{{y}}^{rms}/D \approx 0.10\).
The values of \(f_\mathrm{y}/f_\mathrm{n}\) and \(f_\mathrm{{C_L}}/f_\mathrm{n}\) in terms of \(U^*\) for the isolated and tandem cases at \(Re = 100\) are provided in Fig.~\ref{Response_Characteristics-postcritical-Re100}b. In the isolated configuration, the cantilever experiences a single-frequency response, where there is a match between the frequency of the transverse oscillation and the frequency of the lift coefficient across all examined $U^*$.
Compared to the tandem cases at $Re=60$, a noticeable secondary frequency component is present in the frequency of the transverse oscillation and lift coefficient frequency at $Re=100$. Specifically, we find that within $U^*\in[9,17]$, the dominant frequency of the transverse oscillation remains close to the cantilever's first-mode natural frequency, while the secondary frequency component of the transverse oscillation matches the vortex-shedding frequency of a rigid stationary cylinder at $Re=100$, as illustrated in Fig.~\ref{Response_Characteristics-postcritical-Re100}b. These observations indicate that for the flexible cantilever in the tandem configuration, as \(Re\) increases, i.e., increasing inertial dominance in the flow, the cantilever becomes more susceptible to a multi-frequency response at high $U^*$ values and tends to oscillate predominantly at frequencies close to its first-mode natural frequency over a wider range of \(U^*\).

Based on the results provided in Fig.~\ref{Response_Characteristics-postcritical-Re100}, the response dynamics of the cantilever in the tandem configuration at $Re=100$ could be classified into two main categories in terms of $U^*$; VIV-dominated and WIV-dominated regimes. For $U^*<8$, the phenomenon underlying the cantilever's oscillatory response is similar to that of an isolated cantilever. As shown in Fig.~\ref{Response_Characteristics-postcritical-Re100}b, a single frequency component is observed in the transverse force, which is induced by vortex shedding from the cantilever and synchronizes with the transverse oscillation frequency. Consequently, this vibration regime is characterized by a VIV process, where the vortex shedding frequency matches the cantilever's oscillation frequency. For $U^*\in[9,17]$, two dominant
frequencies exist in the transverse fluid loading and the oscillatory response, as shown in Fig.~\ref{Response_Characteristics-postcritical-Re100}b. The oscillatory response is primarily driven by the low-frequency component of the transverse load, which approximates the cantilever's first-mode natural frequency. Notably, the low-frequency component in the transverse loading is approximately half the vortex-shedding frequency of the upstream cylinder. Within this $U^*$ range, the transverse oscillation frequency exceeds the cantilever's first-mode natural frequency but remains below the vortex shedding frequency.

\begin{figure*}
\begin{subfigure}{0.45\textwidth}
\centering
\includegraphics[width=1\linewidth]{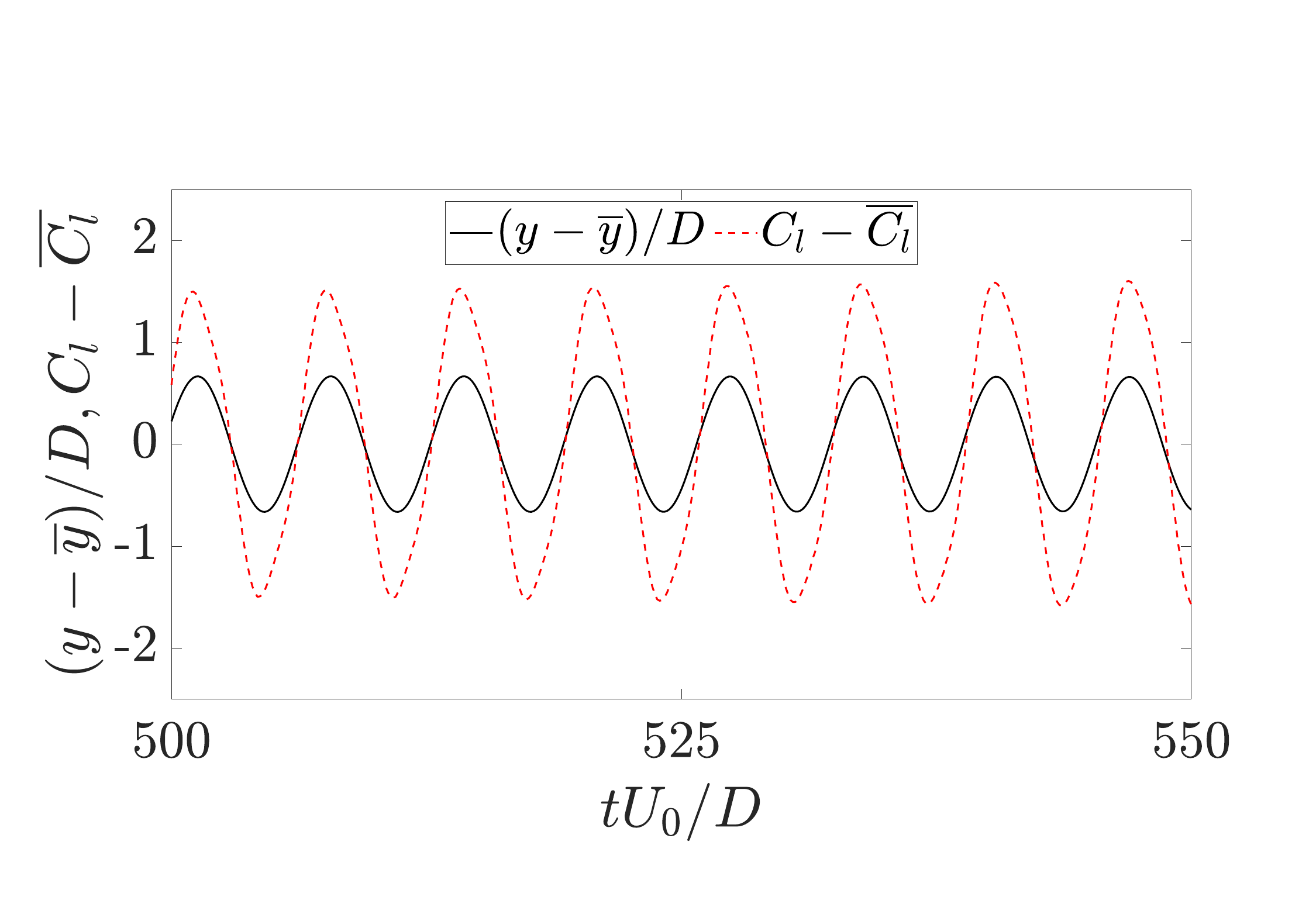}
\caption{}
\end{subfigure}%
\begin{subfigure}{0.37\textwidth}
\centering
\includegraphics[width=1\linewidth]{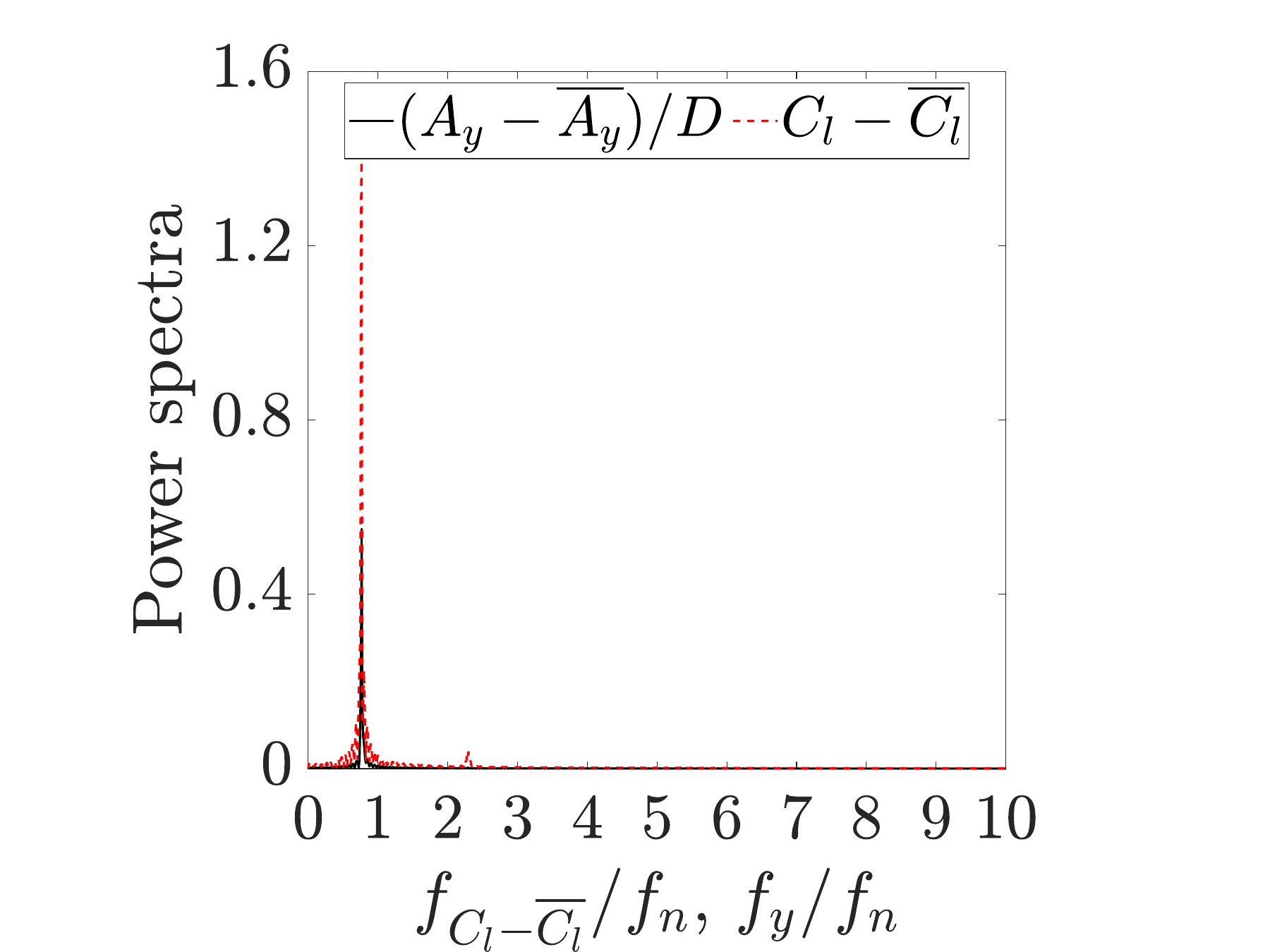}
\caption{}
\end{subfigure}
\begin{subfigure}{0.45\textwidth}
\centering
\includegraphics[width=1\linewidth]{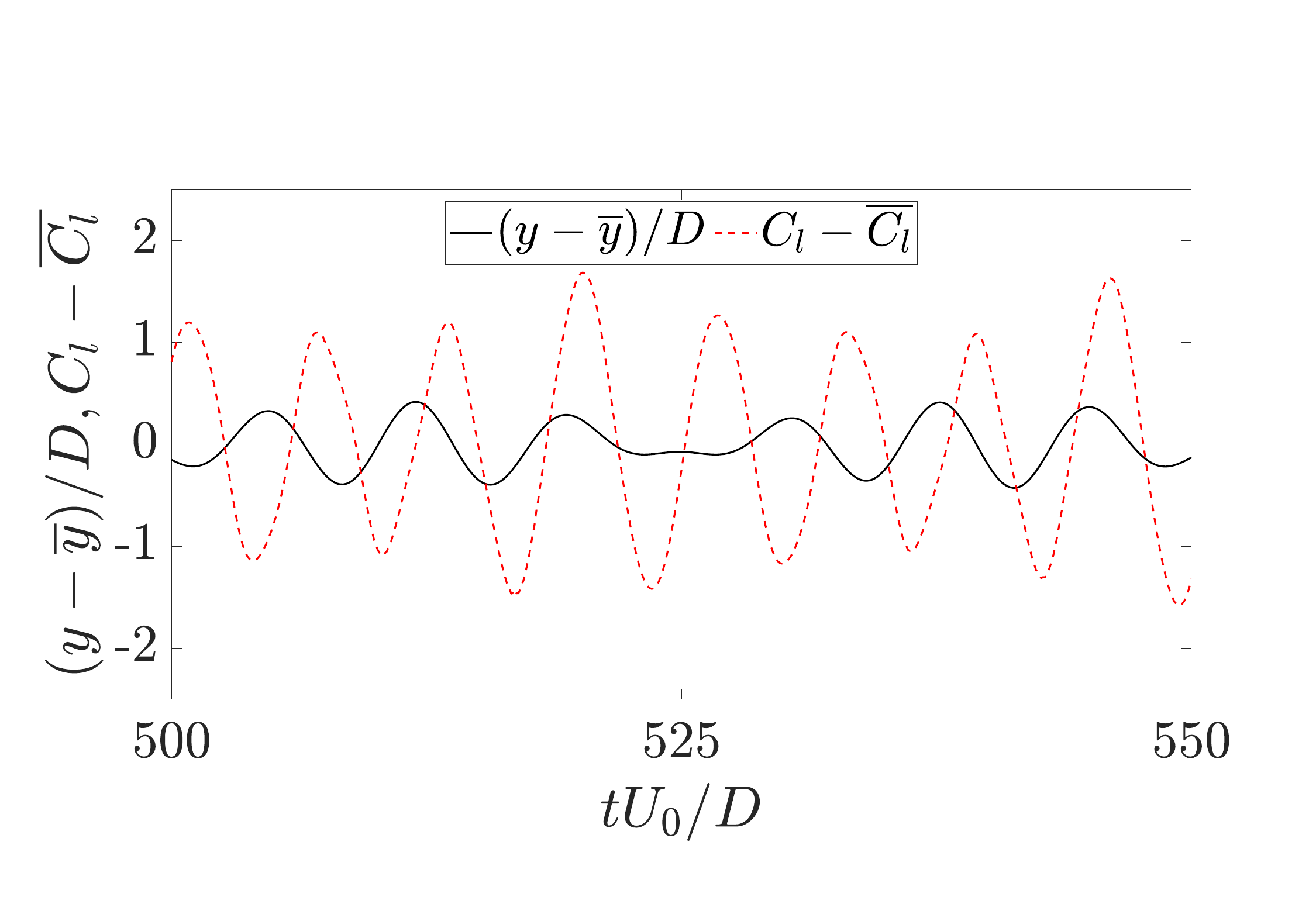}
\caption{}
\end{subfigure}%
\begin{subfigure}{0.37\textwidth}
\centering
\includegraphics[width=1\linewidth]{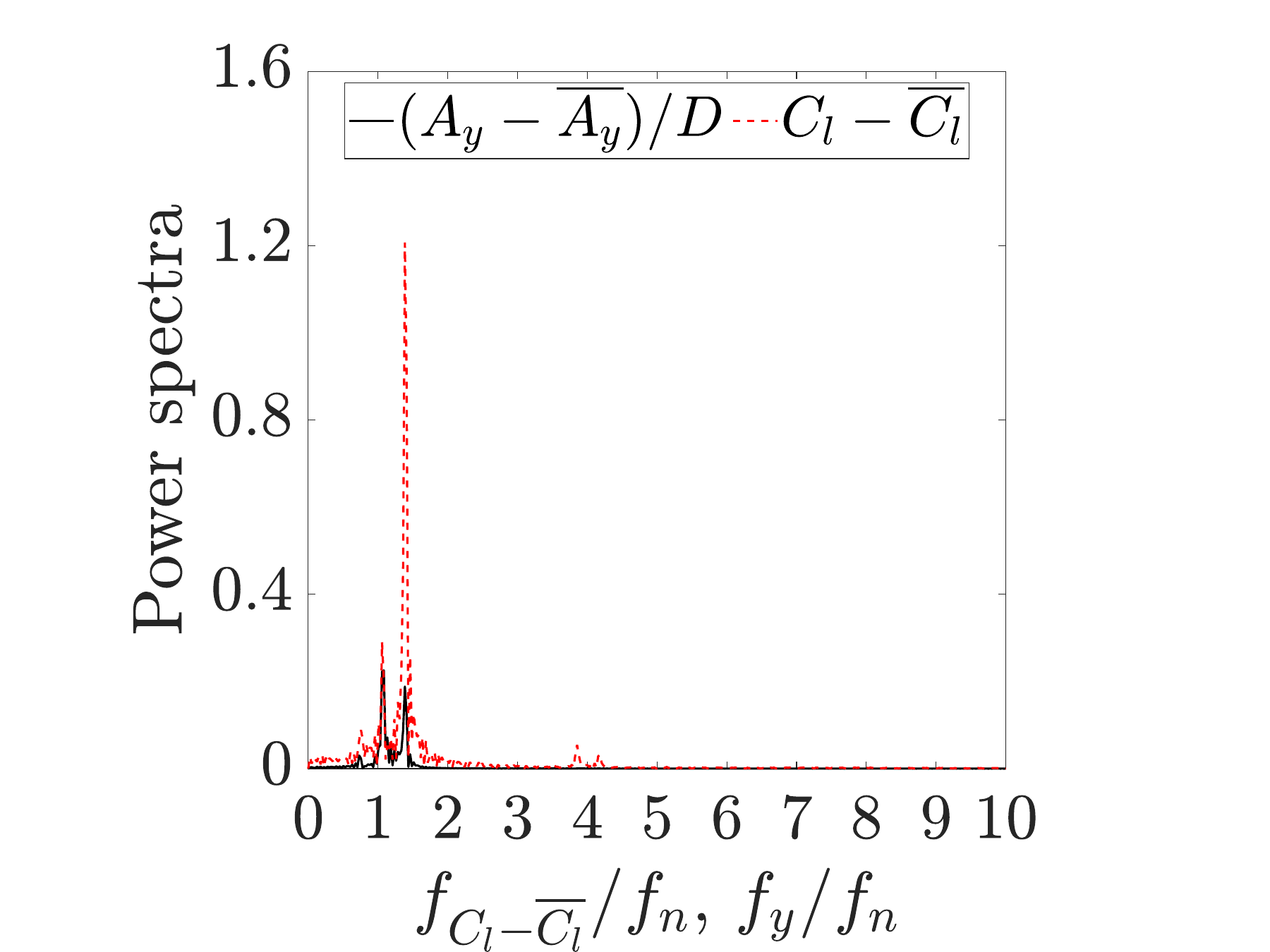}
\caption{}
\end{subfigure}
\caption{(a) Variations of $C_{l}-\overline{C_{l}}$ and $(y-\overline{y})/{D}$ in the time domain at $U^* = 5$; (b) power spectra of the nondimensional transverse oscillation amplitude and the cross-sectional lift coefficient at $U^* = 5$; (c) variations of $C_{l}-\overline{C_{l}}$ and $(y-\overline{y})/{D}$ in the time domain at $U^* = 9$; (d) power spectra of the nondimensional transverse oscillation amplitude and the cross-sectional lift coefficient at $U^* = 9$; The results are gathered at the free end of the cantilever within time window $tU_{0}/D\in[500, 550]$ at $m^* = 1$ and $Re=100$.}
\label{Scalogram-FFT-Spanwise-Re100}
\end{figure*}

To highlight the differences between the cantilever's oscillatory response and the transverse fluid forces in the tandem configuration for the VIV- and WIV-dominated regimes, we have provided the time history of $C_{l}-\overline{C_{l}}$ and $(y-\overline{y})/{D}$ at the free end of the cantilever at $U^*=5$, representative of VIV-dominated regime, and $U^*=9$, representing the WIV-dominated regime, in Fig.~\ref{Scalogram-FFT-Spanwise-Re100}. The results presented in Fig.~\ref{Scalogram-FFT-Spanwise-Re100} further illustrate the cantilever's multi-frequency response in the WIV-dominated regime at \(Re=100\), compared to the VIV-dominated range, where the oscillations exhibit a single-frequency response. For the tandem cases at $Re=100$, when $U^{*}\in[18, 19]$, the response dynamics and the transverse fluid forces converge to a single-frequency response. The presented results suggest that at higher reduced velocities than those investigated in this study, the cantilever could exhibit sustained oscillations at its higher harmonic natural vibration modes, as experimentally observed in flexible cylinders in tandem arrangement~\cite{SEYEDAGHAZADEH2021103276}. While a comprehensive examination of these vibrations is beyond the scope of our current work, it represents a promising direction for future investigations. In the following, we focus on the wake interference and its impact on the cantilever's oscillatory response in the post-critical $Re$ regime.

\subsubsection{\label{subsubsec:wakeDynamics-postcritical} Wake interference and vorticity dynamics in post-critical Re regime}
\begin{figure}
\includegraphics[width=0.65\linewidth]{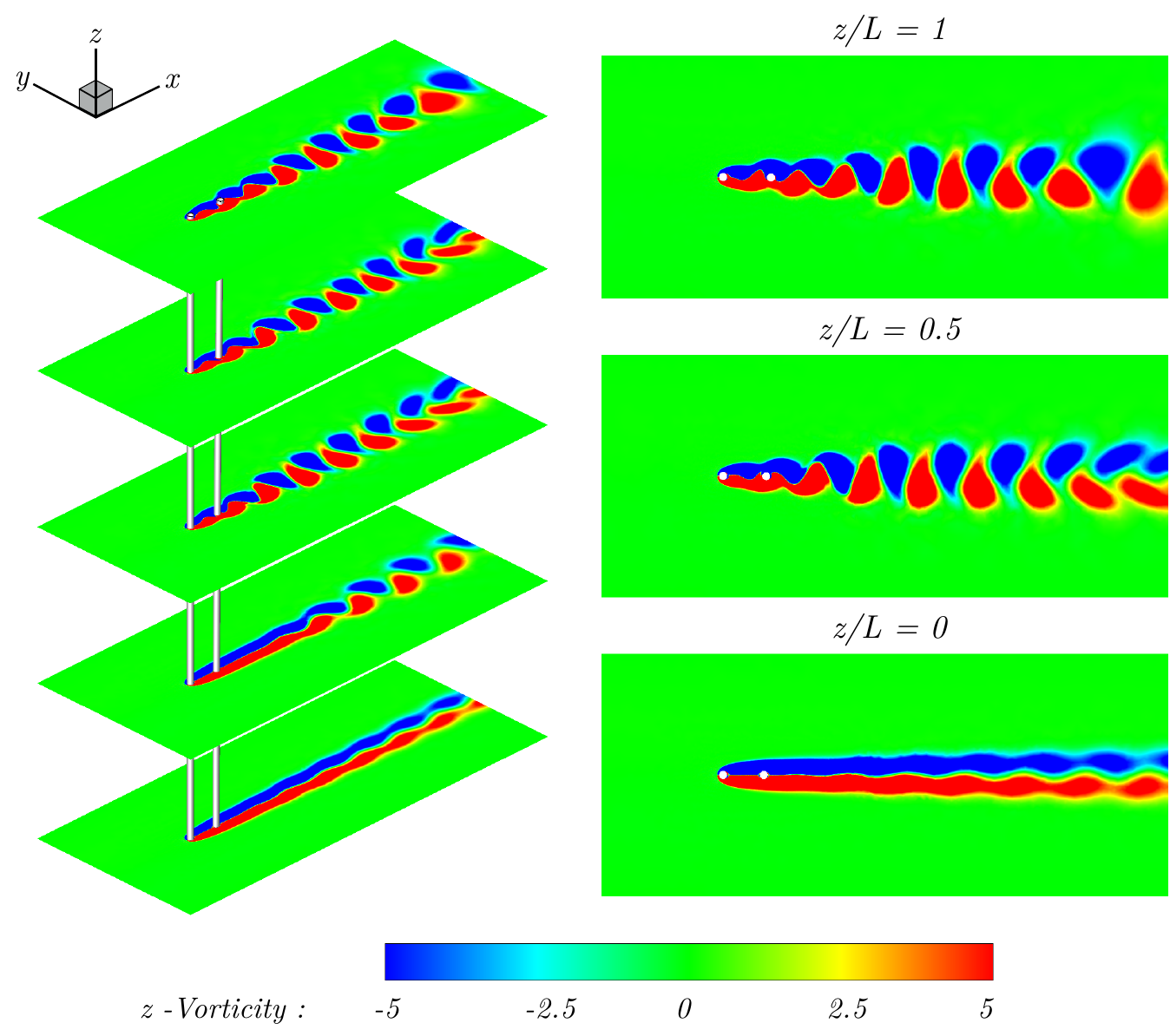}
\caption{\label{Z-Vorticity-5D-Re60-Re100}Isometric view of the $z$-plane slices of the $z$-vorticity contour for the flexible cantilever in the tandem configuration at $Re=60$, $m^*=1$, and $U^* = 11$; The $z$-plane slices of the $z$-vorticity contour at $z/L=1$ (free end), $0.5$ (mid-section), and $0$ (fixed end) are shown in the right-hand side.}
\end{figure}

Next, we present the vorticity dynamics for the isolated and tandem configurations in the post-critical Reynolds number regime. Figure~\ref{Z-Vorticity-5D-Re60-Re100} provides an isometric view of the \(z\)-plane slices of the \(z\)-vorticity contour at \(Re = 60\) for the tandem arrangement at a representative $U^*=11$. We observe that a fully developed vortex street is present at the free end, i.e., $z/L = 1$. Progressing to the mid-span, i.e., $z/L = 0.5$, the vortex street remains evident with a gradual modification of the wake structure. At the fixed end, i.e., $z/L = 0$, a notable transformation in the wake structure is observed, wherein the flow pattern transitions to elongated, parallel vorticity layers. This quasi-steady flow regime near the cantilever's fixed end signifies the influence of the cantilever's motion and flow inertia in determining wake dynamics in the tandem cylinder arrangement. Specifically, we observe that for an isolated flexible cantilever under similar conditions, the flow exhibits unsteady dynamics characterized by a mature von Kármán vortex street along the entire cantilever length. However, in the tandem arrangement, the wake structures exhibit spatial variations: a well-defined vortex street at the free end and a more stable flow near the fixed end.

\begin{figure}\centering
\begin{subfigure}{0.35\textwidth}
\centering
\includegraphics[width=1\linewidth]{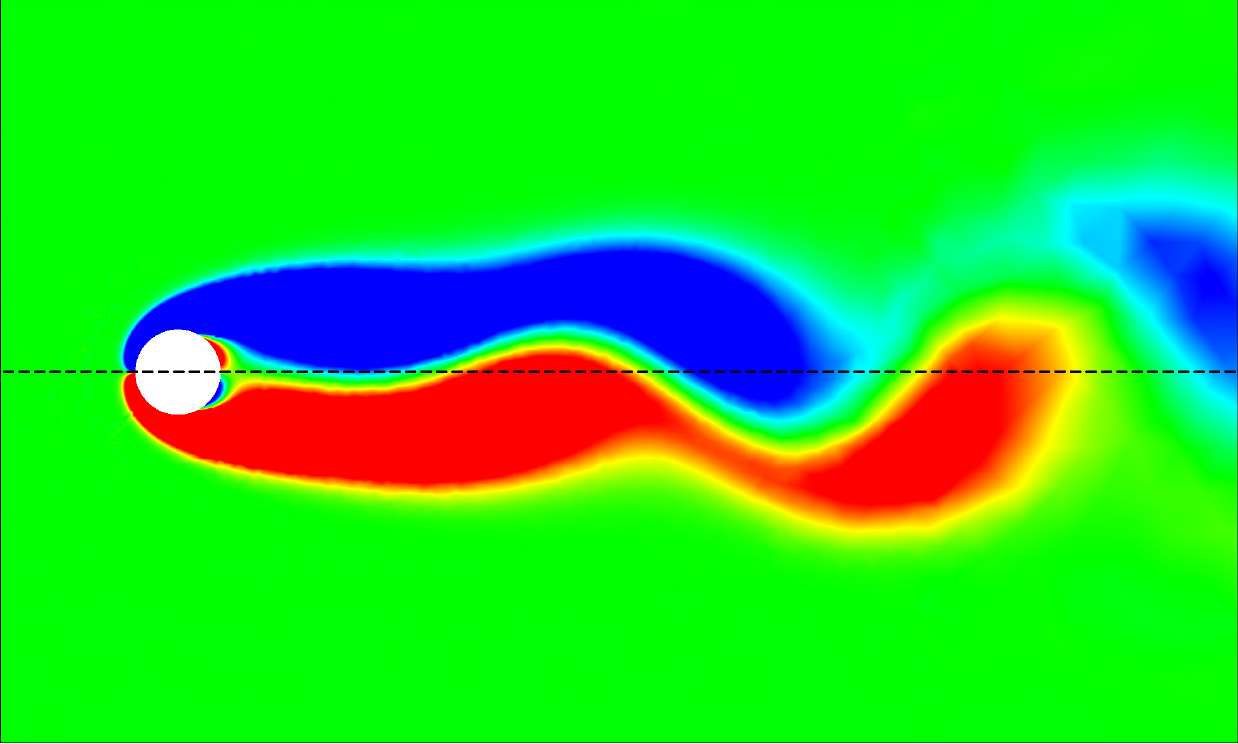}
\caption{}
\end{subfigure}
\begin{subfigure}{0.35\textwidth}
\centering
\includegraphics[width=1\linewidth]{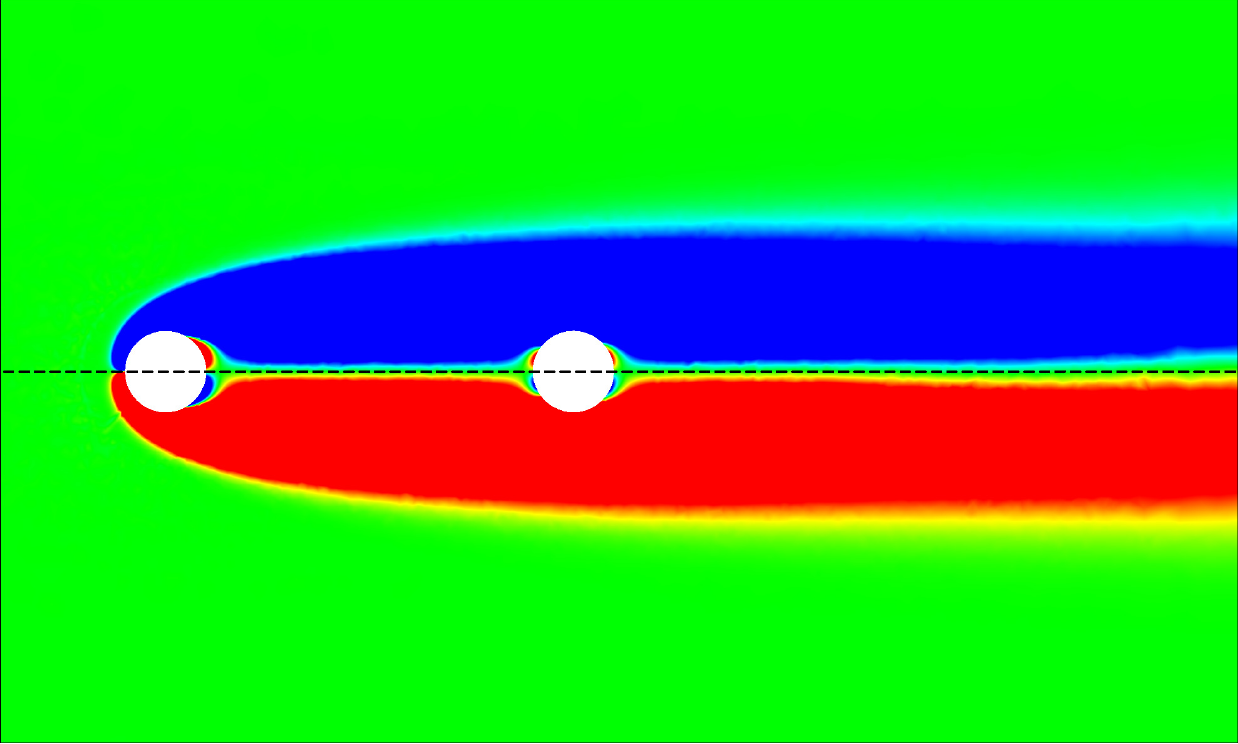}
\caption{}
\end{subfigure}\\ \bigskip
\begin{subfigure}{0.35\textwidth}
\centering
\includegraphics[width=1\linewidth]{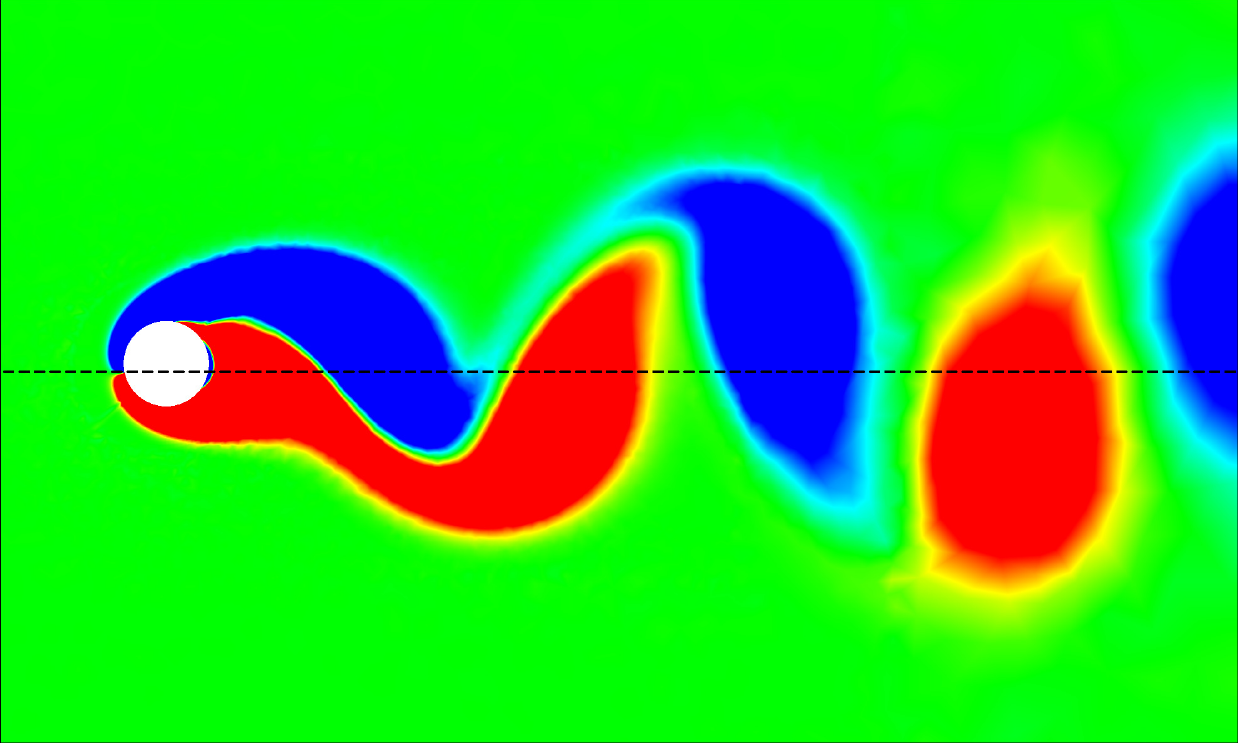}
\caption{}
\end{subfigure}
\begin{subfigure}{0.35\textwidth}
\centering
\includegraphics[width=1\linewidth]{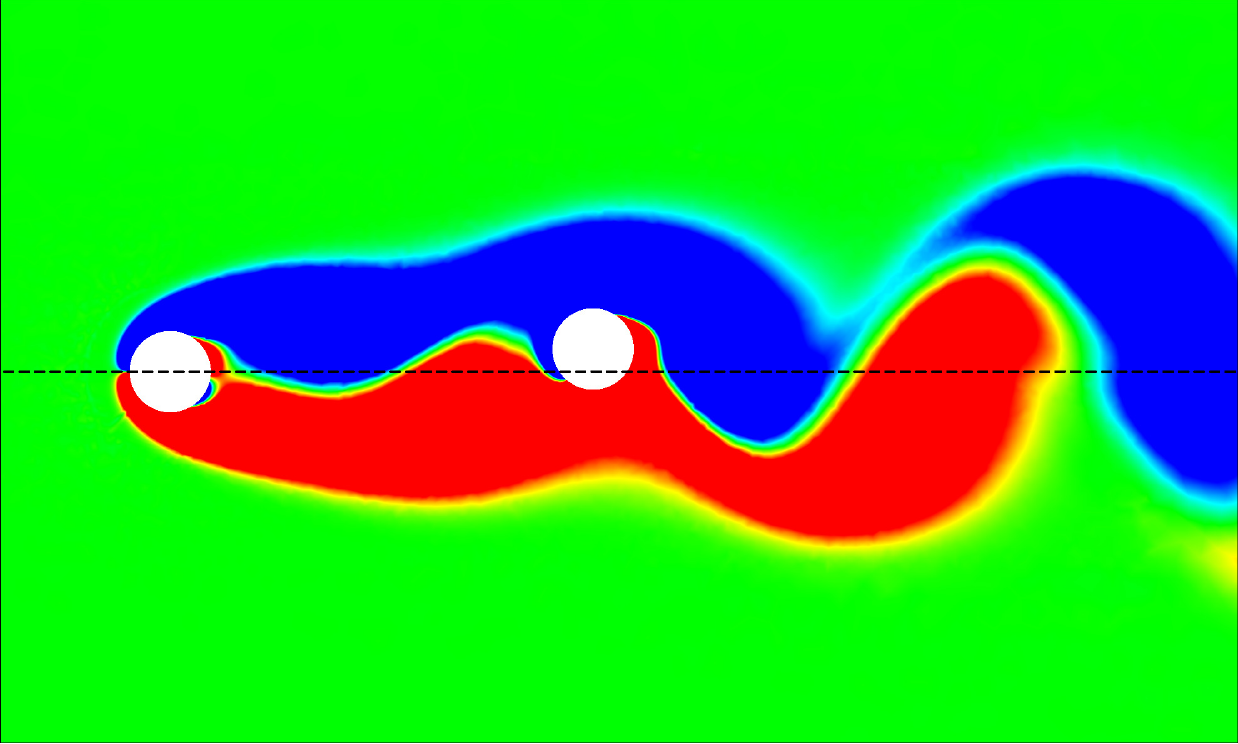}
\caption{}
\end{subfigure}\\ \bigskip
\begin{subfigure}{0.35\textwidth}
\centering
\includegraphics[width=1\linewidth]{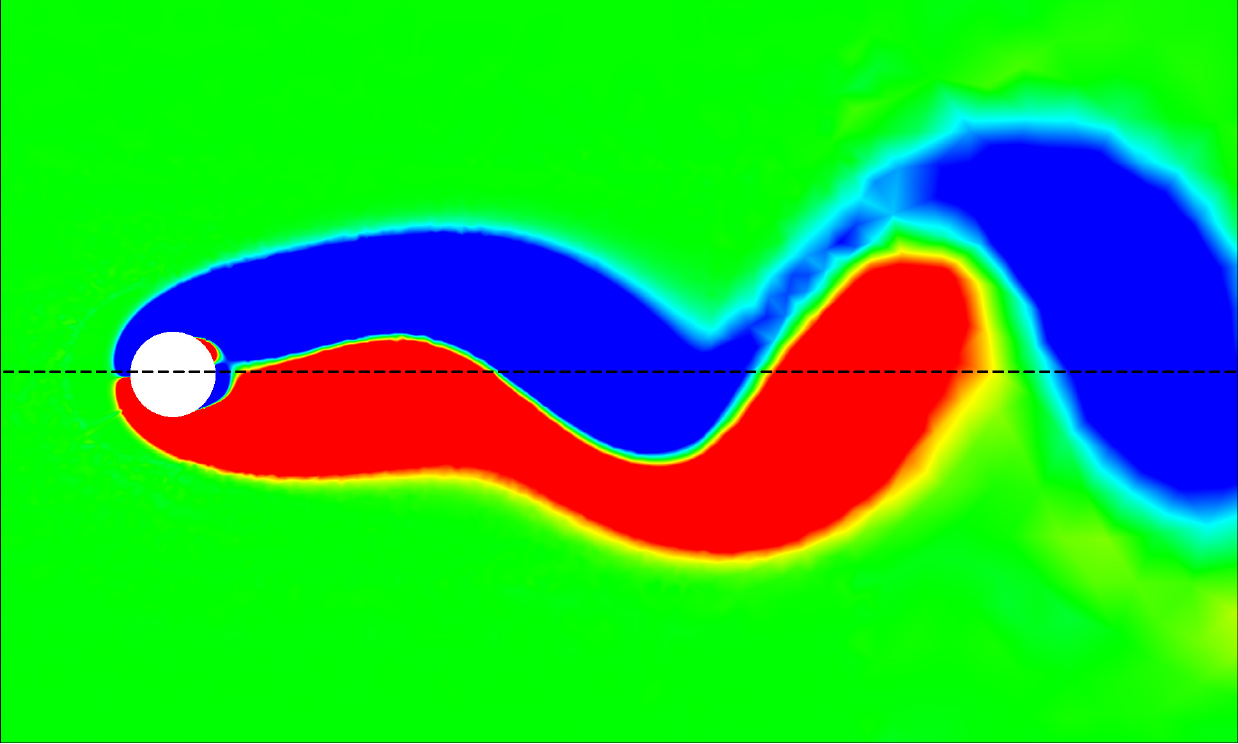}
\caption{}
\end{subfigure}
\begin{subfigure}{0.35\textwidth}
\centering
\includegraphics[width=1\linewidth]{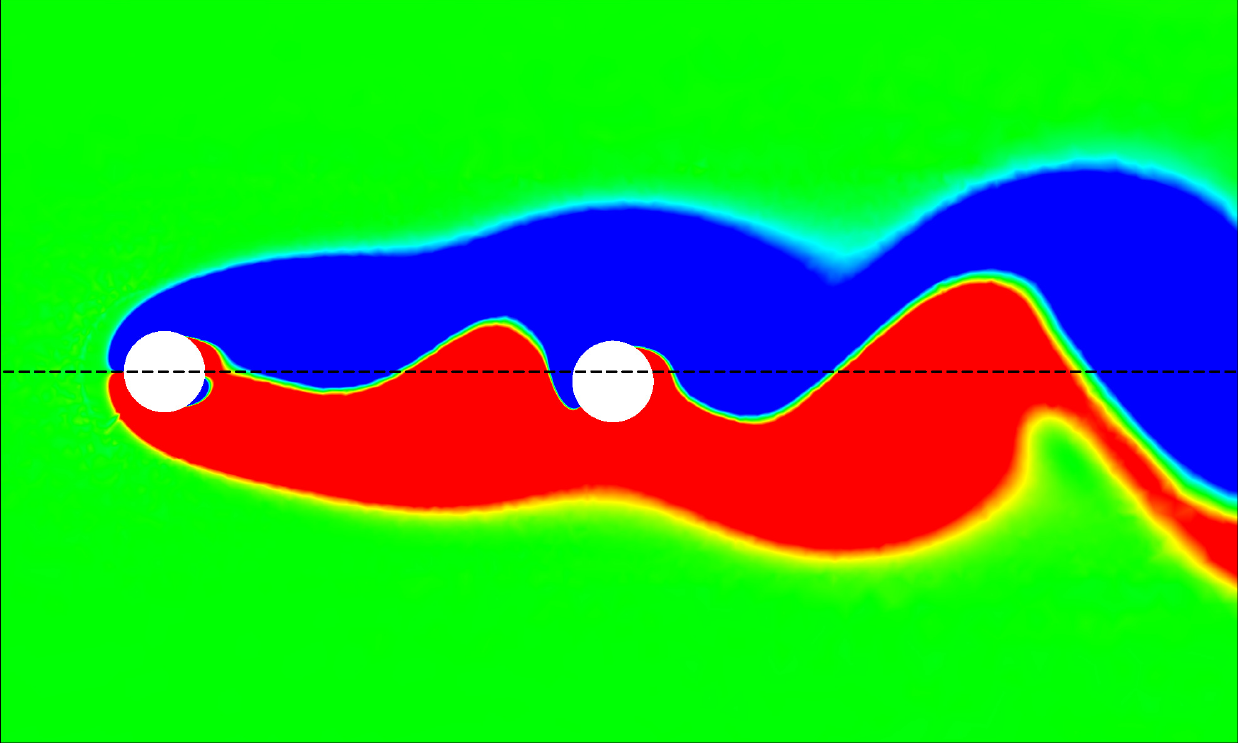}
\caption{}
\end{subfigure}
\caption{\label{fig:zVort-Tandem-Re60-comparison}Comparison of the $z$-plane slices of the $z$-vorticity contour at $z/L=0.5$ for isolated and tandem configurations at (a-b) $U^*=4$, (c-d) $U^*=7$, and (e-f) $U^*=14$. The results are gathered at $Re =60$ and $m^* = 1$.}
\end{figure}

\begin{figure}\centering
\begin{subfigure}{0.35\textwidth}
\centering
\includegraphics[width=1\linewidth]{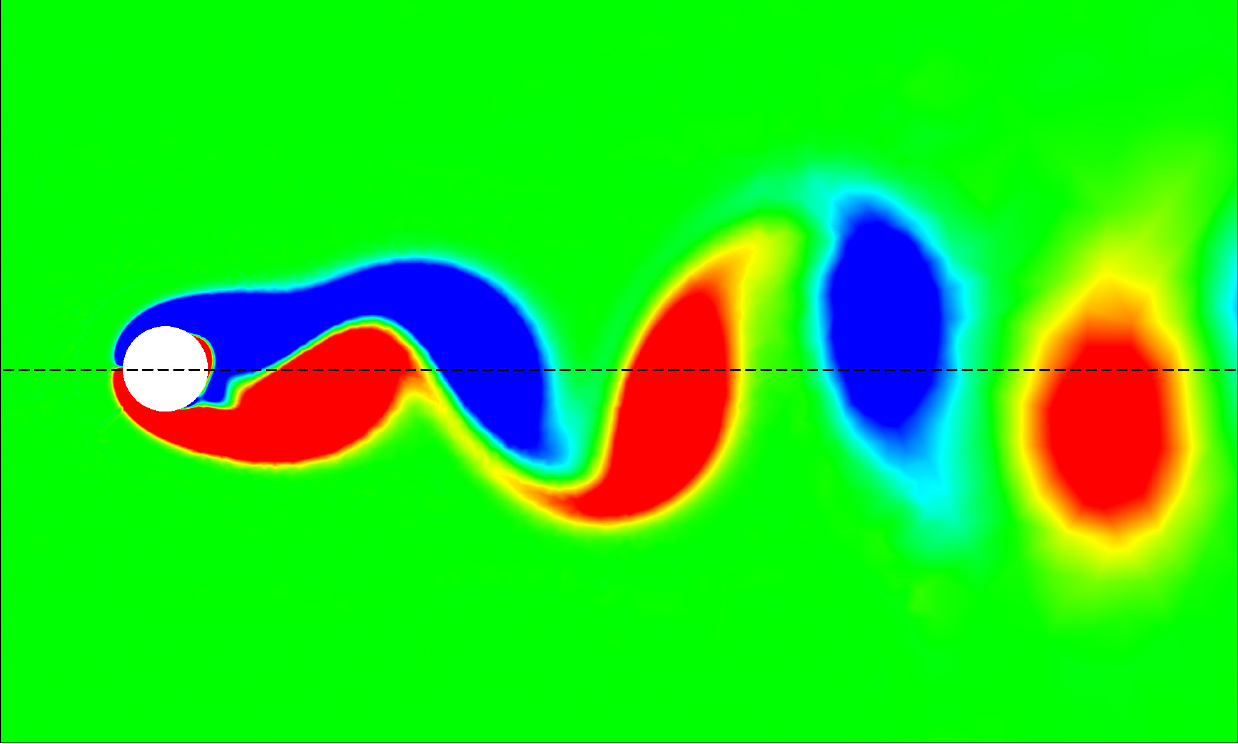}
\caption{}
\end{subfigure}
\begin{subfigure}{0.35\textwidth}
\centering
\includegraphics[width=1\linewidth]{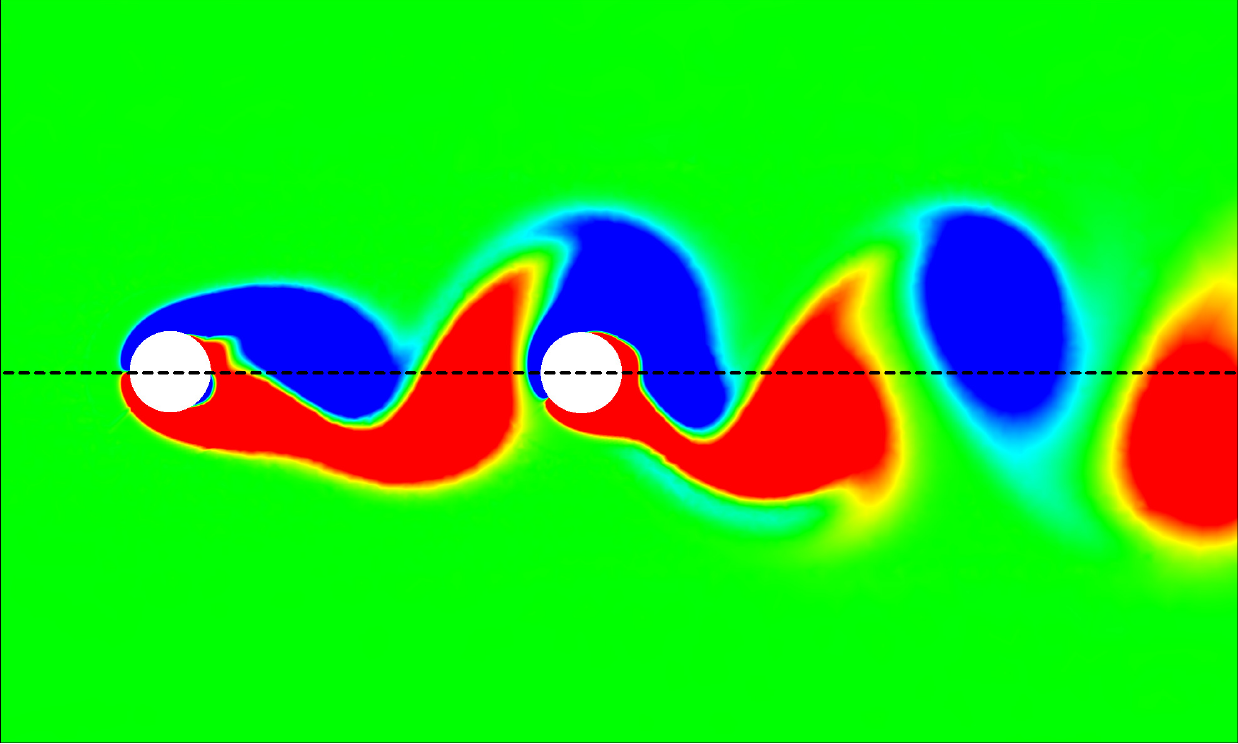}
\caption{}
\end{subfigure}\\ \bigskip
\begin{subfigure}{0.35\textwidth}
\centering
\includegraphics[width=1\linewidth]{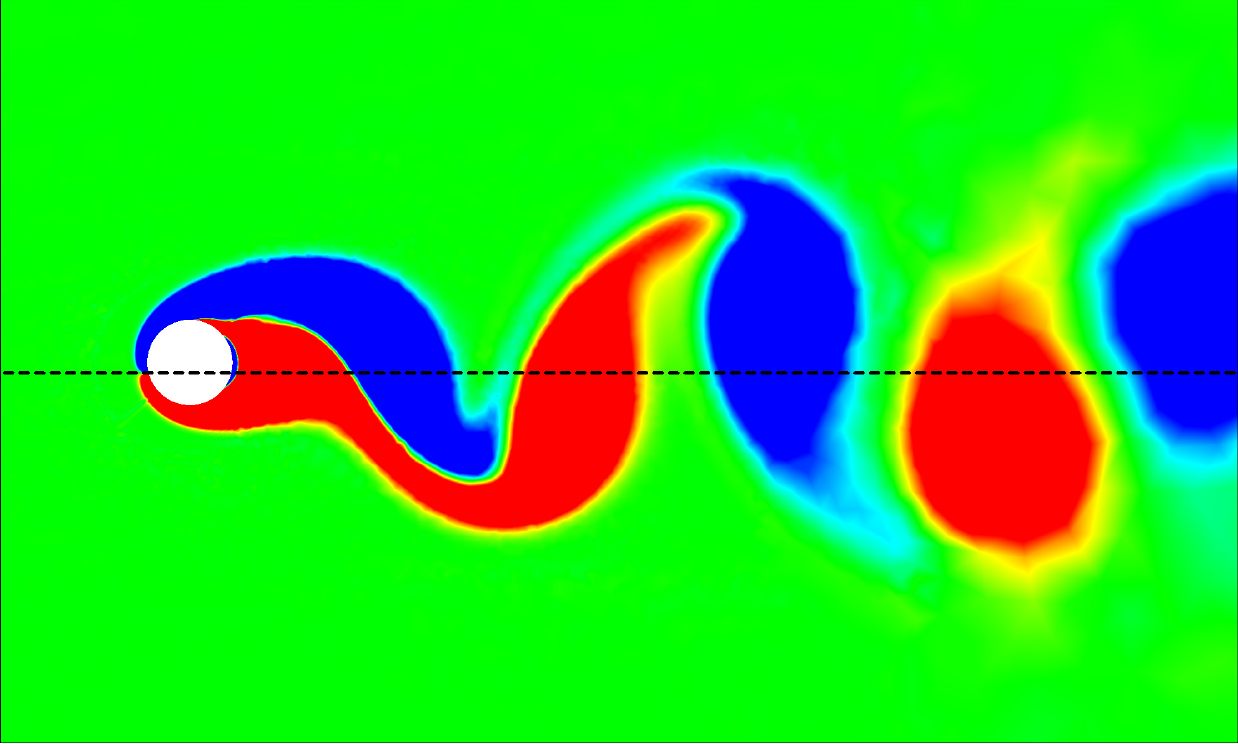}
\caption{}
\end{subfigure}
\begin{subfigure}{0.35\textwidth}
\centering
\includegraphics[width=1\linewidth]{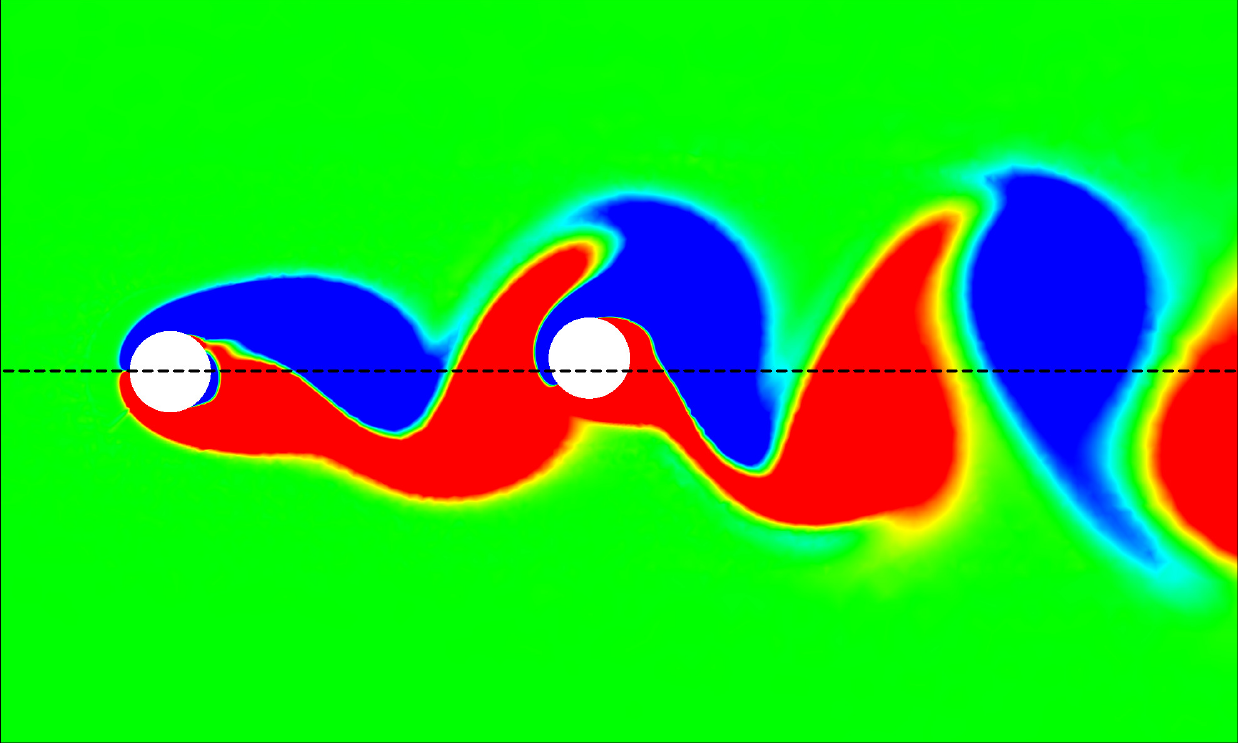}
\caption{}
\end{subfigure}\\ \bigskip
\begin{subfigure}{0.35\textwidth}
\centering
\includegraphics[width=1\linewidth]{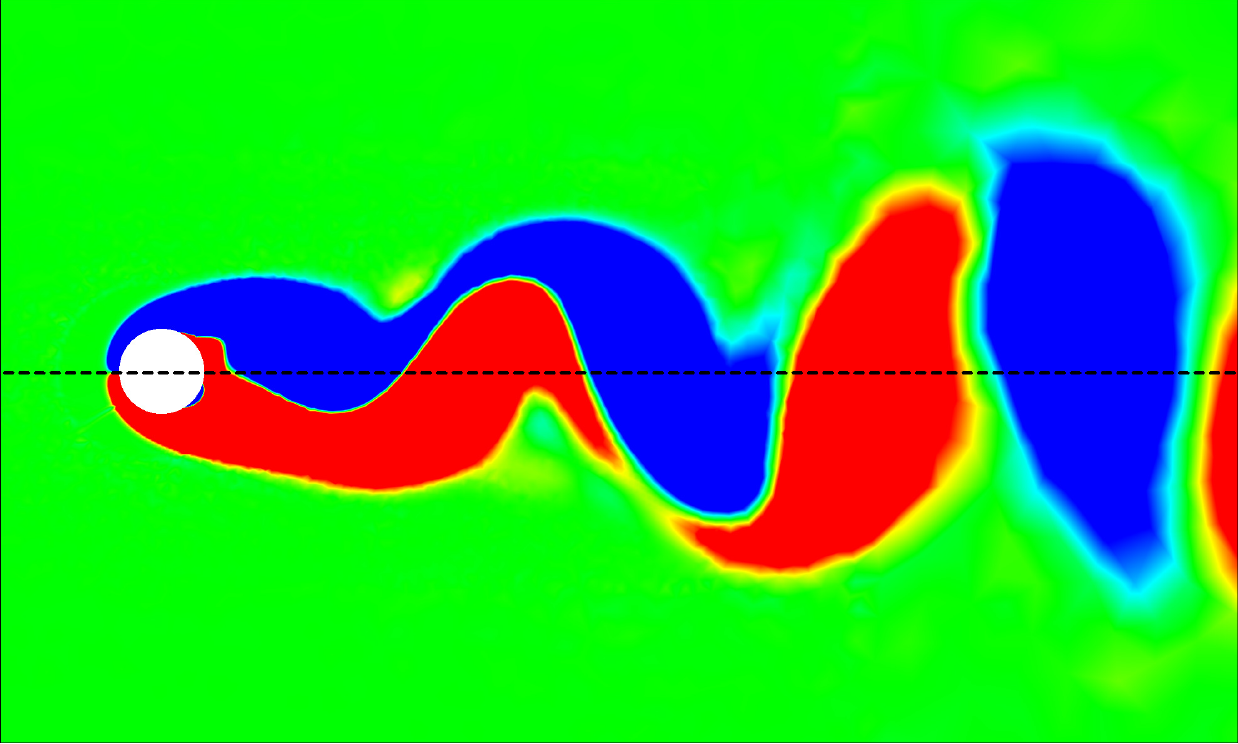}
\caption{}
\end{subfigure}
\begin{subfigure}{0.35\textwidth}
\centering
\includegraphics[width=1\linewidth]{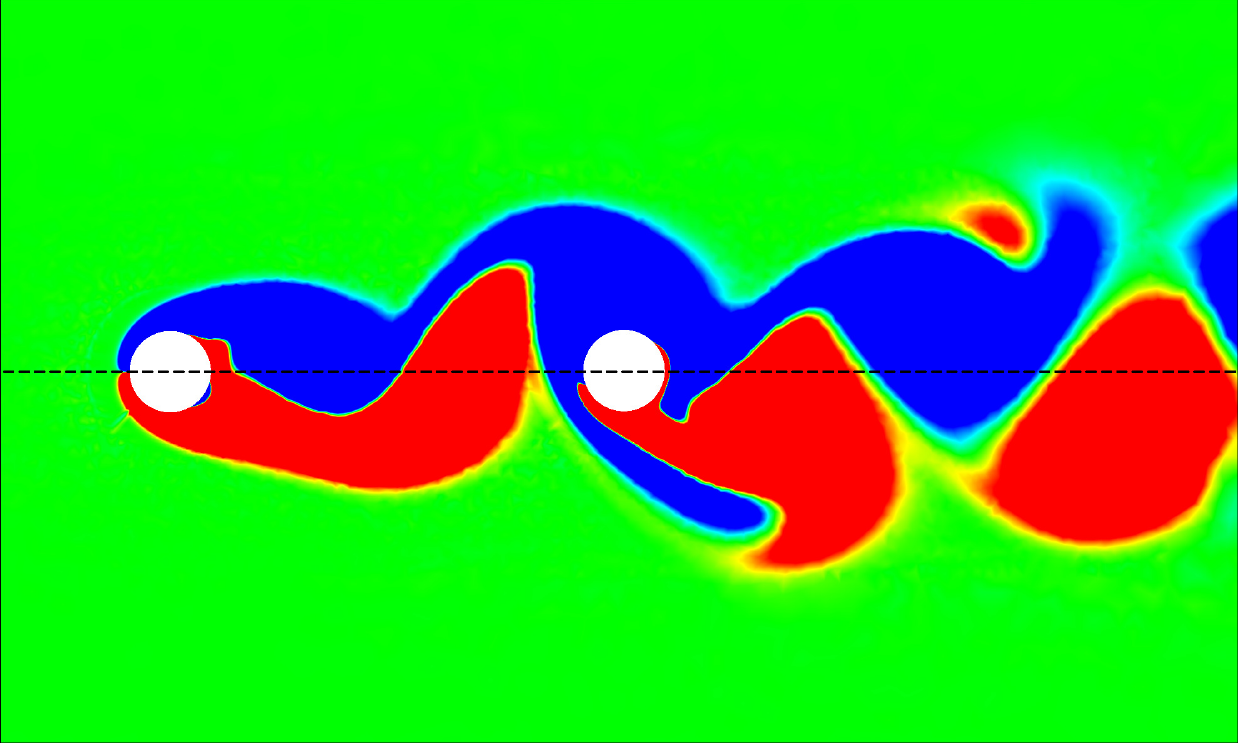}
\caption{}
\end{subfigure}
\caption{\label{fig:zVort-Tandem-Re100-comparison}Comparison of the $z$-plane slices of the $z$-vorticity contour at $z/L=0.5$ for isolated and tandem configurations at (a-b) $U^*=4$, (c-d) $U^*=7$, and (e-f) $U^*=14$. The results are gathered at $Re =100$ and $m^* = 1$.}
\end{figure}

To compare the wake dynamics between the isolated and tandem configurations, we have presented the \(z\)-plane slices of the \(z\)-vorticity contour at the mid-span of the cantilever at \(Re = 60\) for three distinct reduced velocities $U^{*} = 4$, 7, and 14 in Fig.~\ref{fig:zVort-Tandem-Re60-comparison}. At \( U^* = 4 \), the wake of the cantilever in the isolated configuration exhibits unsteady dynamics with clear vortex-shedding patterns, while the tandem configuration shows a nearly steady, streamlined flow with no distinct vortex shedding. The presence of the upstream cylinder delays vortex formation and roll-up, leading to a more stable flow regime compared to the isolated case. As the reduced velocity increases, we observe the onset of wake instability and asymmetry in the tandem arrangement. At \( U^{*} = 7 \) and 14, both configurations exhibit unsteady wakes with vortex-shedding patterns, as shown in Fig.~\ref{fig:zVort-Tandem-Re60-comparison}. 
A similar comparison between the wake dynamics of the isolated and tandem configurations is presented in Fig.~\ref{fig:zVort-Tandem-Re100-comparison} for \( U^{*} = 4 \), 7, and 14 at \( Re = 100 \). For all the studied \( U^* \) values, we observe unsteady vortex-shedding patterns in both isolated and tandem configurations. At \( Re = 100 \) and \( U^* = 4 \), the tandem configuration shows periodic vortex shedding, contrasting with the steady wake observed at \( Re = 60 \) at the same $U^*$. The higher Reynolds number amplifies flow instabilities, resulting in vortex shedding even at lower reduced velocities. As \( U^* \) increases, we see intensified vortex merging and splitting in the tandem arrangement, as shown in Fig.~\ref{fig:zVort-Tandem-Re100-comparison}.

\begin{figure*}
\centering
\includegraphics[width=0.65\linewidth]{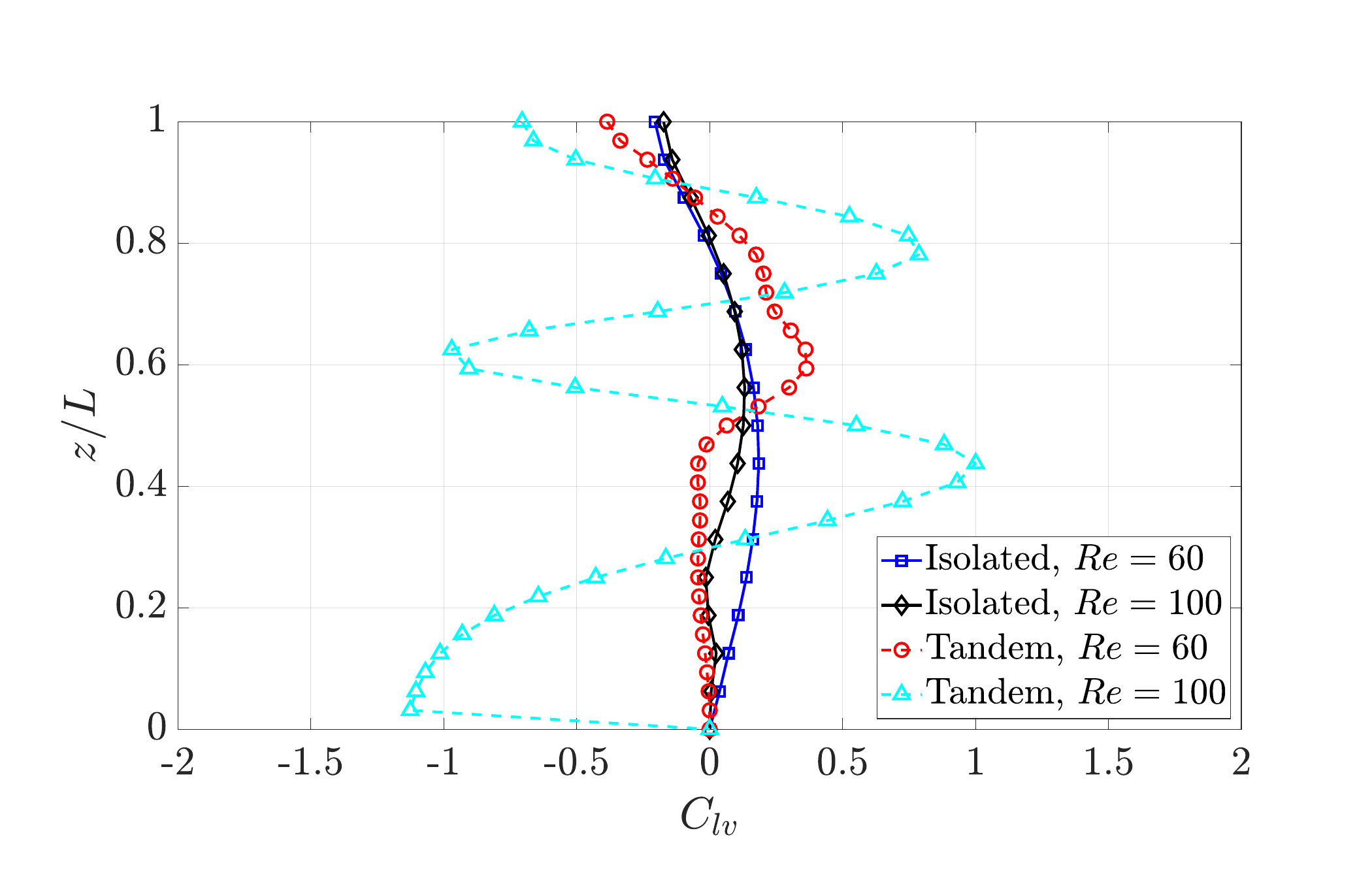}
\caption{Comparison between values of $C_{\mathrm{lv}}$ along the cantilever length in isolated versus tandem configurations at $Re=60$ and 100. The results are gathered at $m^* = 1$ and $U^* = 9$.}
\label{EnergyTransfer-scalogram-postcritical}
\end{figure*}

The presented results underscore the critical roles of both the Reynolds number \( Re \) and reduced velocity \( U^* \) in determining the wake dynamics in the tandem configuration. At lower values of \( Re \) and \( U^* \), such as in the tandem configuration at \( Re = 60 \) and \( U^* = 4 \), the wake remains nearly steady and symmetric, indicating minimal disturbances in the flow. Conversely, as \( Re \) and \( U^* \) increase, the flow becomes significantly more unstable. Higher \( Re \) values enhance flow instabilities, while increased \( U^* \) adds to the wake complexity by intensifying the vortex-body interactions within the vortex-shedding process. As a result, the combined influence of higher \( Re \) and \( U^* \) contribute to unsteady wake behavior in the tandem cylinder configuration.
The vortex-body interaction process could be further investigated by quantifying the \(C_{\mathrm{lv}}\) values in different flow regimes. For the isolated configuration at \(Re = 60\), the \(C_{\mathrm{lv}}\) values are found to be generally positive and close to zero along the entire span of the cantilever, as shown in Fig.~\ref{EnergyTransfer-scalogram-postcritical}. The relatively uniform distribution of $C_{\mathrm{lv}}$ indicates a stable hydrodynamic loading in phase with the cantilever’s transverse velocity. In the isolated configuration at \(Re = 100\), the $C_{\mathrm{lv}}$ values exhibit slightly more variation compared to \(Re = 60\). While the trend still centers around zero, with the values of $C_{\mathrm{lv}}$ being generally positive, there are more noticeable peaks and troughs along the cantilever's length. 
For the tandem configuration at \(Re = 60\), the $C_{\mathrm{lv}}$ values show a more distinct pattern compared to the isolated cases. There are noticeable peaks and troughs along the length of the cantilever, indicating significant in-phase hydrodynamic excitation. In the tandem configuration at \(Re = 100\), the $C_{\mathrm{lv}}$ values exhibit the most significant variations among all cases. 
The continuous introduction of upstream vortices, i.e., angular momentum, to the flexible cantilever, coupled with the vortex-induced motion of the cantilever, results in complex vortex-structure interactions. This interplay culminates in sustained vibrations of the cantilever, particularly at high reduced velocities within the WIV-dominated regime. With all these findings, we summarize the response amplitude of the cantilever as a function of $Re$ and $U^*$ in isolated and tandem configurations in the following subsection.
\subsection{Map of transverse oscillation amplitude}
\begin{figure*}
\begin{subfigure}{0.5\textwidth}
\centering
\includegraphics[width=1\linewidth]{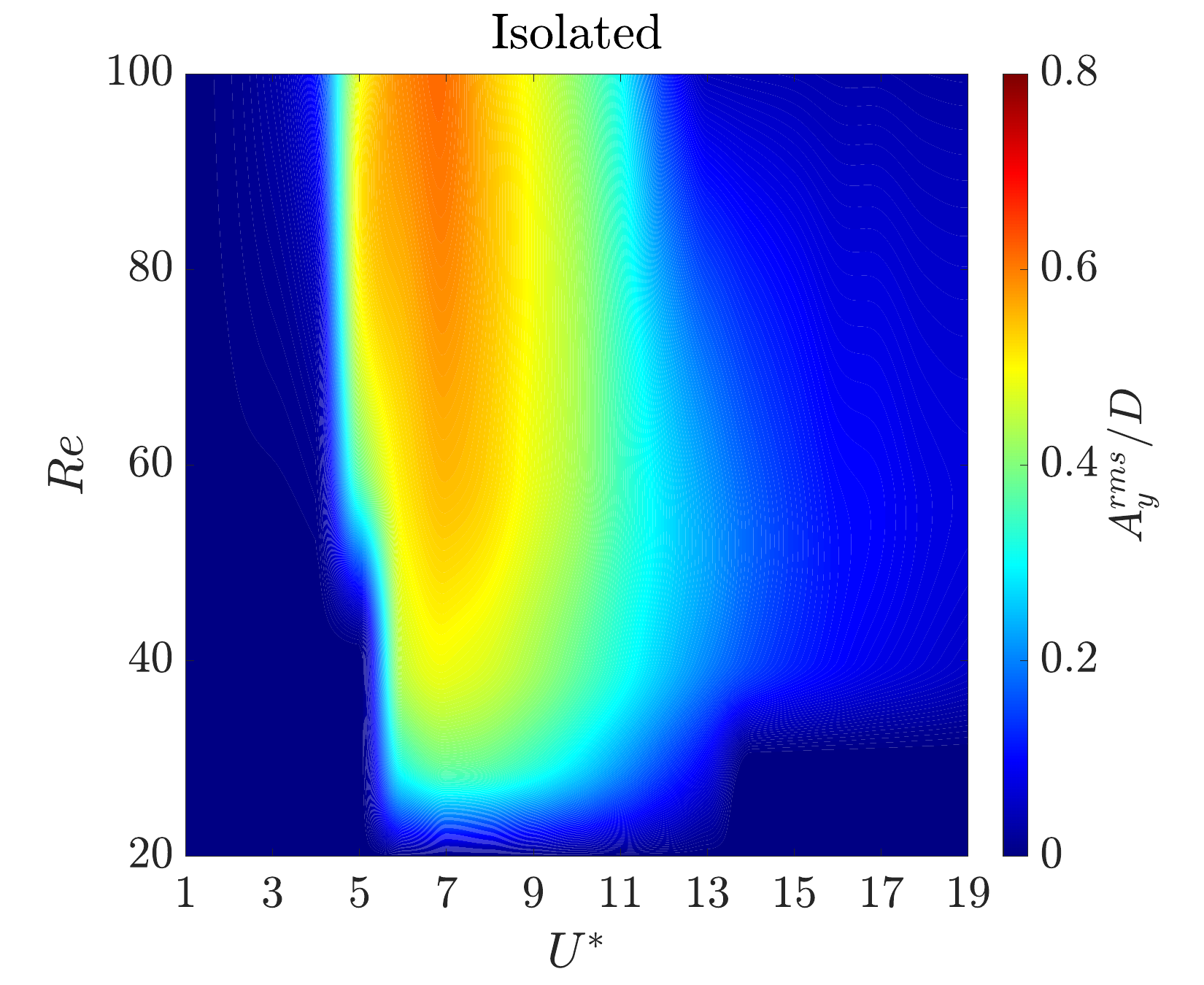}
\caption{}
\label{Ay-Cl}
\end{subfigure}%
\begin{subfigure}{0.5\textwidth}
\centering
\includegraphics[width=1\linewidth]{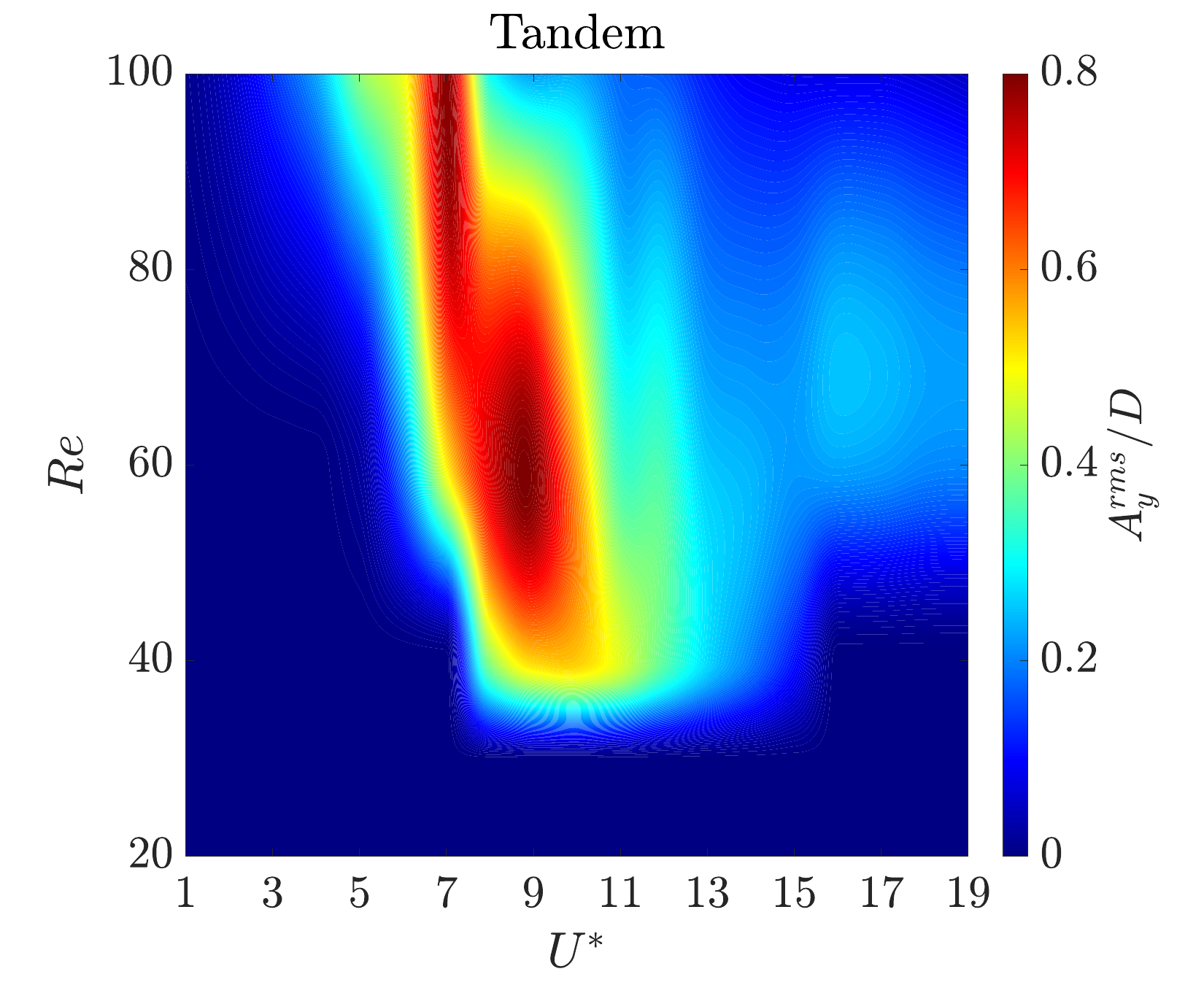}
\caption{}
\label{FFT-CLAy}
\end{subfigure}
\caption{\label{PhaseDiagram}Map of the flexible cantilever's transverse oscillation amplitude with respect to $Re$ and $U^*$ at $m^* = 1$ in (a) isolated and (b) tandem configurations. In the tandem configuration, the cylinders are spaced at a streamwise distance of $x_{0}=5D$.}
\end{figure*}

Figure~\ref{PhaseDiagram} illustrates the cantilever’s response characteristics, represented by the value of $A_{\mathrm{y}}^{rms}/D$, as a function of $Re$ and $U^*$ for both isolated and tandem configurations. In the isolated configuration, a distinct region of high amplitude transverse oscillation, peaking at around \(U^* \approx 7\), is observed. This peak region indicates a strong lock-in phenomenon in the isolated configuration. The upper bounds of \(A_{\mathrm{{y}}}^{rms}/D\) in this region reach up to approximately 0.60. The lower bounds of \(A_{\mathrm{y}}^{rms}/D\) remain close to zero across most regions of the contour, suggesting minimal transverse oscillations outside the lock-in regime. No sustained oscillations are present at $Re=20$ or lower in the isolated configuration.
In the tandem configuration, Fig.~\ref{PhaseDiagram}b reveals a region of high transverse oscillation amplitude, with notable differences compared to the isolated case. The peak region occurs approximately at \(U^* \in [7,9]\) for \(Re \in [60,100]\). The peak amplitude in the tandem configuration reaches values of \(A_{\mathrm{{y}}}^{rms}/D\approx0.80\). A critical difference in the tandem configuration is the extension of the high amplitude region to lower \(U^*\) values at higher Reynolds numbers. For Reynolds numbers close to \(Re = 100\), significant oscillations are observed for \(U^* < 5\), which contrasts with the isolated configuration where low \(U^*\) values generally correspond to minimal transverse vibrations. As shown in Fig.~\ref{PhaseDiagram}b, the tandem configuration demonstrates a broader range of \(U^*\) values with sustained oscillations, particularly at higher Reynolds numbers. No sustained oscillations are present for $Re\leq30$ in the tandem configuration, indicating more effective damping mechanisms between the cylinders within the subcritical $Re$ regime.
%
\section{\label{sec:conclusions} Conclusions}
In this paper, we utilized a high-fidelity numerical framework to investigate the coupled dynamics of a long flexible cylindrical cantilever in a tandem configuration. The dynamics of the cantilever were studied at a mass ratio \(m^*=1\) for reduced velocities \(U^* \in [1, 19]\) in both subcritical and post-critical regimes of \(Re\). The findings reveal distinct oscillatory responses and wake dynamics dependent on \(Re\) and \(U^*\). In the subcritical \(Re\) regime, the oscillatory responses were characterized by synchronization between the vortex shedding frequency and the cantilever’s transverse oscillation frequency. This lock-in behavior, observed within a specific \(U^*\) range, was shown to drive a significant transverse oscillatory response, highlighting the dominant influence of transverse fluid forces on the cantilever's dynamics. In the post-critical \(Re\) regime, the tandem configuration exhibited both single- and multi-frequency responses classified as vortex- and wake-induced vibrations. The power transfer analysis revealed a cyclic interaction between the fluid forces and the cantilever, with positive energy transfer contributing to the sustained oscillations. 
The wake dynamics were shown to differ notably between isolated and tandem configurations. The presence of the upstream cylinder in the tandem configuration was found to delay vortex formation behind the cantilever, resulting in an extended attached near-wake at the subcritical $Re$ regime. At higher \(Re\) values, the wake exhibited unsteady dynamics with pronounced vortex-shedding patterns in both isolated and tandem configurations. The systematic analysis presented in this study helps broaden the knowledge of FIVs in flexible cylindrical cantilevers and is relevant to the design and development of next-generation cantilever flow sensors. Future research will extend the investigations to multi-body FSI problems, involving multiple flexible structures in group arrangements. Additionally, we will explore the application of deep learning-based reduced order models (DL-ROMs) for spatio-temporal prediction of the dynamics, aiming to enable real-time predictive capabilities for engineering applications.

\begin{acknowledgments}
The authors would like to acknowledge the Natural Sciences and Engineering Research Council of Canada (NSERC) for funding the project. The research was enabled in part through computational resources and services provided by the Advanced Research Computing facility at the University of British Columbia (\url{https://arc.ubc.ca/}).
\end{acknowledgments}
%
\bibliography{main}
\end{document}